\newcommand{\planck}{{\em Planck}}
\newcommand{\herschel}{{\em Herschel}}

\newcommand{\blastpol}{BLASTPol}

\newcommand{\um}{$\mu$m}                            % Microns
\newcommand{\dpsi}{$S$}

\newcommand{\ppol}{$\ppol$}

\newcommand{\av}{$A_{\mathrm{V}}$}
\def\3he{$^3{\rm He}$}
\def\4he{$^4{\rm He}$}

\newcommand{\toast}{{\tt TOAST}}

\newcommand{\mickey}{IRAS\,08470$-$4243}
\newcommand{\velac}{Vela\,C}

\newcommand{\rcw}{RCW\,36}
\newcommand{\pdecN}{$p^{[N]}$}
\newcommand{\pdecdpsi}{$p^{[S]}$}
\newcommand{\ith}{$i$th}

\newcommand{\alpNerr}{$-$0.58\,$\pm$\,0.02}

\newcommand{\alpSerr}{$-$0.67\,$\pm$\,0.02}

\newcommand{\alpNdSerr}{$-$0.46\,$\pm$\,0.01}

\newcommand{\alpSdNerr}{$-$0.58\,$\pm$\,0.01}

\newcommand{\alpbNerr}{$-$0.45\,$\pm$\,0.01}
\newcommand{\alpbbNerr}{$-0.45\,\pm\,0.10$}

\newcommand{\alpbSerr}{$-$0.60\,$\pm$\,0.01}
\newcommand{\alpbbSerr}{$-$0.60\,$\pm$\,0.01}

\newcommand{\RsqpN}{0.30}
\newcommand{\RsqpS}{0.47}
\newcommand{\RsqpNdS}{0.35}
\newcommand{\RsqpSdN}{0.50}

\newcommand{\alpNerrRCW}{$-$0.78$\pm\,0.06$}

\newcommand{\betlTerr}{0.125$\pm\,0.005$}

\newcommand{\betT}{0.28}
\newcommand{\nVrcw}{143}
\newcommand{\nVisrfNT}{2235}

\newcommand{\nVboth}{2378}

\def\,{\thinspace}
\def\lsim{\mathrel{\raise .4ex\hbox{\rlap{$<$}\lower 1.2ex\hbox{$\sim$}}}}
\def\gsim{\mathrel{\raise .4ex\hbox{\rlap{$>$}\lower 1.2ex\hbox{$\sim$}}}}

\def\simprop{\mathrel{\raise .4ex\hbox{\rlap{$\propto$}\lower 1.2ex\hbox{$\sim$}}}}
\def\deg{\ifmmode^\circ\else$^\circ$\fi}
\def\pdeg{\ifmmode $\setbox0=\hbox{$^{\circ}$}\rlap{\hskip.11\wd0 .}$^{\circ}
          \else \setbox0=\hbox{$^{\circ}$}\rlap{\hskip.11\wd0 .}$^{\circ}$\fi}
\def\arcs{\ifmmode {^{\scriptstyle\prime\prime}}
          \else $^{\scriptstyle\prime\prime}$\fi}
\def\arcm{\ifmmode {^{\scriptstyle\prime}}
          \else $^{\scriptstyle\prime}$\fi}
\newdimen\sa  \newdimen\sb
\def\parcs{\sa=.07em \sb=.03em
     \ifmmode \hbox{\rlap{.}}^{\scriptstyle\prime\kern -\sb\prime}\hbox{\kern -\sa}
     \else \rlap{.}$^{\scriptstyle\prime\kern -\sb\prime}$\kern -\sa\fi}
\def\parcm{\sa=.08em \sb=.03em
     \ifmmode \hbox{\rlap{.}\kern\sa}^{\scriptstyle\prime}\hbox{\kern-\sb}
     \else \rlap{.}\kern\sa$^{\scriptstyle\prime}$\kern-\sb\fi}

\documentclass[iop,numberedappendix,twocolappendix]{emulateapj}

\usepackage[morefloats=7]{morefloats}
\usepackage[backref,breaklinks,colorlinks,citecolor=blue]{hyperref}
\usepackage[all]{hypcap}

\providecommand{\sorthelp}[1]{}
\newcommand{\lmfup}{} %updated from the original submitted version
\shorttitle{\blastpol~Polarimetry of the Vela C Molecular Cloud}
\shortauthors{Fissel et al.}

\begin{document}

\title{Balloon-Borne Submillimeter Polarimetry of the Vela C Molecular Cloud: Systematic Dependence of Polarization Fraction on Column Density and Local Polarization-Angle Dispersion}

\author{Laura M. Fissel\altaffilmark{1},
Peter A. R. Ade\altaffilmark{2},
Francesco E. Angil\`e\altaffilmark{3},
Peter Ashton\altaffilmark{1},
Steven J. Benton\altaffilmark{4,5},
Mark J. Devlin\altaffilmark{3},
Bradley Dober\altaffilmark{3}, 
Yasuo Fukui\altaffilmark{6},
Nicholas Galitzki\altaffilmark{3}, 
Natalie N. Gandilo\altaffilmark{7,8},
Jeffrey Klein\altaffilmark{3},
Andrei L. Korotkov\altaffilmark{9},
Zhi-Yun Li\altaffilmark{10},
Peter G. Martin\altaffilmark{11},
Tristan G. Matthews\altaffilmark{1},
Lorenzo Moncelsi\altaffilmark{12},
Fumitaka Nakamura\altaffilmark{13},
Calvin B. Netterfield\altaffilmark{4,7},
Giles Novak\altaffilmark{1},
Enzo Pascale\altaffilmark{2},
Fr{\'e}d{\'e}rick Poidevin\altaffilmark{14,15},
Fabio P. Santos\altaffilmark{1},
Giorgio Savini\altaffilmark{16},
Douglas Scott\altaffilmark{17},
Jamil A. Shariff\altaffilmark{7,18},
Juan Diego Soler\altaffilmark{19},
Nicholas E. Thomas\altaffilmark{20},
Carole E. Tucker\altaffilmark{2},
Gregory S. Tucker\altaffilmark{9},
Derek Ward-Thompson\altaffilmark{21}} 

\altaffiltext{1}{Center for Interdisciplinary Exploration and Research in Astrophysics (CIERA) and Department\ of Physics \& Astronomy, Northwestern University, 2145 Sheridan Road, Evanston, IL 60208, U.S.A.}
\altaffiltext{2}{Cardiff University, School of Physics \& Astronomy, Queens Buildings, The Parade, Cardiff, CF24 3AA, U.K.} 
\altaffiltext{3}{Department of Physics \& Astronomy, University of Pennsylvania, 209 South 33rd Street, Philadelphia, PA, 19104, U.S.A.} 
\altaffiltext{4}{Department of Physics, University of Toronto, 60 St. George Street Toronto, ON M5S 1A7, Canada}
\altaffiltext{5}{Department of Physics, Princeton University, Jadwin Hall, Princeton, NJ 08544, U.S.A.}
\altaffiltext{6}{Department of Physics and Astrophysics, Nagoya University, Nagoya 464-8602, Japan}
\altaffiltext{7}{Department of Astronomy \& Astrophysics, University of Toronto, 50 St. George Street Toronto, ON M5S 3H4, Canada}
\altaffiltext{8}{{\lmfup Department of Physics and Astronomy, Johns Hopkins University, 3701 San Martin Drive, Baltimore, Maryland, U.S.A.}}
\altaffiltext{9}{Department of Physics, Brown University, 182 Hope Street, Providence, RI, 02912, U.S.A.}
\altaffiltext{10}{Department of Astronomy, University of Virginia, 530 McCormick Rd, Charlottesville, VA 22904, U.S.A.}
\altaffiltext{11}{CITA, University of Toronto, 60 St. George St., Toronto, ON M5S 3H8, Canada}
\altaffiltext{12}{California Institute of Technology, 1200 E. California Blvd., Pasadena, CA, 91125, U.S.A.}
\altaffiltext{13}{National Astronomical Observatory, Mitaka, Tokyo 181-8588, Japan}
\altaffiltext{14}{Instituto de Astrofísica de Canarias, E-38200 La Laguna, Tenerife, Spain}
\altaffiltext{15}{Universidad de La Laguna, Dept. Astrof\'{i}sica, E-38206 La Laguna, Tenerife, Spain}
\altaffiltext{16}{Department of Physics \& Astronomy, University College London, Gower Street, London, WC1E 6BT, U.K.}
\altaffiltext{17}{Department of Physics \& Astronomy, University of British Columbia, 6224 Agricultural Road, Vancouver, BC V6T 1Z1, Canada}
\altaffiltext{18}{{\lmfup Department of Physics, Case Western Reserve University, 2076 Adelbert Road, Cleveland Ohio, 44106-7079, U.S.A.}}
\altaffiltext{19}{Institute d'Astrophysique Spatiale, CNRS (UMR8617) Universit\'{e} Paris-Sud 11, B\^{a}timent 121, Orsay, France}
\altaffiltext{20}{NASA/Goddard Space Flight Center, Greenbelt , MD 20771, U.S.A.}
\altaffiltext{21}{Jeremiah Horrocks Institute, University of Central Lancashire, PR1 2HE, U.K.}

\begin{abstract}
We present results for Vela C obtained during the 2012 flight of the Balloon-borne Large Aperture Submillimeter Telescope for Polarimetry (BLASTPol).  We mapped polarized intensity across almost the entire extent of this giant molecular cloud, in bands centered at 250, 350, and 500\,\um.  In this initial paper, we show our 500\,\um~data smoothed to a resolution of 2\farcm5 (approximately 0.5\,pc).  We show that the mean level of the fractional polarization $p$~and most of its spatial variations can be accounted for using an empirical three-parameter power-law fit, $p\,\lmfup{\propto\,N^{-0.45}} S^{-0.60}$, where $N$~is the hydrogen column density and $S$~is the polarization-angle dispersion on 0.5\,pc scales.  The decrease of $p$~with increasing $S$~is expected because changes in the magnetic field direction within the cloud volume sampled by each measurement will lead to cancellation of polarization signals.  The decrease of $p$~with increasing $N$~might be caused by the same effect, if magnetic field disorder increases for high column density sightlines.  Alternatively, the intrinsic polarization efficiency of the dust grain population might be lower for material along higher density sightlines.  We find no significant correlation between $N$~and $S$.  Comparison of observed submillimeter polarization maps with synthetic polarization maps derived from numerical simulations provides a promising method for testing star formation theories.  Realistic simulations should allow for the possibility of variable intrinsic polarization efficiency.  The measured levels of correlation among $p$, $N$, and $S$~provide points of comparison between observations and simulations.  
\end{abstract}

\keywords{instrumentation: polarimeters, ISM: dust, extinction, ISM: magnetic fields, ISM: individual objects (Vela C), stars: formation, techniques: polarimetric}

\section{Introduction}\label{sect:intro}
The Balloon-borne Large Aperture Submillimeter Telescope for Polarimetry (BLASTPol; \citealt{galitzki_2014}) is sensitive to magnetic field structure ranging from scales of entire giant molecular clouds (GMCs) down to cores (for nearby clouds).  In this paper we present a very sensitive survey of the star-forming region Vela C from the 2012 flight of BLASTPol.  Our goal is to quantify the dependence of polarization fraction on column density, temperature, and local magnetic field disorder, in order to provide empirical formulae that can be used to test numerical simulations of molecular clouds. These observations are timely because the role that magnetic fields play in the formation of molecular clouds and their substructures persists as an outstanding question in the understanding of the detailed mechanics of star formation \citep{mckee_2007}.  Strong magnetic fields that are well coupled to the gas can inhibit or slow down gravitational collapse of gas in the direction perpendicular to the cloud magnetic field lines \citep{mous_1999}.  This in turn can contribute to the low observed star formation efficiency seen in molecular clouds. 
Numerical simulations of molecular clouds show that magnetized clouds differ from unmagnetized clouds in cloud density contrasts \citep{kowal_2007} and in star formation efficiency  \citep{myers_2014}.  However, obtaining detailed measurements of magnetic fields in molecular clouds over a wide range of relevant spatial and density scales remains challenging.

Zeeman splitting in molecular lines can be used to measure the component of magnetic field strength parallel to the line of sight directly \citep{crutcher_2012}. However this technique is challenging as the Doppler broadening of molecular lines is generally much larger than the Zeeman splitting.  After many years of careful observations there are now several dozen detections of Zeeman splitting in molecular lines that trace dense gas \citep{crutcher_2012}.

An alternative method for studying magnetic fields in molecular clouds is to use polarization maps to infer the orientation of the magnetic field projected on the plane of the sky ($\Phi$).
Dust grains are believed to align with their long axes on average perpendicular to the local magnetic field (see \citealt{lazarian_2007} for a review).  
Current evidence suggests that radiative torques (RATs) from 
anisotropic radiation fields might be the dominant alignment mechanism \citep{lazarian_hoang_2007,andersson_2015}.  Optical and near-IR light from stars that passes through a foreground cloud of aligned dust grains becomes polarized parallel to $\Phi$. 
This method has long been used to study the magnetic field orientation in the diffuse ISM \citep{hall_1949,hiltner_1949,heiles_2000}, but is not easily applicable for high extinction cloud sightlines.  However, dust grains emit radiation preferentially polarized parallel to their long axes, so that the resulting far-infrared/submillimeter thermal emission is polarized orthogonal to $\Phi$~\citep{hildebrand_1988}.  The emission is generally optically thin.

The fraction of dust emission that is polarized ($p$), does not give any direct 
estimate of the magnetic field strength.  However, it can encode information about the dust grain {\lmfup shape and alignment efficiency}, angle of the
 field with respect to the line of sight, and changes in field direction. 
\cite{hildebrand_1988} reviews the factors that affect $p$~for optically-thin thermal emission from a population of grains.  First, consider the case of perfect spinning alignment of an ensemble of identical grains in a uniform magnetic field oriented orthogonally to the line of sight ($\gamma\,=\,0$\deg).  In this case $p$~will be determined by the grains' optical constants and shape (e.g., ratio of axes for the case of oblate spheroids).  Next, if the grain spin axes are not all exactly parallel to the ambient field, the polarization will be reduced by what is known as the Rayleigh reduction factor \citep{greenberg_1968,lazarian_2007}.  
For this paper we define the ``intrinsic polarization efficiency''
as the polarization $p$~of the emission from such an ensemble of imperfectly aligned grains.  
The measured polarization fraction can be less than this intrinsic polarization efficiency if there are variations in magnetic field direction within the conical volume being sampled by an observation.  Finally, for arbitrary values of $\gamma$, the polarization is proportional to $\cos^2(\gamma)$.

Comparisons between statistical properties of observed polarization maps and synthetic observations of 3-D numerical models of star formation are a promising method for constraining molecular cloud physics.
Examples include \cite{fg_2008} as well as histograms of relative orientations (HRO, \citealt{soler_2013,planck2015-XXXV}).  If the intrinsic polarization efficiency varies within the cloud, then measurements of the inferred magnetic field orientation will be weighted toward the field orientation in regions along the line of sight where the intrinsic polarization efficiency is high.  
To use polarization observations to constrain the structure of the magnetic field in star-forming clouds it is therefore important to understand how the intrinsic polarization efficiency varies within molecular clouds.

The Vela\,C GMC was discovered by \cite{murphy_1991} via CO observations of a larger scale structure known as the Vela Molecular Ridge.  Vela\,C was later observed in the submillimeter by \cite{netterfield_2009} and was found to be a cool molecular cloud in an early evolutionary state.
At a distance of \,700$\,\pm$\,200\,pc \citep{liseau_1992}, the cloud subtends 3\deg\ on the sky (35\,pc), and contains a large quantity of dense gas ($M\,\approx\,5\,\times\,10^4$~$M_{\sun}$~as traced by C$^{18}$O 1-0 observations from \citealt{yamaguchi_1999}).  A \herschel\footnote{\herschel~is an ESA space observatory with science instruments 
provided by European-led Principal Investigator consortia and with important participation 
from NASA.}~survey of Vela C by \cite{hill_2011} showed that the cloud could be divided 
into five subregions at an \av\,=\,7\,mag threshold as shown in Figure \ref{fig:ref_regs}.  These subregions show a range of cloud substructures, 
from the apparently cold network of overlapping filaments in the South-Nest subregion, to the high mass Centre-Ridge, which contains a compact \ion{H}{2} region, \rcw.  

This paper presents an overview of the \blastpol~500\,\um~maps of Vela\,C from the 2012 flight.  
\blastpol~polarization data at 250\,\um~and 350\,\um~{\lmfup are discussed in a separate paper} on the polarization spectrum of Vela\,C \citep{gandilo_2016}.
In Section 
\ref{sect:obs} we describe the \blastpol~telescope and 
\blastpol~2012 science flight.  Section \ref{sect:data_analysis}~gives an overview of the 
data analysis pipeline {\lmfup and} Section \ref{sect:maps} 
presents the \blastpol~500\,\um~polarization maps.  For comparison with the \blastpol~polarization data we used spectral energy distribution fits to the well-calibrated, higher resolution \herschel~SPIRE {\lmfup and PACS} maps to produce maps of Vela\,C column density ($N$) and dust temperature ($T$) 
 as described in Section \ref{sect:n_and_t_maps}. We then examine the correlations 
 between the polarization fraction $p$ and $N$~and $T$~in Section \ref{sect:p_N_T}, and develop a two-variable power-law model of $p$~as a function of $N$ and the local polarization-angle dispersion $S$ in Section \ref{sect:p_N_dpsi}.  Finally, in Section \ref{sect:discussion},
 we discuss the implications of our power-law model and we place a rough upper 
limit on the degree to which reduced intrinsic polarization efficiency at high $N$~might bias our polarization measurements toward lower density cloud regions.  Our findings are 
  summarized in Section \ref{sect:conclusions}.  

\section{Observations}\label{sect:obs}
\begin{figure*}
\epsscale{.80}
\plotone{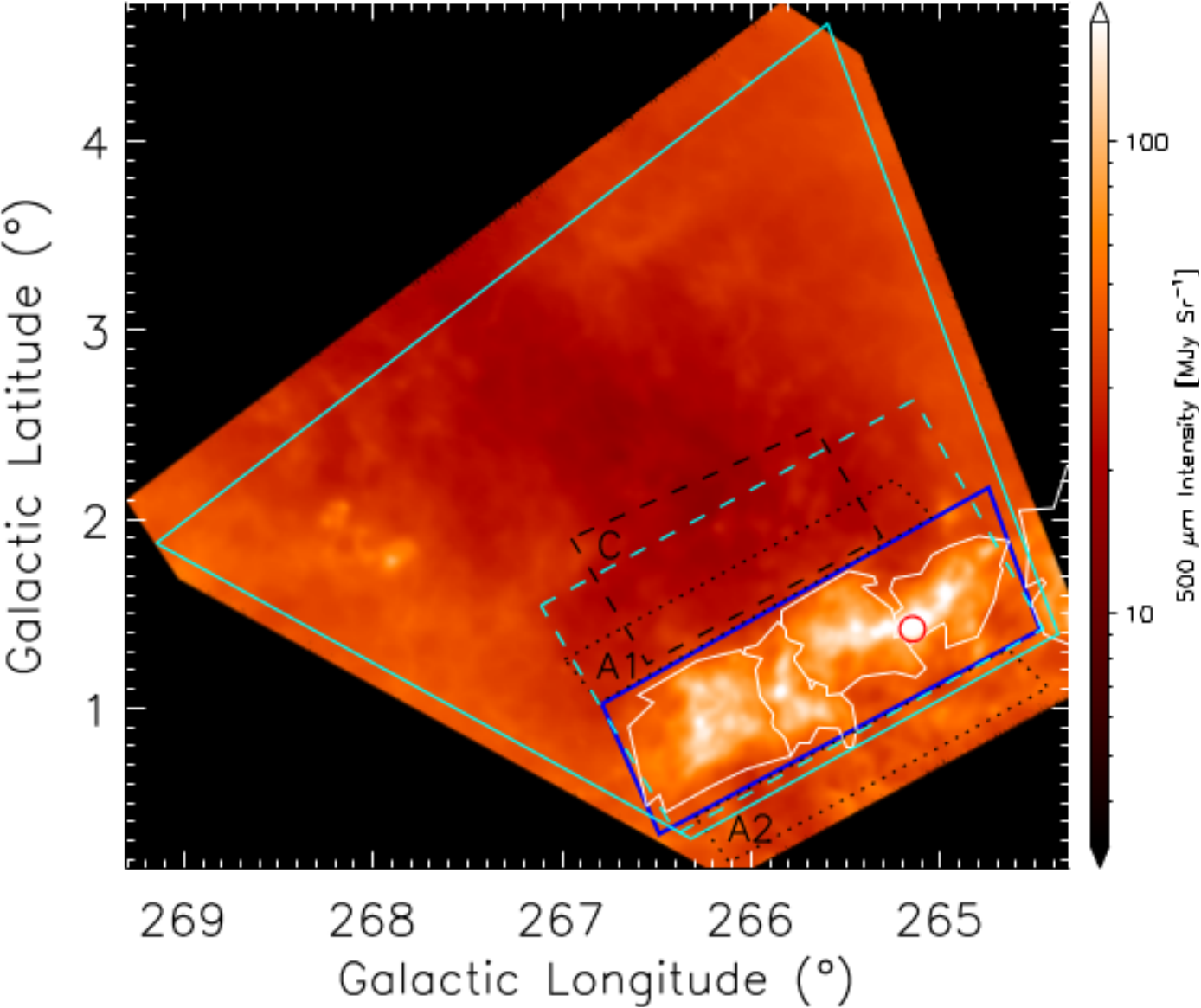}
\caption{\blastpol~500\,\um~$I$~map of Vela\,C and surroundings.  Cyan lines show the boundaries of the two raster scan regions used to make the maps in this paper: the region marked with cyan dashes was observed for 43 hours, while the solid cyan lines show a larger region covering Vela\,C and surrounding diffuse emission, which was observed for 11 hours in total. Also shown are the locations of the regions used in the diffuse emission subtraction for the BLASTPol $I$, $Q$, and $U$~maps as described in Section \ref{sect:ref_reg}.  The region labeled {\em C} is 
used in the ``conservative'' diffuse emission subtraction method.  The ``aggressive'' method used the two regions labelled {\em A1} and {\em A2} . Cloud subregions defined by \cite{hill_2011} are indicated in white contours.  The region outlined in blue shows the ``validity region'' where the null tests discussed in Section \ref{sect:null_tests} were passed, and where both diffuse emission subtraction methods discussed in Section \ref{sect:ref_reg} are valid; only
polarization measurements within this validity region are used for science analysis in later sections.  The red circle shows the area near \rcw~excluded from our polarization analysis because of large Stokes $I$, $Q$, and $U$~null test residuals.
\label{fig:ref_regs}}
\end{figure*}

\blastpol\ is a high altitude balloon-borne polarimeter that utilizes a 1.8-m diameter aluminium parabolic primary mirror, and a 40-cm aluminum 
secondary mirror.  Incoming light is directed onto a series of reimaging optics cooled to 1\,K 
in a liquid nitrogen-helium cryostat \citep{galitzki_2014}. A series of 
dichroic filters direct light onto focal-plane arrays of bolometers (cooled below 300\,mK),  
which are similar to those used by \herschel~SPIRE \citep{grif02,grif03}.  

The use of dichroic filters allows \blastpol\ to
observe simultaneously in three frequency bands centered at 250, 350, and 500\,\um.
Unlike ground based telescopes it is not restricted to observe through narrow windows in the atmospheric transmission spectrum.  Instead, \blastpol~observes in three wide frequency bands ($\Delta f/f\,\simeq\,30\%$), which bracket the peak of 10$-$20\,K thermal dust emission.  
A metal mesh polarizing 
grid is mounted in front of each detector array.  The polarizing grid is
patterned so that each adjacent detector samples only the vertical or horizontal 
component of the incoming radiation.  In this way a single Stokes parameter ($Q$~or $U$) can be measured in 
the time it takes light from a source to move from one bolometer to an adjacent bolometer ($<$1 second for typical scan speeds).
A sapphire achromatic half-wave plate (hereafter HWP) provides additional 
polarization modulation \citep{moncelsi_2014}.
A detailed description of the instrument and summary of the observations will appear in a forthcoming publication (F. Angil{\`e} et al.\ 2016, in preparation).\footnote{See also \cite{matthews_2014} for a description of the 2010 BLASTPol flight.}
  The present paper refers only to observations of Vela C and 
surrounding regions made during the 2012 flight.

BLASTPol was launched on 26 December 2012.  The payload rose to an average altitude of 
38.5\,km above sea level and began science operations, taking data until cryogens 
were depleted 12 days and 12 hours after launch.  Our selection of target molecular clouds was informed by target distance, visibility from Antarctica, and cloud brightness.  The nearby GMC Vela\,C was our highest priority target.  The observations discussed in this paper include two types of {\lmfup scans} as shown in Figure \ref{fig:ref_regs} (cyan lines).  Most of the integration time (43 hours) was used to map a ``deep" (3.1 deg$^2$) quadrilateral region covering four of the five cloud subregions defined by \cite{hill_2011}.\footnote{The North region, as defined by \cite{hill_2011}, has a significant spatial offset from the other four subregions and so was not included in our deep scan region. }  A further 11 hours were spent mapping a larger ($\sim$10\,deg$^2$) area that includes significant regions of low dust column where $A_V\,\sim\,1$ according to extinction maps from \cite{dobashi_2005}.   The larger region was observed to reconstruct the map zero-intensity levels. 

  Observations were made in sets of four raster scans,
where the HWP was rotated to one of four angles (0\deg, 22\fdg5, 45\deg, 67\fdg5) after every 
completed raster scan.  A complete set of four scans typically required one hour.
Scans were made while the source was rising and setting to maximize the range of parallactic angle.

As discussed in Section \ref{beam_smooth}, the telescope beam was much larger than predicted by our optics model, with significant non-Gaussian structure.  We smoothed our data to achieve an approximately round beam having a FWHM of 2\farcm5 at 500\,\um .

\section{Data Analysis} \label{sect:data_analysis}
In this section we give a brief overview of the \blastpol\ data analysis pipeline and highlight differences from the 2010 data pipeline described in \cite{matthews_2014}.
An in-depth description of the data reduction pipeline and iterative mapmaker \toast~will be given in a forthcoming paper (S. Benton et al.\ 2016, in preparation). 

\subsection{Time Domain Preprocessing}

Standard techniques were applied to the bolometer time-ordered data (hereafter TOD) to remove detector spikes (mostly due to 
cosmic rays), deconvolve the bolometer time constant, and remove an elevation-dependent feature (see \citealt{matthews_2014}). The data were further preprocessed by fitting and removing an 
exponential function fit to each detector's TOD in the first 30 seconds after a HWP rotation
or a telescope slew.  A high-pass filter with power-law cutoff was used to whiten noise in the TOD below 5 mHz.
Temporal gain variations were removed using the DC voltage level of each detector and periodic measurements of an internal calibration source.  Pixel-to-pixel detector gain variations were corrected by frequent observations of the bright compact source \mickey.

Telescope attitude was reconstructed using pointing solutions generated from the BLASTPol optical star camera,\footnote{BLASTPol flew two redundant star-boresight optical star cameras during the 2012 flight, but one experienced a harddrive failure six hours after the launch \citep{galitzki_2014}.} with payload rotational velocities from gyroscopes used to interpolate between pointing solutions \citep{pascale_2008}.  Data having pointing uncertainties $>5$\arcsec~were discarded.  The final on-sky pointing solution was calibrated to match
the astrometry of publicly available \herschel~SPIRE maps\footnote{\url{http://www.cosmos.esa.int/web/herschel/science-archive}} at the same wavelength.

\subsection{Beam Analysis} \label{beam_smooth}
The BLASTPol 2012 beam differs from the beam predicted by our optics model.  It has multiple elongated peaks, and the relative power in each peak varies from detector to detector.  BLASTPol filters were designed for near-space conditions and the telescope far field is several kilometers away so it was not possible to map the far-field beam at sea level \citep{galitzki_2014}.  Instead the beam had to be inferred from in-flight measurements of astronomical objects.  
\begin{figure*}
\plotone{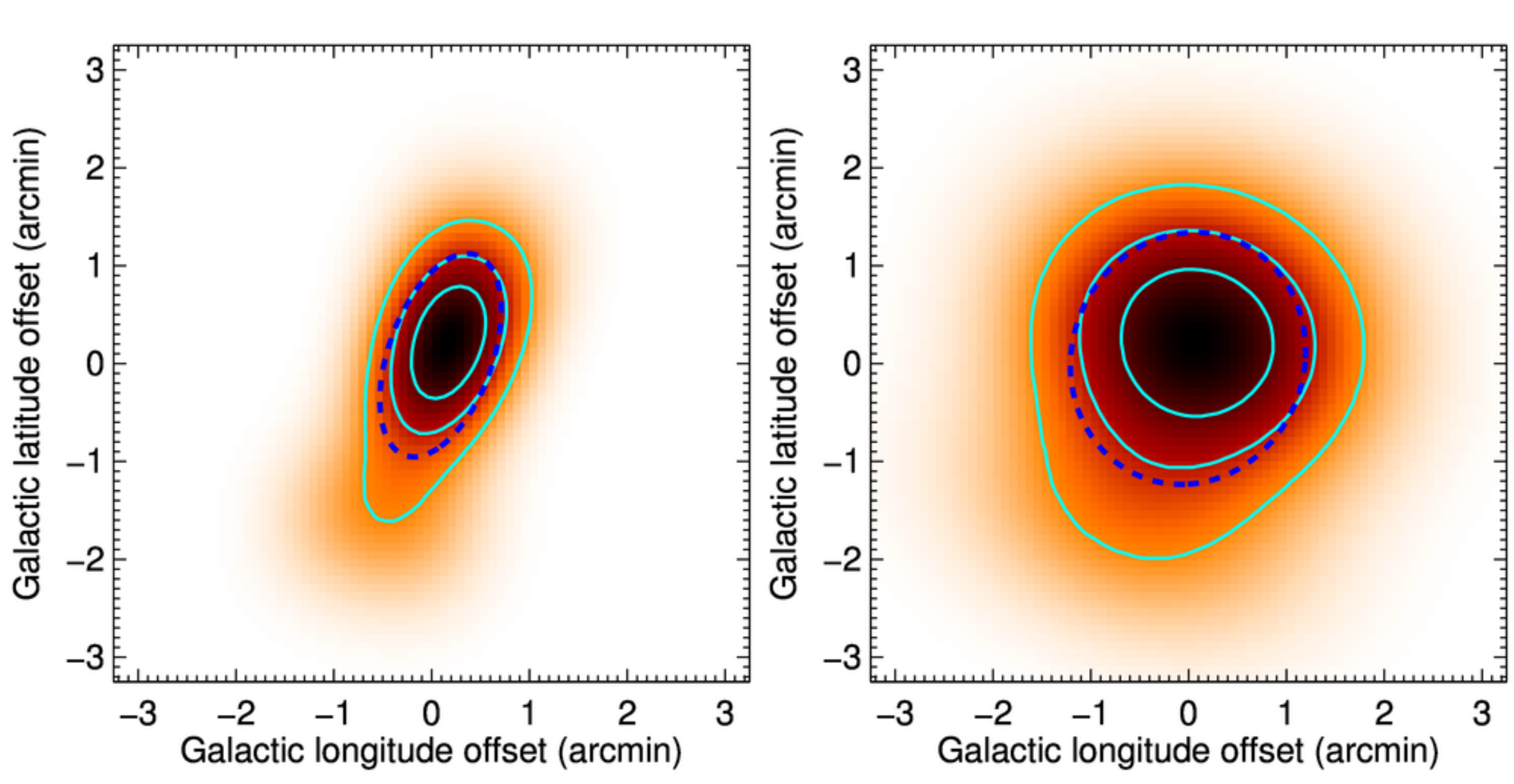}
\caption{{\em Left:} \blastpol~500\,\um~beam model for the Vela\,C map.  {\em Right:} \blastpol~500\,\um~beam model for the Vela\,C map after convolution with the smoothing kernel discussed in Section \ref{beam_smooth}.  Contour levels (cyan) are 25, 50, and 75 $\%$~of the peak brightness.  The dashed blue lines in each image show the FWHM of a fit to an {\lmfup elliptical} Gaussian.  FWHMs of the fitted Gaussians are 130\arcsec~by 64\arcsec~for the Vela\,C beam model (left panel) and 144\arcsec\ by 157\arcsec~for the smoothed beam model (right panel).
\label{fig:vela_beam}}
\end{figure*}

Our 2012 instrument beam model was defined in telescope coordinates and was informed by observations of two objects: \mickey, a warm compact dust source
located in the Vela Molecular Ridge; and limited observations of the planet Saturn made on 27 December 2015.  \mickey~was observed every 4$-$8 hours, with reasonable coverage for all detectors, but it is not a point source at BLASTPol resolution.  BLAST observations of \mickey~in 2006 found a FWHM of $\sim$40\arcsec~\citep{netterfield_2009}.  Saturn has a radius of 6$\,\times\,10^4$\,km, which corresponds to an angular size of $<$20\arcsec, considerably smaller than the \blastpol~2012 beam. Saturn was only observed early in the flight at telescope elevations of $<$25\deg~and was only fully mapped by the bolometers near the center of the focal plane arrays.  
Three {\lmfup elliptical} Gaussians were fit to the Stokes $I$~maps of \mickey~and Saturn. The free parameters were the Gaussian amplitudes, centroids, widths, and position angles. Only pixels above an intensity threshold of 20$\%$~of the peak intensity for \mickey~and 7.5$\%$~of the peak intensity for Saturn were used in the fits.  The final 2012 instrument beam model used the centroid positions, amplitudes, and position angles from the fits to \mickey~and the Gaussian
widths calculated from the fits to Saturn.

Next, an on-sky beam model was computed for the Vela\,C observations. The on-sky beam model is a time-weighted average of the instrument beam model rotated by the angle between the telescope vertical direction and Galactic north for each raster scan. 
The resulting average beam model for Vela\,C is shown in the left panel of Figure \ref{fig:vela_beam}.  A single elliptical Gaussian fit to this beam model gives FWHMs of  130\arcsec~by 64\arcsec, at position angle $-$51\deg.  This beam is significantly larger than the expected diffraction limit of the telescope (FWHM=60\arcsec).  

Lucy-Richardson (LR) deconvolution was previously used to correct a larger-than-expected beam from the BLAST 2005 flight \citep{roy_2011}. Here we used an iterative LR method and our beam model to deconvolve a simulated map consisting of a single Gaussian source with FWHM\,=\,145\arcsec. The deconvolved map from this step when applied as a smoothing kernel to convolve the original beam model should restore the 145\arcsec~Gaussian.  The success of this step can be judged from the right panel in Figure \ref{fig:vela_beam}: it does produce a single-peaked source that is approximately round (FWHM=144\arcsec\ by 157\arcsec).  This same smoothing kernel was used to convolve the $I$, $Q$, and $U$~maps of Vela C, with a resulting resolution of approximately 2\farcm5.  

\subsection{Instrumental Polarization}\label{sect:inst_pol}
To determine the {\lmfup instrumental polarization (the polarization signal introduced by the instrument hereafter referred to as IP)} we followed methods described in \cite{matthews_2014}. In brief, the set of observations of Vela C was split into two bins based on parallactic angle, and maps were produced for each detector individually using the {\tt naivepol} mapmaker \citep{moncelsi_2014}. The measured polarization is a superposition of one component fixed with respect to the sky and {\lmfup the IP, which is }fixed with respect to the telescope. 
{\lmfup By comparing the polarization measurements at different parallactic angles the IP of each bolometer could be reconstructed.  These IP terms were then removed during the mapmaking stage (Section \ref{sect:toast}).  The 500\,\um~array has an average IP amplitude of 0.53$\%$.}
{\lmfup To check the effectiveness of the IP correction the Vela\,C data were divided into two halves and IP estimates derived from the first half of the data were used to correct the second half of the data. By measuring the IP of the ``corrected'' second half of the Vela\,C data} we estimate that the minimum value of the fractional polarization $p$~measurable by \blastpol~at 500\,\um~is 0.1$\%$.  

\subsection{TOAST}\label{sect:toast}
Maps were made using {\tt TOAST}~(Time Ordered Astrophysics Scalable Tools),\footnote{\url{http://tskisner.github.io/TOAST}} a collection of serial and OpenMP/MPI parallel tools for simulation and map making.
Specifically, the generalized least-squares solver was used, which iteratively inverts the map-maker equation using the preconditioned conjugate gradient method (see \citealt{cantalupo2010},  a predecessor to TOAST).
The map-maker's noise model was estimated using power spectra from observations of faint dust emission in the constellation Puppis (map centered at $l\,=\,239.0$\deg, $b\,=\,-1.7$\deg), with simulated astrophysical signal subtracted.
The noise model is consistent with white noise plus $\left(1/f\right)^\alpha$ correlations that level out at low frequency due to data preprocessing.
Correlations between detectors and non-stationarity of the noise were not required by the model.
Instrumental polarization (Section \ref{sect:inst_pol}) was removed as per \cite{matthews_2014}.
Per-pixel covariance matrices for Stokes $I,~Q$,~and~$U$ were estimated as the $3 \times 3$ diagonal block of the full pixel-pixel covariance matrix.
Noise-only maps, both simulations and null tests (see Section \ref{sect:null_tests}), are consistent with the estimated covariances, up to a constant scaling factor due to unmodeled noise.
A pixel size of 10\arcsec~was used for all signal and covariance maps.

\subsection{Diffuse Emission Subtraction} \label{sect:ref_reg}
To study the polarization properties and magnetic field morphology of Vela\,C it is necessary to isolate polarized dust emission originating in Vela\,C from the diffuse polarized emission associated with Galactic foreground and background dust as well as the Vela Molecular Ridge.\footnote{It should also be noted that observations made by BLASTPol are inherently differential measurements, and thus the map zero-intensity level is uncertain.}  This separation requires extra care as previous studies show that diffuse sightlines, which may be used to estimate the foreground/background polarized emission, tend to have a higher average polarization fraction than dense cloud sightlines. {\lmfup In particular, \planck~observations show that in the more diffuse clouds there is a range of $p$~values with the maximum reaching 15 to 20$\%$, while such high values are not seen in the higher column density clouds (e.g., Orion, Ophiuchus, Taurus), where the average $p$~is consistently lower (\citealt{planck2014-XX}).  
Polarized emission from diffuse dust along dense cloud sightlines could therefore contribute significantly to the overall polarization measured.}
In this section we present two different diffuse emission subtraction methods, one  conservative and one more aggressive with respect to diffuse emission removal.

In the conservative method for diffuse emission subtraction, we considered most of the emission surrounding Vela C as defined by \cite{hill_2011} to be associated with the cloud.  The \cite{hill_2011} Vela C cloud subregions are overplotted (solid white lines) on a map of 500\,\um~total intensity in Figure \ref{fig:ref_regs}. The zero-point for the intensity of the cloud emission was set by specifying a region with relatively low intensity near Vela C and assuming that emission in this reference region also contributes to sightlines on the cloud with spatial uniformity. This low flux region is labeled ``C'' in Figure \ref{fig:ref_regs}. We calculated the average Stokes $I$, $Q$, and $U$ in that region, and the appropriate mean flux was then subtracted from each of the maps. The result was a set of maps of Vela\,C emission isolated from the Galaxy, assuming a minimal, uniform contribution from foreground and background emission.

In the aggressive method we considered most of the diffuse emission surrounding Vela C to be unassociated with the cloud, and accordingly a higher level of flux needed to be removed to isolate the cloud. Furthermore, we noted that there is significantly more emission to the south of Vela C than to the north; thus it is reasonable to assume that the true map of the region consists of the Vela C cloud superimposed on a varying Galactic emission profile. In this method of referencing, we defined two regions closely surrounding the cloud (marked ``A1" and ``A2" in Figure \ref{fig:ref_regs}) and performed 2-dimensional linear{\lmfup-plane} fits to the Stokes $I$, $Q$, and $U$ {\lmfup maps excluding all map pixels except those located in regions A1 and A2. The three free parameters in these fits were the linear slopes of the plane in the directions tangent to $l$~and $b$ and a map offset}. The equations for each of the resulting {\lmfup plane-fits to the $I$, $Q$, and $U$ maps were used to specify} the intensity to be subtracted from each pixel in the maps. Note that for regions far from Vela C, the linear approximation of the Galactic emission profile breaks down, leading to an inappropriate extrapolation. Therefore we defined an area within which the linear fit referencing method is valid (blue quadrilateral in Figure \ref{fig:ref_regs}), bounded on the north and south by the reference regions A1 and A2, and on the east and west by the edges of the well-sampled portion of the map . This ``validity" area roughly coincides with the four southernmost regions of \cite{hill_2011}.  We note that some of the emission in A1 and A2 might in fact be associated with Vela\,C, so this method is likely to over-subtract the diffuse dust emission.

The true $I$, $Q$, and $U$~maps of Vela C probably exist somewhere between our most extreme physically reasonable assumptions corresponding to the conservative and aggressive diffuse emission subtraction methods.  In this paper we present results for an ``intermediate'' diffuse emission subtraction method, derived by taking the arithmetic mean of the $I$, $Q$,~and $U$~maps from the aggressive and conservative methods.  
Most of our analyses are then repeated using the aggressive and conservative methods as a gauge of the uncertainties associated with the diffuse emission subtraction.

\subsection{Null Tests}\label{sect:null_tests}
To characterize possible systematic errors in our data, we performed a series of null tests, which are described in detail in Appendix \ref{sect:null_append}.  In these, we split the 250\,\um~observations\footnote{As discussed in Appendix \ref{sect:null_append} the \blastpol\,250\,\um\ observations of Vela C are better suited than the 500\,\um~data for performing null tests.} into two mutually exclusive sets. If the polarization parameters from the two independent data sets agree, we can conclude that the impact of systematics is small, and the uncertainties are properly characterized by Gaussian errors produced by {\tt TOAST}.  The four methods of splitting are: data from the left half of the array vs.~the right half; data from the top half of the array vs.~the bottom half; data from earlier in the flight vs.~later in the flight;~and alternating every other scan set sequentially throughout the flight. 

For each null test we made separate maps of $I$, $Q$, and $U$, which were then used to calculate residual maps of the polarized intensity $P$, the polarization fraction $p$,~and the polarization-angle $\psi$~as described in Appendix \ref{sect:null_append}.
{\lmfup(The quantities $P$, $p$, and $\psi$~are defined in Appendix \ref{sect:pol_conventions}).}
If our data had no systematic errors we would expect to see uncorrelated noise in the residual maps.
For $P$~and $p$~if the residuals were less than one-third of the signal in a given map pixel then that pixel was said to pass the null test.  For $\psi$~the residuals had to be less than 10\deg~to pass the null test.
We examined each of the four null tests listed above for each of the two methods of diffuse emission subtraction (Section \ref{sect:ref_reg}) in $P$, $p$~and $\psi$~giving a total of 24 checks that our measured polarization signal is significantly above the systematic uncertainty level.

We found that our map passed these tests for the majority of sightlines inside the “cloud” region shown in Figure \ref{fig:ref_regs} (blue solid line). The exceptions occurred in regions where the fractional polarization was small, so that a comparison of the scale of the polarization signal to the scale set by residuals in the null tests resulted in an apparent failure. The fact that we saw null test failures correlating with low $p$, but not with absolute difference in the null test $I$~maps led us to the interpretation that the apparent low signal level compared to the null test residual is due to decreased signal and not increased systematic uncertainties.  
We did see significant structure in the null test residual maps of $Q$~and $U$~near the compact \ion{H}{2}~region RCW\,36, which coincided with null test residuals of one-fourth $p$, though the residuals in $\psi$~were much smaller than 10\deg.  These $p$~measurements technically pass the null test criteria, but the systematic errors are larger than the statistical errors. 
We conclude that for the validity region shown in Figure \ref{fig:ref_regs} the null tests are passed, with the exception of a circular area centered on \rcw~($l=$265.15\deg, $b=1.42$\deg within a radius of 4\arcmin).

\section{BLAST-Pol Polarization Maps}\label{sect:maps}
\begin{figure*}
\epsscale{1.0}
\plotone{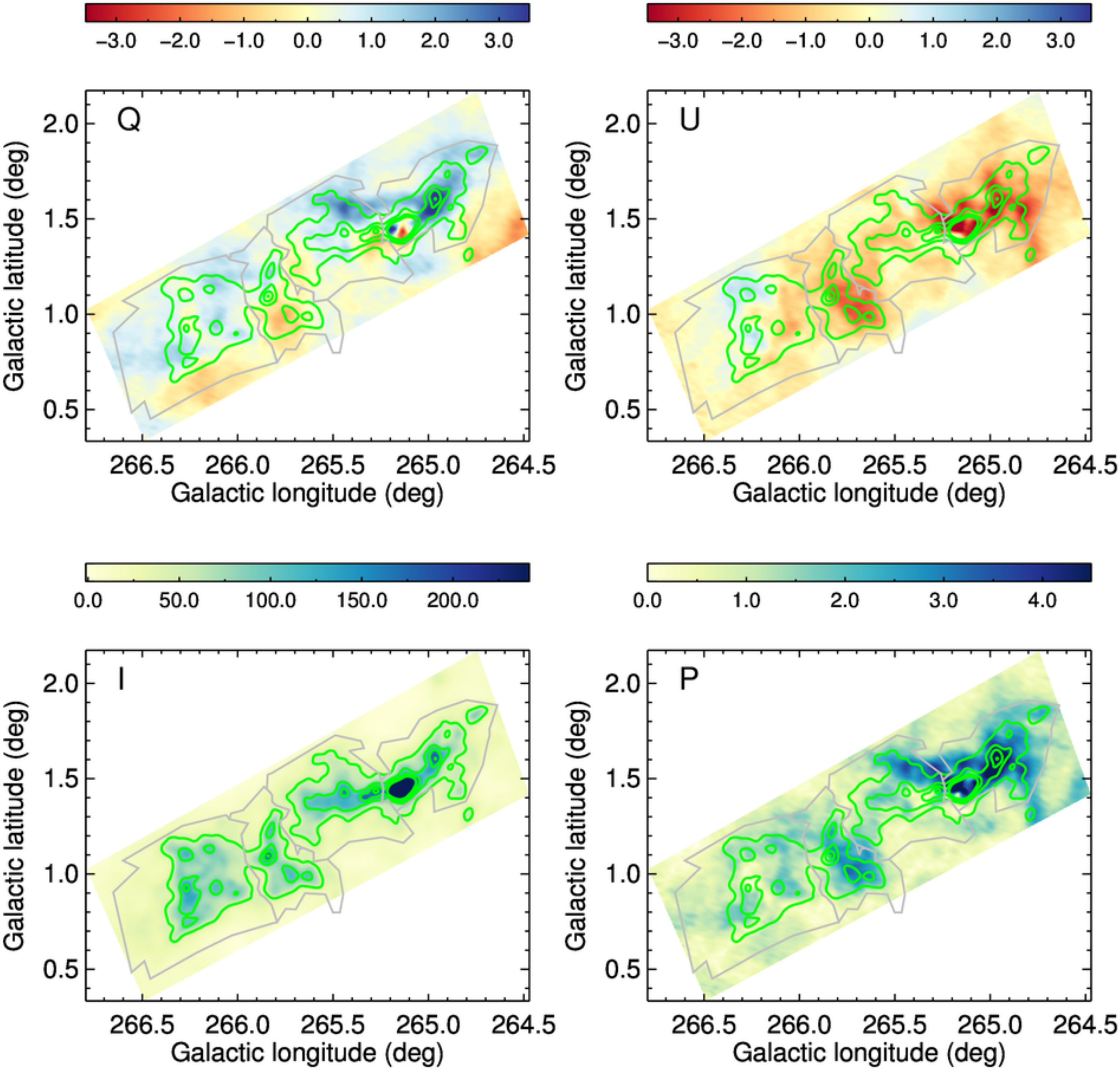}
\caption{ \blastpol~500\,\um~\velac~maps of $I$, $Q$, $U$,~and the total polarized
intensity $P$ {\lmfup(which is debiased as described in Section \ref{sect:maps}). The color scale units are MJy\,sr$^{-1}$.  Contours indicate $I$~levels of 46, 94, 142, and 190 MJy\,sr$^{-1}$}, and the gray outlines indicate cloud subregions covered by our observations as in Figure \ref{fig:ref_regs}.  
\label{fig:maps_IQUP}}
\end{figure*}

In this section we present maps of the Stokes parameters $I$, $Q$,~and $U$, linearly polarized intensity ($P$), and polarization fraction ($p\,=\,P/I$). The polarization descriptors and covariances used in our analysis are summarized in Appendix \ref{sect:pol_conventions}.  We also present maps of $\Phi$, the inferred orientation of magnetic field projected onto the plane of the sky, which is assumed to be the orientation of the polarization of the dust emission (described by $\psi$) rotated by $90^{\circ}$, and the localized dispersion in the polarization-angle ($S$).  
 
Because $p$~and $P$~are constrained to be positive any noise in the $Q$~and $U$~maps will tend to increase the measured polarization.  Accordingly, we crudely debias $p$~and $P$~according to
\begin{equation}
p_{db}\,=\,\sqrt{p^2\,-\,\sigma_p^2},
\end{equation}
and 
\begin{equation}
P_{db}\,=\,\sqrt{P^2\,-\,\sigma_P^2},
\end{equation}
\citep{wardle_1974}.
This method of debiasing is appropriate only where $\sigma_p$~is small compared with $p$ \citep{montier_2015}. We note that the median value of $p/\sigma_p$~in our map is $\sim25$, so for most of our map this debiasing method is applicable.

\subsection{Diffuse Emission Subtracted Maps of $I$, $Q$, and $U$, and Derived Maps of $P$, and $p$}\label{sect:pol_maps}

Figure \ref{fig:maps_IQUP} shows Vela\,C 500\,\um\,maps for the three Stokes parameters $I$, $Q$, and $U$.  The maps have been smoothed to 2\farcm5~resolution, as
described in Section \ref{beam_smooth}, and use the intermediate diffuse emission subtraction method (Section \ref{sect:ref_reg}).  Overlaid in 
gray are the outlines of the subregions of Vela\,C as defined in \cite{hill_2011} and labelled in Figure \ref{fig:tyr_n_t_maps}.  The \blastpol~$I$~map peaks at the location of
\rcw.

\begin{figure}
\epsscale{1.2}
\plotone{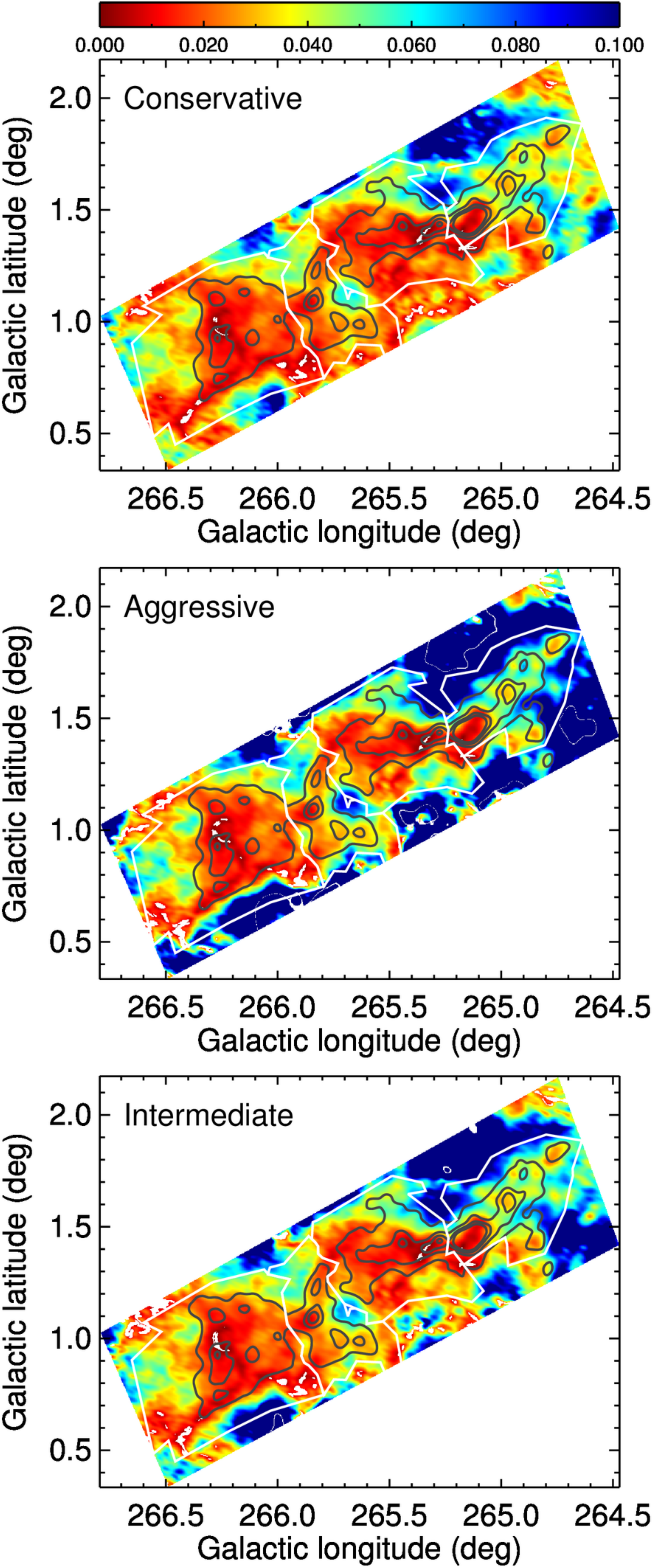}
\caption{\blastpol~500\,\um~maps of $p$ obtained using different 
methods for separating the polarized emission of Vela\,C from that of diffuse background/foreground dust (Section \ref{sect:ref_reg}): conservative method (top panel); aggressive method (middle panel), and intermediate method (bottom panel).
  Only sightlines where $p\,>\,3 \sigma_p$ {\lmfup and $I\,>\,0$} are shown. {\lmfup The $p$~maps shown have been debiased using the methods described in Section \ref{sect:maps}.   Gray contours indicate $I$~levels of 46, 94, 142, and 190 MJy\,sr$^{-1}$}, and the white outlines indicate the four Vela\,C cloud subregions as in Figures \ref{fig:ref_regs} and \ref{fig:tyr_n_t_maps}.
\label{fig:maps_littlep}}
\end{figure}

\begin{figure*}[t]
\epsscale{0.9}
\plotone{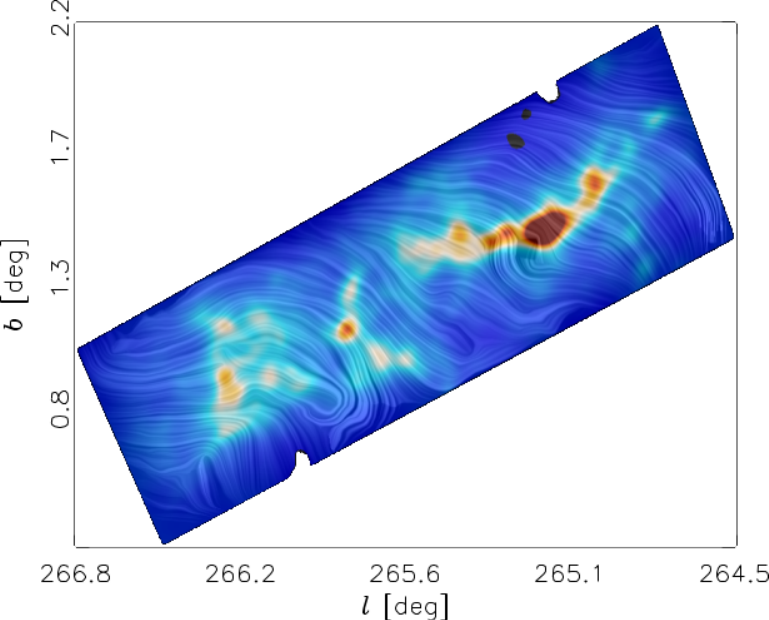}
\caption{\blastpol~500\,\um~$I$~map with the inferred plane of the sky magnetic field component ($\Phi$) overlaid as a ``drapery'' image {\lmfup(only regions where $I\,>\,0$~are shown)}. The drapery pattern is produced using the line integral convolution method \citep{cabral_1993} and indicates the orientation of the magnetic field as projected on the plane of the sky.  Note that this drapery pattern was made from all of the $\Phi$~data with no masking of sightlines having large uncertainties in $\Phi$.  This image is meant show the level of detail available in the \blastpol~$\Phi$~maps, but should not be used for quantitative analysis.
\label{fig:maps_drapery}}
\end{figure*}

Also included in Figure \ref{fig:maps_IQUP} is the derived map of the polarized intensity ($P$), which generally shows some signal where there is cloud emission.
  However, the correspondence is certainly 
not perfect, and varies considerably across the map.  For example, along most of the 
Centre-Ridge there is a corresponding peak in the $P$~map along the main ridge.  In the South-Ridge there are peaks at similar locations in the $P$~and $I$~maps.  
But in the South-Nest, prominent areas of polarized emission are only seen around the edge of the cloud structure seen 
in $I$.  There are also some regions of significant $P$~that stand out less in $I$, for example, along the north edge of the Centre-Ridge.

Figure \ref{fig:maps_littlep} shows the polarization fraction ($p\,=\,P/I$) for each of the three different 
diffuse emission subtraction choices discussed in Section \ref{sect:ref_reg}.  The conservative diffuse emission subtraction (top panel) results in $p$~that is lower on average than from the aggressive diffuse emission subtraction (middle panel).  
This is expected as $p$~is commonly observed to increase for regions of low dust emission.  Thus, compared to the aggressive subtraction method that uses regions closer to the cloud with lower average $p$, the conservative method removes more $P$~relative to $I$.
The bottom panel shows the $p$~map resulting from the intermediate diffuse emission subtraction method.  Unless otherwise specified, for the remainder of the paper $p$, $\psi$, and $\Phi$~are calculated from Stokes parameter maps using the intermediate diffuse emission subtraction method.

The mean value of $p$~in our map is {\lmfup 6.0$\%$~with a median of 3.4 $\%$.  For the map pixels within the dense cloud subregions defined by \cite{hill_2011} the mean polarization fraction is 3.5$\%$~with a median of 3.0$\%$~and a standard deviation of 2.4$\%$}.  {\lmfup Previous submillimeter polarization maps having spatial coverage corresponding to the scales of entire clouds have yielded roughly similar values.  Specifically, after subtracting the background/foreground emission, \cite{planck2014-XXXIII} found mean 850\,\um~polarization fractions of 1.8$\%$, 5.0$\%$, and 6.1$\%$ for three nearby molecular clouds, while 450\,\um~polarization maps of four GMCs made by \cite{li_2006} with SPARO at the South Pole yield a mean polarization fraction of 2.0$\%$.  Our $p$~map shows behavior that is broadly consistent with expectations from the $P$~map.}  Values of $p$~tend to decrease with increasing $I$, but there is not a one-to-one anticorrelation between $p$~and $I$.

\subsection{Inferred Magnetic Field Direction}\label{sect:bpos_maps}

\begin{figure*}
\epsscale{1.0}
\plotone{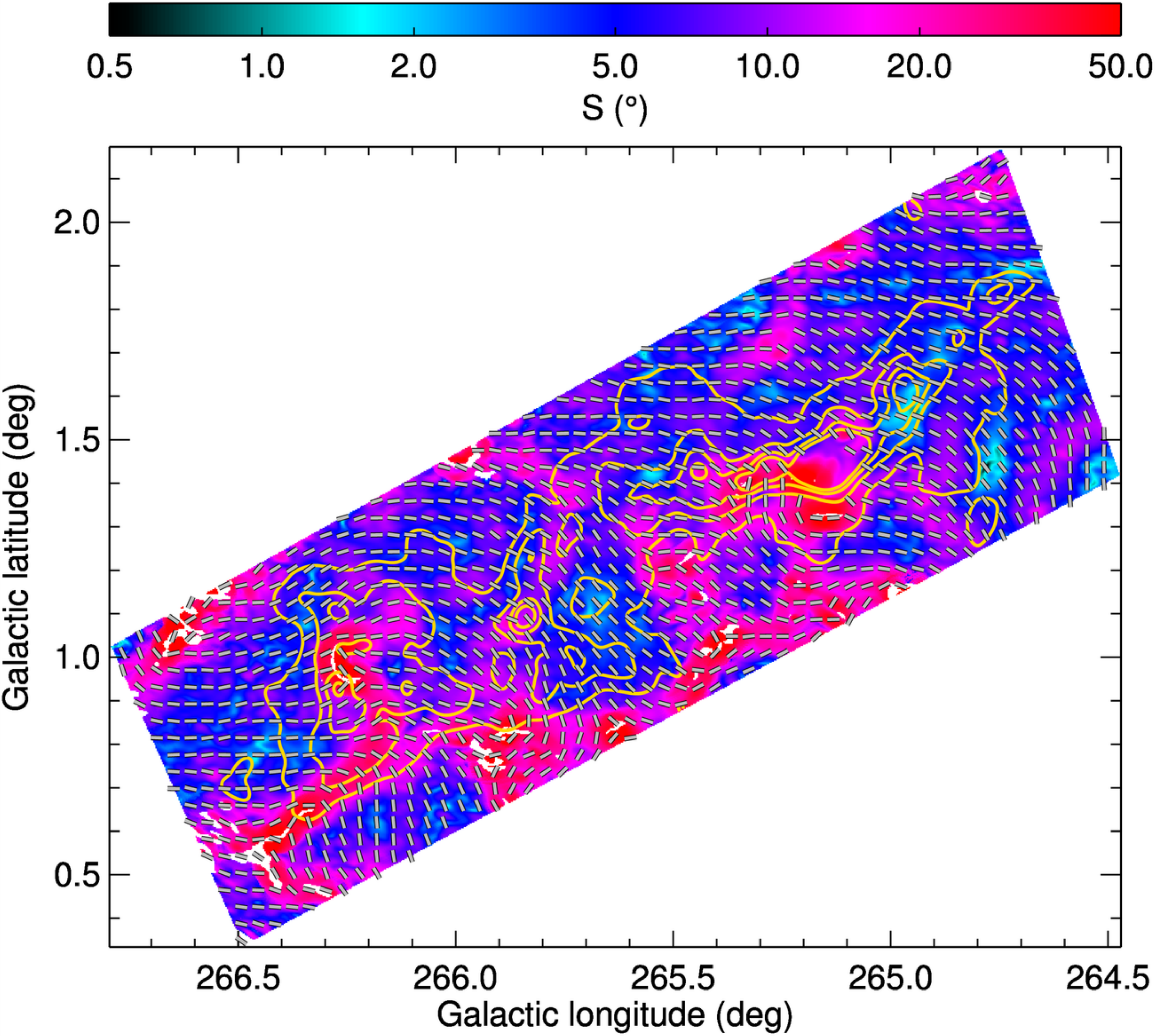}
\caption{\blastpol~500\,\um~map of the dispersion in the polarization-angle (\dpsi)
{\lmfup in degrees} on 0.5\,pc scales ($\delta$\,=2.5\arcmin) as defined in Section \ref{sect:delta_psi}.  The $S$~map is shown where $S\,>\,3 \sigma_S$.
Line segments show the orientation of the magnetic field as projected on the plane of the sky ($\Phi$), derived from the \blastpol~500\,\um~data. The $\Phi$~measurements are
{\lmfup shown} approximately every 2\farcm5.  Contours {\lmfup indicate 500\,\um~$I$~intensity levels of 46, 94, 142, and 190 MJy\,sr$^{-1}$}.
\label{fig:maps_dpsi}}
\end{figure*}

Figure \ref{fig:maps_drapery} shows a detailed view of the magnetic field orientation projected onto the plane of the sky $\Phi$, as inferred from the \blastpol~500\,\um~data. This figure uses a ``drapery'' pattern produced using the line integral convolution method of \cite{cabral_1993} superimposed on the \blastpol~500\,\um~$I$~map.\footnote{This visualization is produced with the same code used in \cite{planck2015-XXXV} and will be further discussed in a forthcoming paper by D. Falceta-Gon{\c c}alves et al.} \cite{dotson_1996} showed that there is significant ambiguity in inferring the magnetic field lines from polarization data, particularly as polarization maps can sample multiple cloud structures along the line of sight, each potentially having a different magnetic field orientation.
  The drapery image is presented solely to show the range of orientations of $\Phi$, and to give a sense of the range of spatial scales probed by \blastpol. Figure \ref{fig:maps_dpsi} shows $\Phi$~as a series of line segments (approximately one line segment per 2\farcm5 \blastpol~beam).  

The projected cloud magnetic field 
direction appears to change across Vela\,C:
at low Galactic latitudes the field is mostly perpendicular to the main cloud elongation 
direction, while at higher Galactic latitudes it bends to run mostly parallel to the cloud 
elongation direction.  We also see some sharp changes in $\Phi$, most noticeably 
in the South Nest, and near the compact \ion{H}{2} region \rcw.  

\begin{figure}
\epsscale{1.2}
\plotone{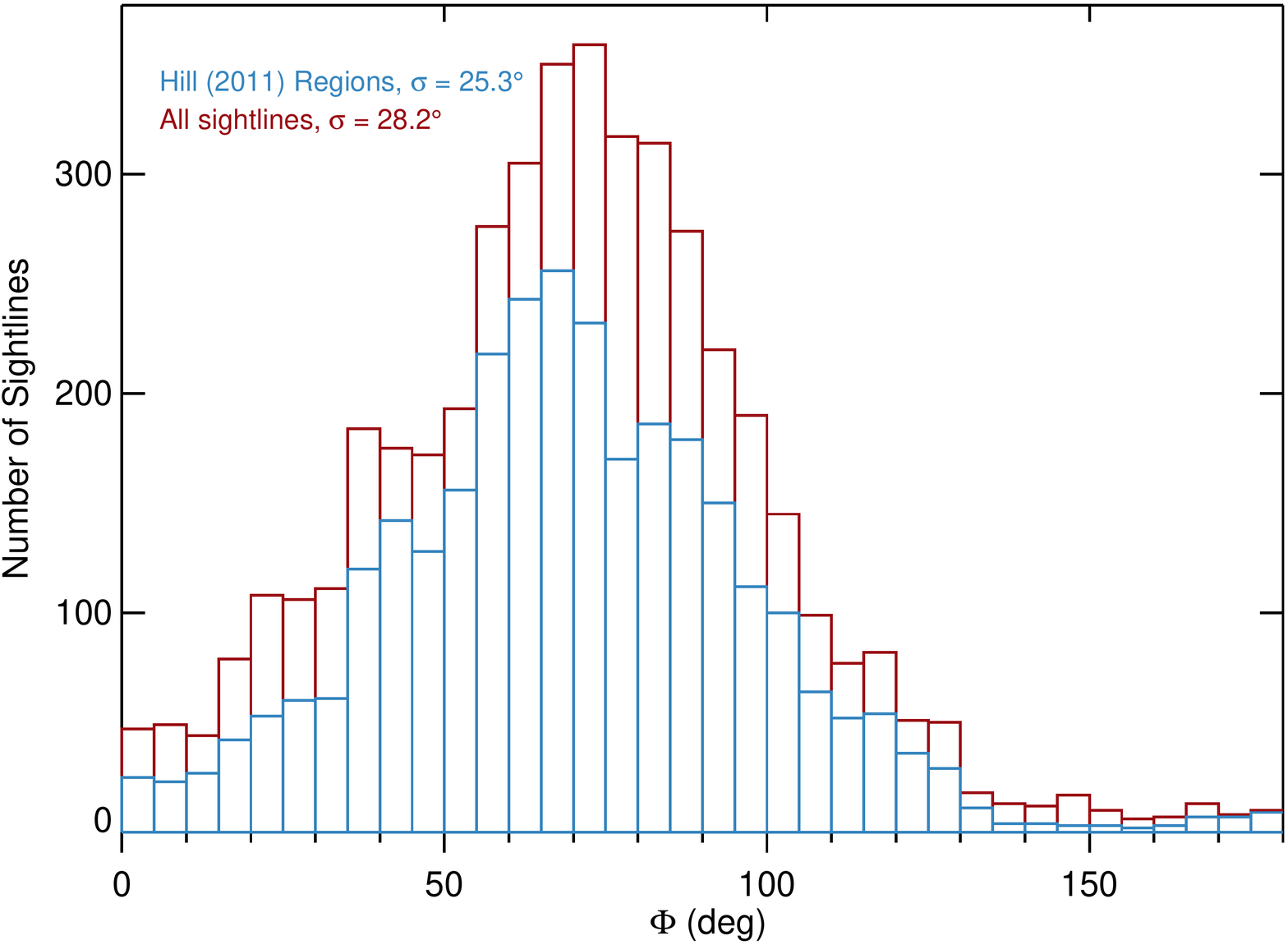}
\caption{Histograms of the \blastpol~500\,\um~inferred magnetic field direction $\Phi$~for  all \velac~sightlines (red) and sightlines lying inside the \cite{hill_2011} subregions (blue). {\lmfup Sightlines included in these histograms have $\sigma_{\Phi}\,<\,10$\deg~and both $|\psi_{\rm int}-\psi_{\rm con}|\,<\,10$\deg~and $|\psi_{\rm int}-\psi_{\rm agg}|\,<\,10$\deg~(see Section \ref{sect:cuts}).}  The standard deviation of each distribution is given at upper-left.
\label{fig:ang_histos}}
\end{figure}

{\lmfup Figure \ref{fig:ang_histos} shows that the dispersion in magnetic field orientation across the Vela C cloud is 28\deg.  \cite{novak_2009} calculated dispersions on similar spatial scales by combining the large-scale GMC polarization maps of \cite{li_2006} with higher angular resolution submillimeter polarimetry data.  They obtained 27\deg-28\deg, nearly the same result.}  In future publications we will present statistical studies of the correlations between magnetic field orientation, filamentary structure, and cloud velocity structure. 

\subsection{Polarization Angle Dispersion Function} \label{sect:delta_psi}

To quantify the disorder of $\Phi$~in our Vela\,C maps at small scales we calculate the polarization-angle dispersion function \dpsi, implementing the formalism described in Section 3.3 of \cite{planck2014-XIX}. For each pixel in our map, \dpsi~is defined as the rms deviation of the polarization-angle $\psi\left(\vec{x}\right)$ for a series of points on 
an annulus of radius $\delta$:
  \begin{equation}
   S^2\left(\vec{x},\delta\right)=\frac{1}{N}\sum_{i=0}^N S_{xi}^2,
  \end{equation}
  where $\delta$\ is the length scale of the dispersion, $\vec{x}$ is the position for which we evaluate the polarization-angle dispersion and
  \begin{equation}
    S_{xi}=\psi\left(\vec{x}\right) - \psi\left(\vec{x}+\vec{\delta_i}\right).
  \end{equation}

  Because \dpsi\ is always positive it is biased due to noise.  We debias using the standard formula
  \begin{equation}
    S_{db}^2\left(\delta\right)=S^2\left(\delta\right)\ -\ \sigma_{S}^2,
  \end{equation}
  where $\sigma_{S}^2$\ is the variance of \dpsi.

Figure \ref{fig:maps_dpsi} shows \dpsi~for $\delta$\,=\,2\farcm5~($\sim$0.5\,pc), the smallest
scale that can be resolved with our smoothed beam.
{\lmfup (Hereafter we refer to $S\left(2\farcm5\right)$~as \dpsi).} The most striking features in the \dpsi~map correspond to regions of sharp changes in $\Phi$, which is indicated with line segments.  These high 
dispersion regions sometimes occur near the locations of dense filaments (for example, 
the sharp bend in the South Nest).  More often they correspond to sightlines of lower than average $p$~and do not appear to be coincident with
any prominent cloud feature in $I$.

\subsection{Polarization Map Sampling and Sightline Selection}\label{sect:cuts}
 Our \blastpol~polarization maps were calculated from Stokes parameters smoothed to a resolution of $\sim$2\farcm5 as discussed in Section \ref{beam_smooth}.  The resulting polarization maps were then sampled every 70\arcsec~to ensure at least Nyquist sampling.  In total there are 
4708 projected magnetic field sightlines over the validity region defined in Section \ref{sect:ref_reg}.

In the following sections we attempt to model the polarization fraction $p$~as a function of $N$, $T$, and $S$.
For these detailed studies we restrict our analysis to sightlines that encompass the dense cloud regions as defined by \cite{hill_2011}.
These sightlines are better probes of the polarization structure in the cloud and are less sensitive to systematic uncertainties in our ability to separate the polarized emission {\lmfup emitted by} diffuse dust foregrounds/backgrounds from the polarized emission {\lmfup emitted by dust grains in} Vela\,C.

To ensure a robust sample, we use only $p$~values 
that are large enough to be unaffected by uncertainties in instrumental polarization removal ($p\,>0.1\%$, see Section \ref{sect:inst_pol}), 
and for which we have at least a 3$\sigma$~detection of polarization ($p\,>\,3\sigma_p$){\lmfup, which corresponds to an uncertainty in the polarization angle  $\sigma_{\psi}\,<\,10$\deg.}  
  To ensure that the polarization values are not dependent on our choice of diffuse emission subtraction method we require that $p_{\rm int}\,>\,3|p_{\rm int}-p_{\rm con}|$\ and $p_{\rm int}\,>\,3|p_{\rm int}-p_{\rm agg}|$, {\lmfup where $p_{\rm con}$, $p_{\rm agg}$, and $p_{\rm int}$~are the polarization fraction values calculated using the conservative, aggressive, and intermediate diffuse emission subtraction methods respectively} (see Section \ref{sect:ref_reg}). {\lmfup Similarly, we require that $|\psi_{\rm int}-\psi_{\rm con}|\,<\,10$\deg~and $|\psi_{\rm int}-\psi_{\rm agg}|\,<\,10$\deg.} 
We also exclude sightlines from a 4\arcmin~radius region near \rcw~as these show residual structure in our null tests (see Section \ref{sect:null_tests}).
In total 2488 out of a 3056 possible \cite{hill_2011}~sightlines meet these criteria.
For our analysis of $p$~vs.\ $N$, $T$, and $S$~in Sections \ref{sect:p_N_T} and \ref{sect:p_N_dpsi} we also require 
at least 3-$\sigma$ measurements of $N$, $T$, and $S$~where the errors on $N$~and $T$~are derived from the SED fit covariance matrices {\lmfup(see Section \ref{sect:n_and_t_maps})}.  This results in a final sample of
{\lmfup \nVboth}~sightlines.

\section{Column Density and Temperature Maps Derived from \herschel~SPIRE Data}\label{sect:n_and_t_maps}
\begin{figure}
\epsscale{1.25}
\plotone{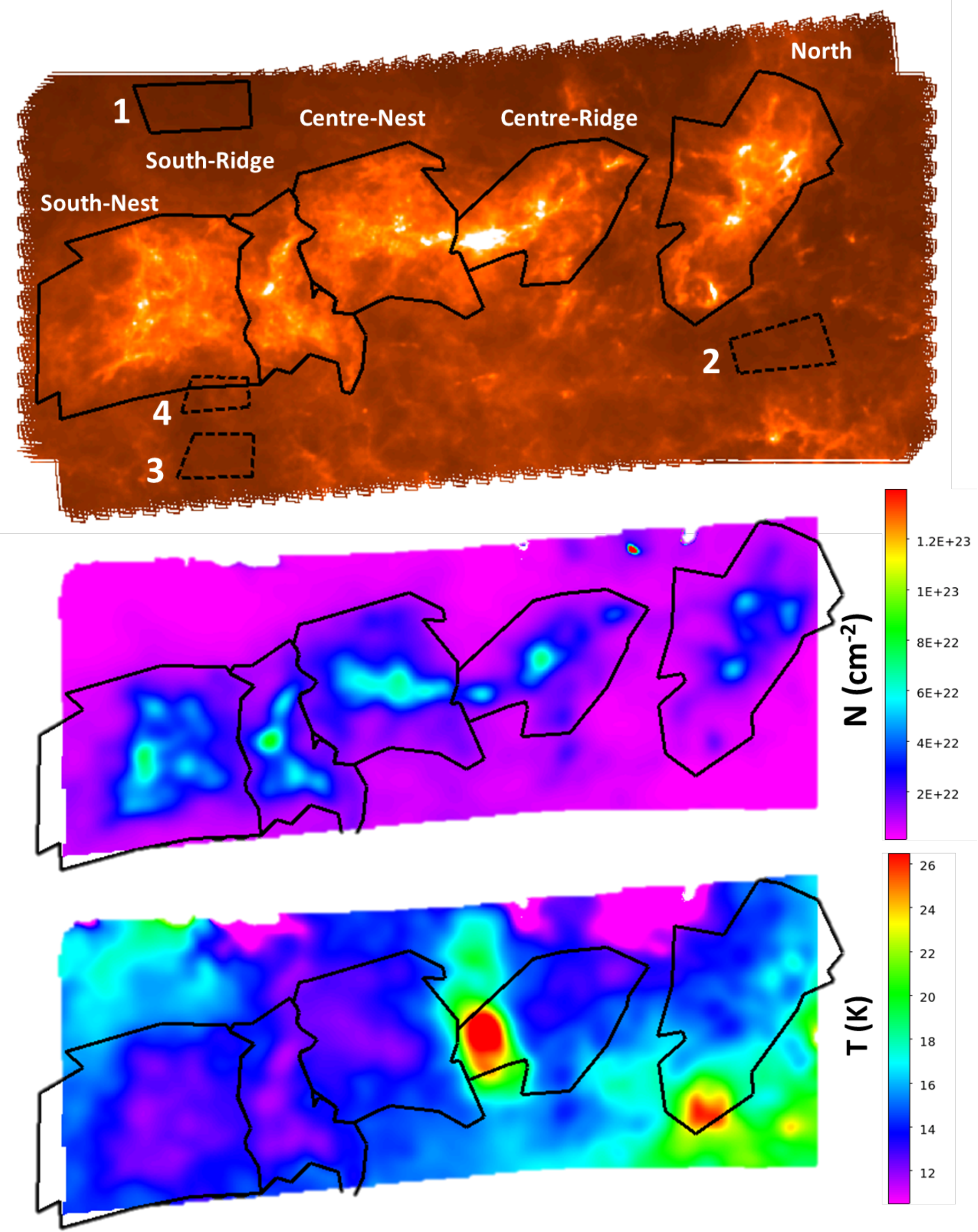}
\caption{{\lmfup \herschel~Vela\,C 500 \um~intensity (top panel, FWHM\,=\,35\farcs2), column density ($N$, middle panel, FWHM\,=\,2\farcm5), and dust temperature 
($T$, bottom panel, FWHM\,=\,2\farcm5). The $N$~and $T$~maps were derived from Herschel data using} the methods described in Section \ref{sect:n_and_t_maps}.  Numbered quadrilaterals correspond to different diffuse emission regions for which the average intensity is indicated in Figure \ref{fig:fl_ratios}.  Note that the mean intensity in Region 1 was subtracted from each of the {\lmfup 160,} 250, 350, and 500\,\um~maps before SED fitting. The solid black polygons {\lmfup (labeled in the top panel)} correspond to the cloud subregions as defined in \cite{hill_2011}. From left to right 
these are: the South Nest, a region of many overlapping filaments; the South Ridge, 
dominated by a single dense filament; the Centre Nest; and Centre Ridge, which contains
the ionizing source(s) powering the compact \ion{H}{2} region associated with RCW\,36.
\cite{hill_2011} also include an additional region, designated North, that was not covered 
in the deep \blastpol~survey of Vela\,C.
\label{fig:tyr_n_t_maps}}
\end{figure}

\begin{figure}[h]
\epsscale{1.2}
\plotone{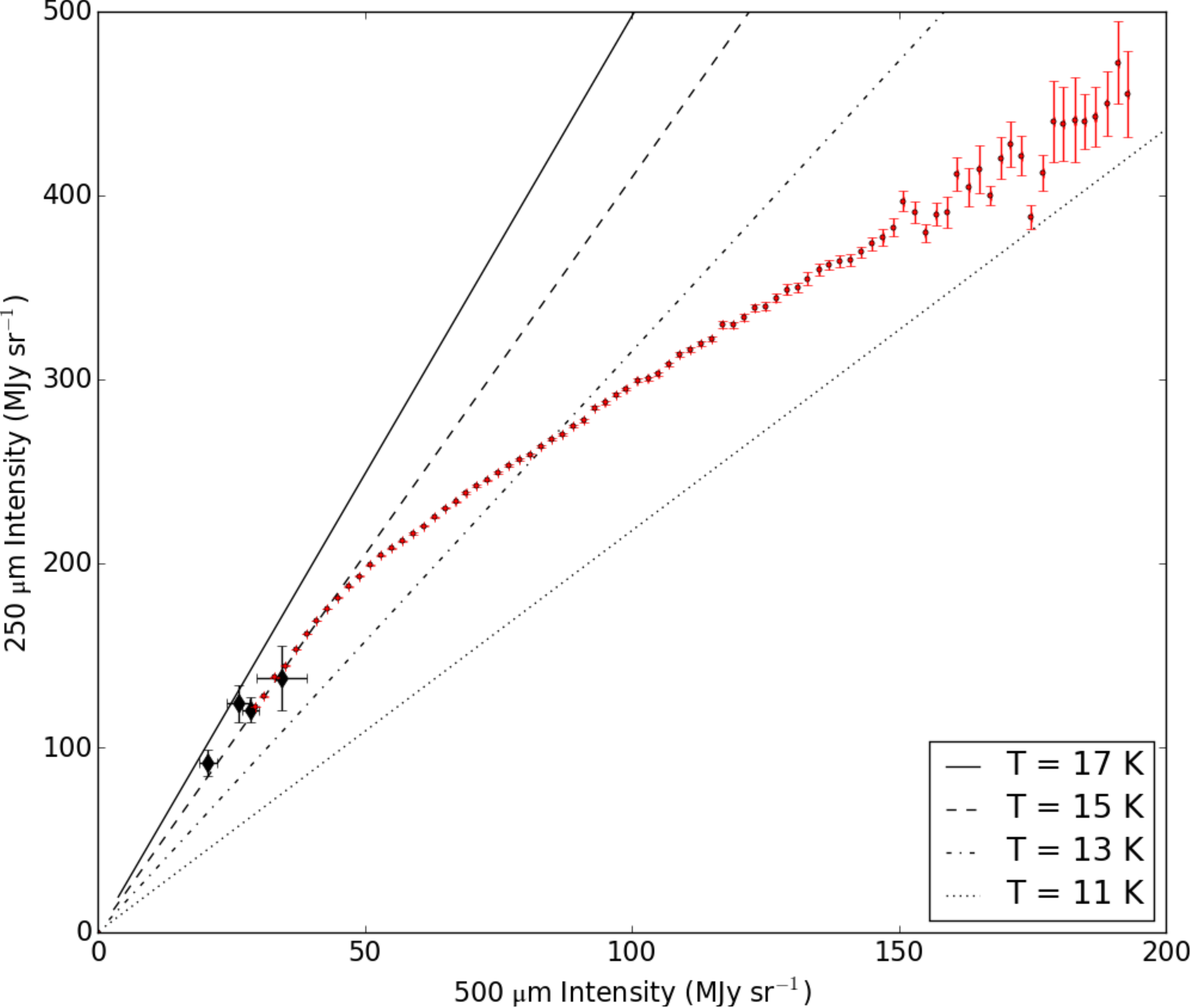}
\caption{Median values of \herschel~250\,\um~intensity in bins of \herschel~500\,\um~intensity for the South Nest region in Vela C (as labeled in Figure \ref{fig:tyr_n_t_maps}).  The error bars correspond to the standard deviation of the 250\,\um~intensity values in each bin.  Black lines correspond to the expected intensity ratios for uniform temperature dust.  {\lmfup Diamonds indicate the average 250\,\um~and 500\,\um~intensities for the four numbered diffuse emission regions indicated in the top panel of Figure \ref{fig:tyr_n_t_maps} (from left to right these indicate regions 1, 2, 3, and 4). Error bars show the standard deviation of intensity values in each region.}
\label{fig:fl_ratios}}
\end{figure}

To derive column density and dust temperature maps we used publicly available \herschel~SPIRE {\lmfup and PACS}~data.  SPIRE uses nearly identical filters to BLASTPol, but has higher spatial resolution (FWHM of 17\farcs6, 23\farcs9, and 35\farcs2 for the 250, 350, and 500\,\um~bands, respectively). {\lmfup Data taken with the PACS instrument in a band centered at 160\,\um~(FWHM of 13\farcs6)~were used to provide additional sensitivity to warm dust.}
\herschel~maps were generated using {\tt Scanamorphos} \citep{roussel_2013} and additional reduction and manipulation was performed in the Herschel
Interactive Processing Environment {\lmfup (HIPE version 11) including the Zero Point Correction function for the SPIRE maps}. The resulting \herschel~maps were smoothed to 35\farcs2~resolution by convolving with Gaussian kernels of an appropriate size and then regridding to match the 500\,\um~map.

Similar to the diffuse emission subtraction described in Section \ref{sect:ref_reg}, we attempted to separate the Galactic foreground and background dust emission from the emission of Vela\,C.  As the regions used to define the diffuse emission subtraction in Section \ref{sect:ref_reg} were not covered by the Herschel map we defined four alternate ``diffuse emission regions'' ({\lmfup see Figure \ref{fig:tyr_n_t_maps} top panel}).  These regions were presumed to contain little emission from dust in Vela\,C and thus they are reasonable representations of the contribution due to diffuse dust emission. For the initial analysis described below, the mean intensity in Region 1 was subtracted from each of the {\lmfup 160}, 250, 350, and 500\,\um~maps{\lmfup, and the maps were then further smoothed to match the 2\farcm5 resolution of the \blastpol~maps.}

Modified blackbody SED fits were made for each map pixel using the methods 
described in \cite{hill_2009,hill_2010,hill_2011} and using the dust opacity law of 
\cite{hildebrand_1983} with a dust spectral index $\beta$\,=\,2.
The resulting column density ($N$)~and dust temperature ($T$) maps are shown in Figure \ref{fig:tyr_n_t_maps} {\lmfup (middle and bottom panels, respectively)}. {\lmfup It should be noted that above a temperature of $\sim$20\,K, the dust emission is expected to peak at wavelengths shorter than 160\,\um.  For these warmest sightlines our estimates will have a higher degree of uncertainty.}
{\lmfup The derived $N$~and $T$}~maps  
were visually compared to the {\lmfup higher resolution} column density and temperature maps from \cite{hill_2011}, which did not include a diffuse emission subtraction. Our maps are in close agreement with the \cite{hill_2011} maps for column density sightlines where Vela\,C emission is strong compared to the diffuse emission component.  Note that we computed maps of 
the column density of hydrogen nuclei while \cite{hill_2011} calculated the column density of H$_2$.

Much of the analysis in the present paper focuses on comparisons between parameters such as polarization fraction $p$, $N$, and $T$. 
From Figure \ref{fig:tyr_n_t_maps} we see that $N$~and $T$~are strongly anti-correlated.
Similar trends were noted by \cite{palm_2013} in 
their \herschel~study of a cold cloud in Taurus. 
We interpret this as a result of 
radiation shielding in the densest parts of the cloud.  
This interpretation can be tested by
examining a plot of 250\,\um\ intensity vs.\ 500\,\um\ intensity, as shown in Figure \ref{fig:fl_ratios}. In 
this figure there is a noticeable bend in the otherwise linear relationship between the 
two intensities. Since submillimeter dust emission in molecular clouds is typically optically-thin, 
larger intensity at either wavelength corresponds to higher column density.
However, beyond the bend we notice that the slope of the 500\,\um~intensity
vs.\ 250\,\um~intensity relation decreases. The simplest explanation is that the dust 
in denser regions of the cloud is colder, due to radiation shielding.

An alternative interpretation of the bend seen in Figure \ref{fig:fl_ratios} is to hypothesize a uniformly cold cloud spatially superimposed on diffuse emission from warmer dust.
To explore this possibility we examined the location of each diffuse region on Figure \ref{fig:fl_ratios} relative to the bend in the observed curve of 250\,\um~vs.~500\,\um~intensity. 
Subtracting the diffuse emission flux essentially sets a new origin for this graph and is 
equivalent to the diffuse emission subtraction discussed in Section \ref{sect:ref_reg}, leaving only emission from dust grains in the Vela C 
cloud. As can be seen from Figure \ref{fig:fl_ratios}, the diffuse emission regions, even very aggressively placed ones, 
reposition the origin to locations significantly below the bend in the curve, indicating  that 
the observed $T$~and $N$~anticorrelation is intrinsic to the Vela\,C molecular 
cloud.
{\lmfup As a further check}
the SED fits described above were redone using diffuse emission Regions 2, 3, and 4 as the reference regions, instead of using Region 1 (see Figure \ref{fig:tyr_n_t_maps}).  The corresponding $N$~maps are very similar to the one shown in Figure \ref{fig:tyr_n_t_maps}, especially for the densest regions.

\section{Dependence of Polarization Fraction on $N$~and $T$} \label{sect:p_N_T}

\begin{figure}
\epsscale{1.2}
\plotone{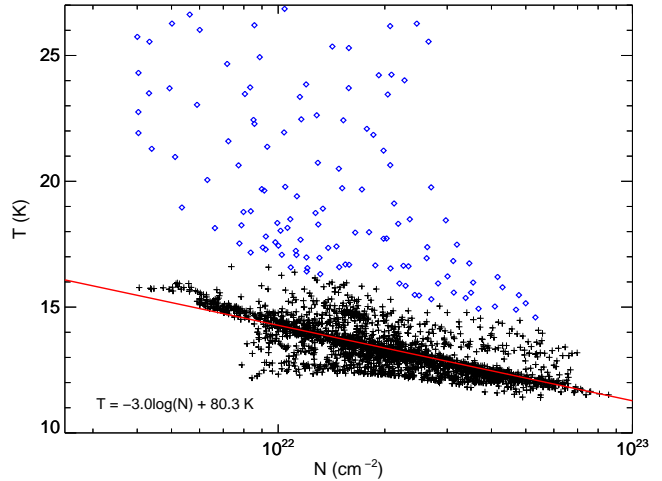}
\caption{Dust temperature ($T$) vs.\ $\log N$ for all Vela C sightlines in the subregions defined by \cite{hill_2011}.  
Blue diamonds show sightlines that were rejected by an iterative application of Chauvenet's criterion. These
{\lmfup 143} sightlines
appear to be heated by the compact \ion{H}{2} region RCW\,36.  The other
{\lmfup 2235} sightlines (crosses) appear to be heated only by the interstellar radiation field (ISRF).  The red line corresponds a fit to all ISRF-heated sightlines, as described in Section \ref{sect:p_N_T}.\label{fig_n_vs_t}}
\end{figure}

\begin{figure}
\epsscale{1.2}
\plotone{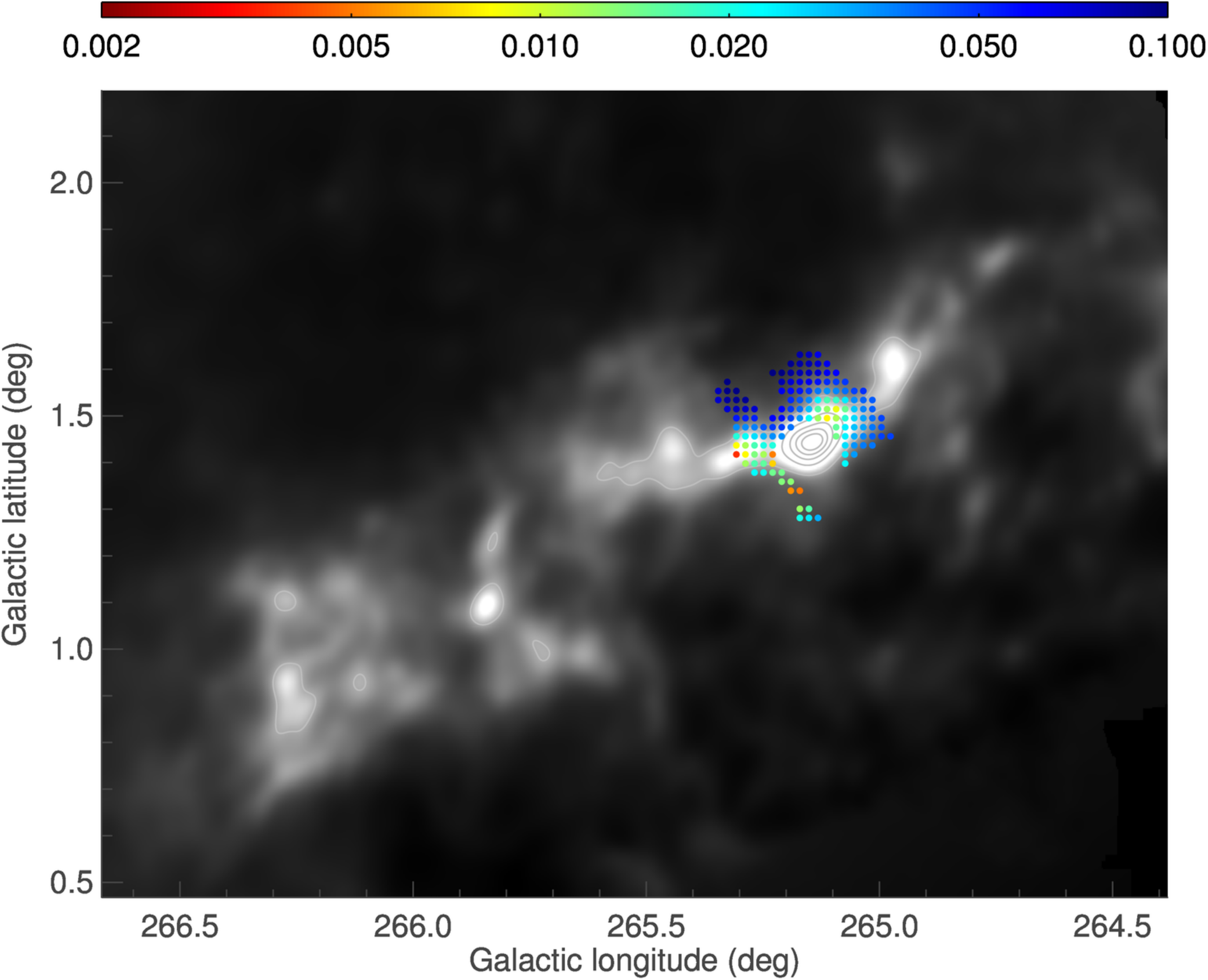}
\plotone{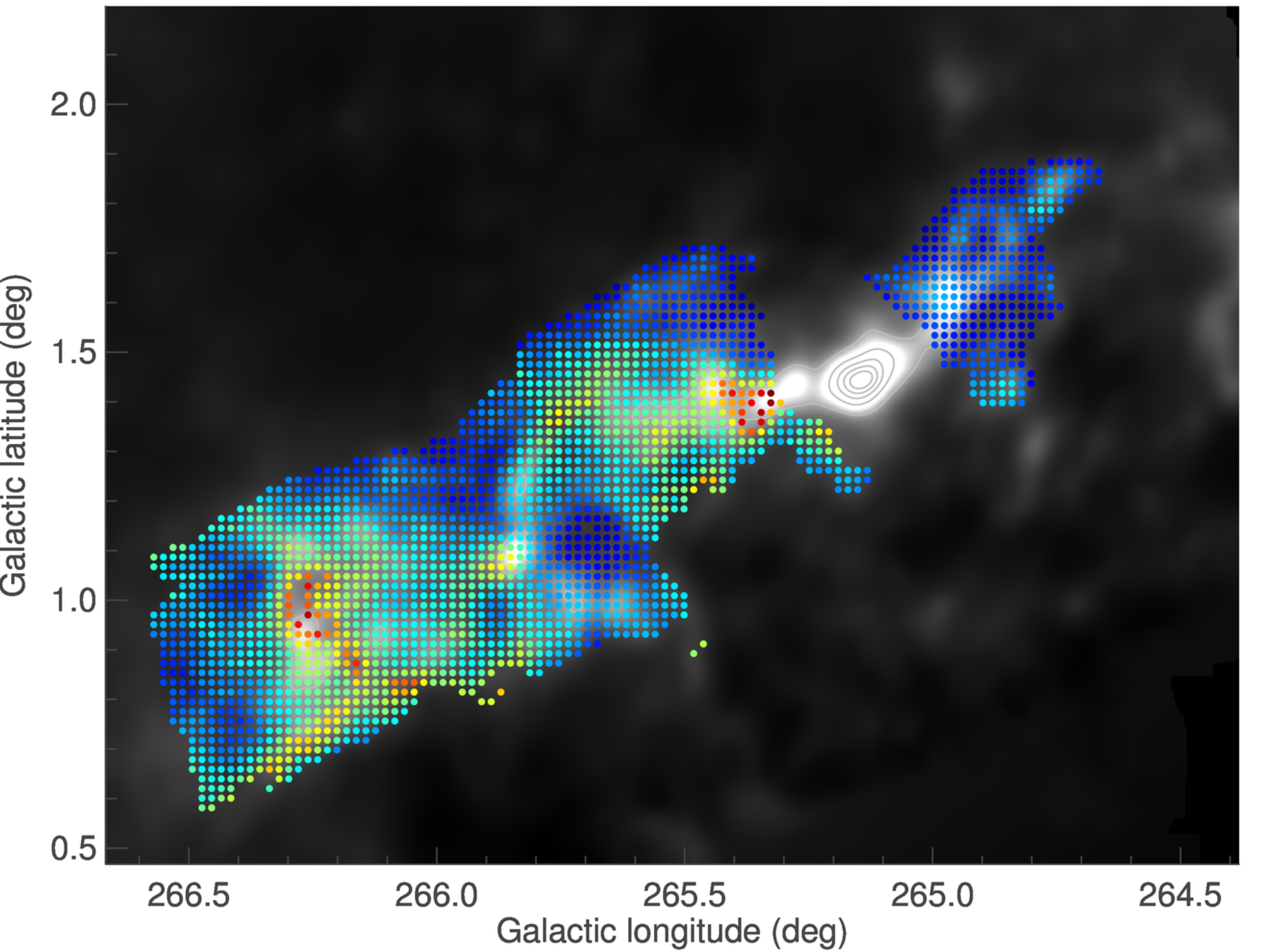}
\plotone{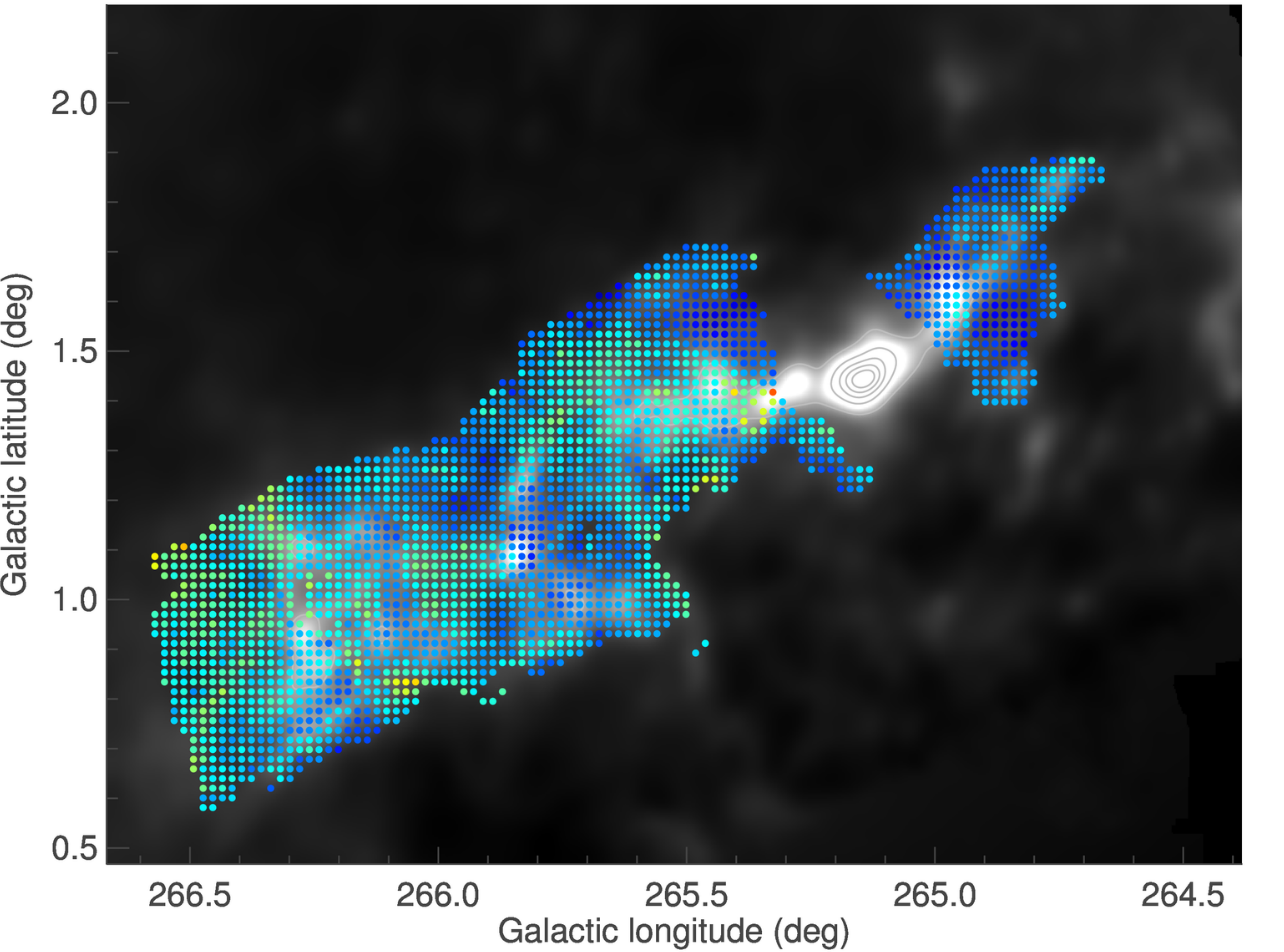}
\caption{Color-coded plot of $p$~over the range of 0.002 to 0.100 for all {\lmfup \rcw~heated (top panel)} and ISRF heated {\lmfup(middle panel)} \blastpol~sightlines that pass the criteria described in Section \ref{sect:cuts}. 
  The bottom panel shows $p^{[N,S]}$, which is $p$~{\lmfup for the ISRF heated sightlines} decorrelated from $N$~and $S$~using Equation (\ref{eqn:pdecorr_dpsi_N_plane}) in Section \ref{sect:p_N_dpsi_2corr}.  If Equation (\ref{eqn:p_dpsi_N_plane}) accounted for the entire variation of $p$, then the value of $p^{[N,S]}$~would be constant at 0.029.
The background image is $I_{500}$.\label{fig_p_xy}}
\end{figure}

{\lmfup Before considering the polarization fraction $p$, we first attempt to separate sightlines that show significant heating from sources internal to Vela\,C from sightlines that appear to be predominantly heated by the interstellar radiation field (ISRF).}
 The polarization properties of sightlines near a source of intense radiation, such as the compact \ion{H}{2} region \rcw~in Vela\,C, might differ from the polarization properties of cloud sightlines where star formation is at an earlier stage.  The presence of a bright radiation source might affect the efficiency of radiative torques in aligning dust grains with respect to the local magnetic field.  Also, the presence of expanding ionized gas in \ion{H}{2} regions can alter the magnetic field geometry, for example as seen in SPARO observations of the Carina Nebula \citep{li_2006}.  

Figure \ref{fig_n_vs_t} shows $T$~vs.\ $\log N$~for sightlines selected as discussed in Section \ref{sect:cuts}.  (Note that throughout this paper $\log$~refers to $\log_{10}$.)  
{\lmfup As discussed in Section \ref{sect:n_and_t_maps}} the ISRF can more easily penetrate sightlines of low column
and therefore average temperatures of low $N$~sightlines tend to be higher.  
{\lmfup Figure \ref{fig_n_vs_t} generally shows decreasing $T$~with increasing $\log N$, however it}
also shows that a minority of sightlines have temperatures lying well above this
{\lmfup approximately} linear trend. 
We fit the equation $T\,=\,a\,\log N\,+\,b$, using Chauvenet's 
criterion \citep{chauvenet_1863} iteratively to remove outliers (diamonds in Figure \ref{fig_n_vs_t}).  The
{\lmfup 143} sightlines rejected as outliers are located near the compact
\ion{H}{2} region \rcw~{\lmfup(upper panel of Figure \ref{fig_p_xy})}. These sightlines appear to be heated by the \ion{H}{2} region, {\lmfup yielding} temperatures {\lmfup lying} above the trend seen for ISRF heated sightlines.  

\begin{figure*}
\epsscale{1.1}
\plotone{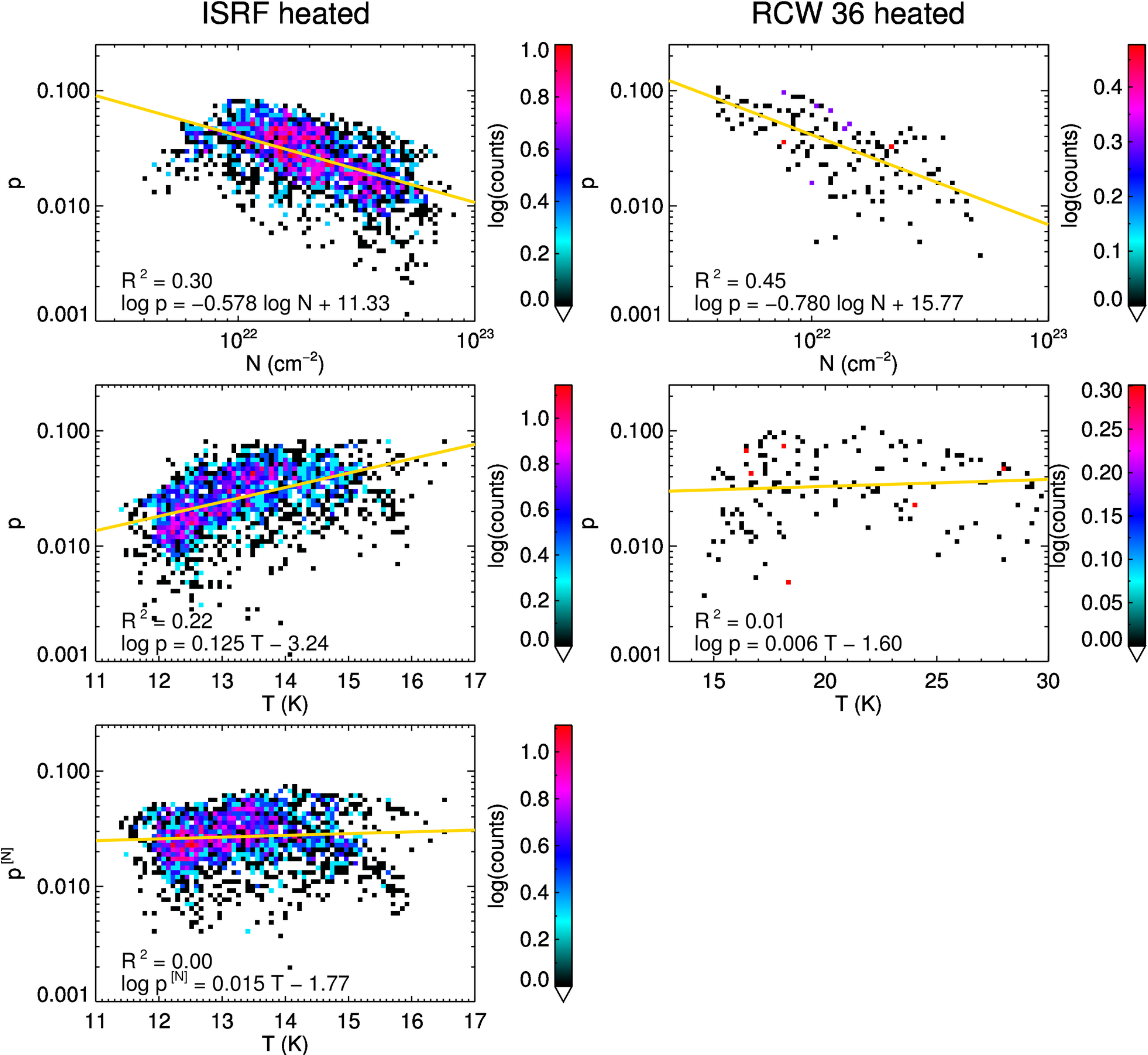}
\caption{Two-dimensional histograms showing the correlations between $p$, $N$,~and $T$~for ISRF-heated sightlines (left) and sightlines that show evidence of heating from \rcw~(right). The correlations shown are: polarization fraction ($p$) vs.\ column density ($N$) (top panels); $p$~vs.\ dust temperature ($T$) (middle panels); and $p^{[N]}$, the polarization fraction with the dependence on column density removed using Equation (\ref{p_decorr_N}) vs.\ $T$~(bottom
{\lmfup left} panel).  All data points used to make these plots passed the selection criteria described in \ref{sect:cuts}.
The color of each pixel is proportional to the logarithm of the number of data points located within the pixel.  The solid lines show fits to the data (Section \ref{sect:p_N_T}).  The best-fit equations are listed on each plot in addition to the coefficient of determination ($R^2$).\label{fig_p_n_t}}
\end{figure*}

Figure \ref{fig_p_n_t} shows the dependence of $p$\ on $N$~and $T$~for ISRF-heated sightlines (left side, top and middle panels respectively).  
In general $p$~decreases with increasing $N$~and increases with increasing $T$.  To quantify the dependence of $p$~on $N$~we fit a model of the form
\begin{equation}\label{eqn:p_vs_N_fit_no_l}
 	p\,=\,C N^{\alpha_N}
\end{equation}
where, $C$~and $\alpha_N$~are constants.  This is equivalent to a linear fit in logarithmic space
\begin{equation}\label{eqn:p_vs_N_fit}
 	\log p\,=\,\alpha_N\,\log N\,+\,C.
\end{equation}
Via a fit to Equation (\ref{eqn:p_vs_N_fit}), we find that $\alpha_N$=\alpNerr.
Each measurement of $\log p$~is given equal weight in our fit.   
By giving each data point equal weight (equal fractional error in $p$) we are assuming that the deviations of the $\log p$~data points from the fit described in Equation (\ref{eqn:p_vs_N_fit}) are caused by additional dependences of $p$~on other quantities, rather than uncertainties in our measurements of $\log p$.  This assumption is reasonable, because our polarization measurement uncertainties are generally quite small.  For example, the median signal-to-noise of our $p$~measurements is 36, and the median signal-to-noise of the $N$~measurements for these same sightlines is even higher.
The uncertainties on our fitted parameters are
calculated using the bootstrapping method with replacement \citep{nr_boot}, repeating the fits for each of 10,000 random selections.  The standard deviation of the
derived power-law exponents is used as an estimate of their uncertainty.

Similarly, for $p$~vs.\ $T$~(Figure \ref{fig_p_n_t} middle left panel),
we fit to the relation $log p\,=\, \beta_T T\,+\,c_1$, and find that $\beta_T\,=\,$\betlTerr, which implies that $p\,\propto\,\exp\left(0.29 T\right)$.  However, Figure \ref{fig_n_vs_t} shows 
that $N$\ and $T$\ are highly correlated for ISRF heated sightlines.  We can remove the correlation of $p$~with $N$~by computing:
\begin{equation}\label{p_decorr_N}
p^{[N]}_i\,=\,p_i\,\left(\frac{\bar{N}}{N_i}\right)^{\alpha_N},
\end{equation}
where $p^{[N]}_i$, $p_i$,~and $N_i$~are the \ith~decorrelated $p$~measurement, {\lmfup and} original $p$~and $N$~measurement{\lmfup s}, respectively, and $\bar{N}$~is the median value of $N$~for our sightlines. {\lmfup The bottom left panel of Figure \ref{fig_p_n_t}} shows $p^{[N]}$ vs.\ $T$. By removing the anticorrelation of $p$~with $N$, we also remove any correlation with $T$.
Thus it appears that there is no correlation between $p$\ and $T$~that is independent of the correlation between $p$~and $N$.
 
For the sightlines that show significant heating from \rcw\ we see a similar decrease in $p$~with increasing $N$~and find a 
power-law exponent of $\alpha_N=$\alpNerrRCW~(Figure \ref{fig_p_n_t} top right panel).  However, for these sightlines there is no 
apparent correlation between $p$~and $T$ (Figure \ref{fig_p_n_t} middle right panel).  

\begin{figure}
\epsscale{1.2}
\plotone{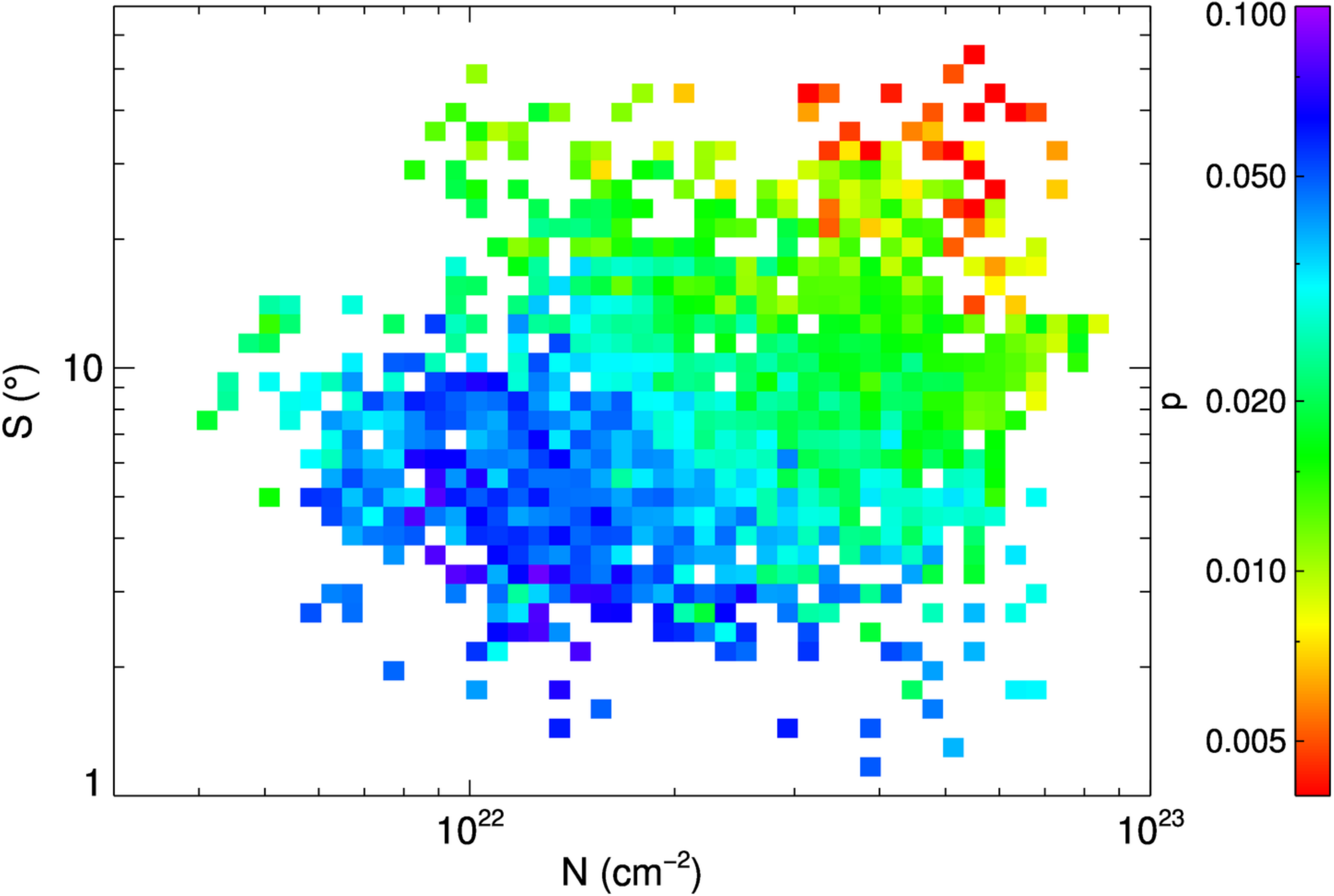}
\caption{Median $p$~color-coded in bins of $S$~and $N$~for all ISRF-heated sightlines.  The use of logarithmic scales for $p$, $N$, and $S$~brings out the systematic relationship suggestive of power-laws.  \label{fig:p_n_dpsi}}
\end{figure}
  
\section{Dependence of Polarization Fraction on $N$~and \dpsi} \label{sect:p_N_dpsi}

{\lmfup In this section our goal is to build an empirical model for the dependence of $p$~on $N$~and the polarization-angle dispersion on 2\farcm5~(0.5\,pc) scales $S$, for an early stage star-forming region. Therefore we only consider ISRF-heated sightlines.  Additionally, we do not include T as a parameter of the empirical model as it was shown in Section \ref{sect:p_N_T} that the $p$~vs.~$N$~and $p$~vs.~$T$~correlations are degenerate.  We choose $N$~rather than $T$~as our independent variable because the most natural explanation for the $N$~vs.~$T$~anticorrelation for ISRF-heated sightlines is that the density structure of the cloud determines the average temperature of the sightlines, rather than $T$~determining $N$.}  

\subsection{Individual Correlations among $p$, $N$, and \dpsi} \label{sect:p_N_dpsi_1corr}
\begin{figure}
\epsscale{1.2}
\plotone{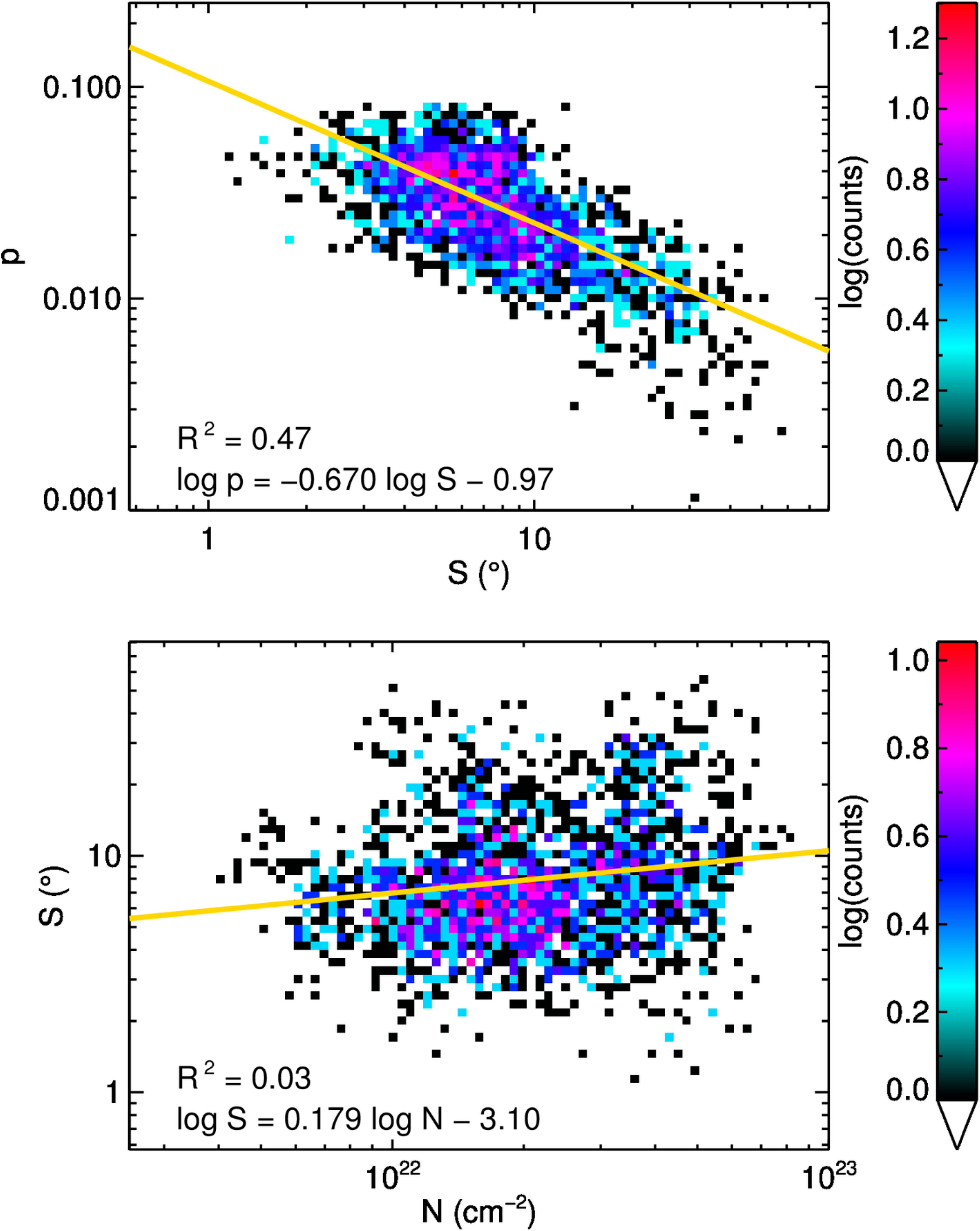}
\caption{Two-dimensional histograms showing correlations between $p$, $N$, and $S$~for ISRF-heated sightlines:  $p$~vs.\ $S$~(top panel); and $S$~vs $N$ (bottom panel).  All data points used to make these plots passed the selection criteria described in \ref{sect:cuts}.
The color is proportional to the logarithm of the number of data points located within each bin.  The solid lines show fits to the data (Section \ref{sect:p_N_dpsi_1corr}).  }\label{fig_p_n_dpsi}
\end{figure}

Figure \ref{fig:p_n_dpsi} shows the median $p$ (color map) for bins of \dpsi\ and $N$~for ISRF-heated sightlines. 
There is a clear decrease of $p$\ with increasing $N$\ and \dpsi. Individual correlations are shown in Figures \ref{fig_p_n_t} and \ref{fig_p_n_dpsi}, and the derived associated power-law exponents are listed in Table \ref{tab:ref_slopes}.

{\lmfup Decreasing} $p$\ with increasing $N$\ has been seen in submillimeter polarization observations of many clouds
{\lmfup and cores (e.g., \citealt{matthews_2001,li_2006}).
The observed decrease in $p$~is often attributed to either cancelation of polarization signal for high-$N$~sightlines due to more disorder in the magnetic field, or to changes in grain alignment efficiency within the cloud.  These possible explanations are discussed further in Section \ref{sect:discussion}.}

In Section \ref{sect:delta_psi}, we showed that Vela\,C has high values of \dpsi~in localized filament-like regions, where there are sharp changes in magnetic field direction. 
High $S$~depends implicitly on spatial changes in the magnetic field locally in the map, and any related changes in the magnetic field direction within the volume sampled by the \blastpol~beam could lead to lower $p$.  The top panel of Figure \ref{fig_p_n_dpsi} shows $p$~vs.\ \dpsi.  There is a clear anticorrelation between $p$~and\ \dpsi~
($\alpha_{S}\,=\,$\alpSerr, the coefficient of determination $R^2\,=\,$\RsqpS).
We see no dependence of $S$~on $N$~(Figure \ref{fig_p_n_dpsi}, lower panel).

\capstartfalse
\begin{deluxetable*}{lrrrr}
\tablecolumns{3}
\tabletypesize{\footnotesize}
\tablecaption{Power-law exponents of $p$~vs.\ $N$~and \dpsi
\label{tab:ref_slopes}}
\tablewidth{0pt}
\tablehead{
\colhead{Diffuse Emission}& \colhead{$\alpha_{N}$} & \colhead{$\alpha_{S}$} & 
\colhead{$\alpha_{p^{[S]}\,N}$} & \colhead{$\alpha_{p^{[N]}\,S}$}  \\
\colhead{Subtraction Method} & & & & 
}
\startdata
Intermediate      & \alpNerr       & \alpSerr       & \alpNdSerr     & \alpSdNerr    \\
Conservative      & $-$0.53\,$\pm$\,0.02 & $-$0.67\,$\pm$\,0.02 & $-$0.39\,$\pm$\,0.01 & $-$0.56\,$\pm$\,0.01 \\
Aggressive        & $-$0.66\,$\pm$\,0.02 & $-$0.66\,$\pm$\,0.02 & $-$0.57\,$\pm$\,0.01 & $-$0.59\,$\pm$\,0.01 \\
\enddata
\tablecomments{The power-law exponents listed in this table are derived from linear fits of $\log p$~to $\log N$~and $\log S$~as described in Sections \ref{sect:p_N_T} and \ref{sect:p_N_dpsi_1corr}.}
\end{deluxetable*}
\capstarttrue

\begin{figure}
\epsscale{1.2}
\plotone{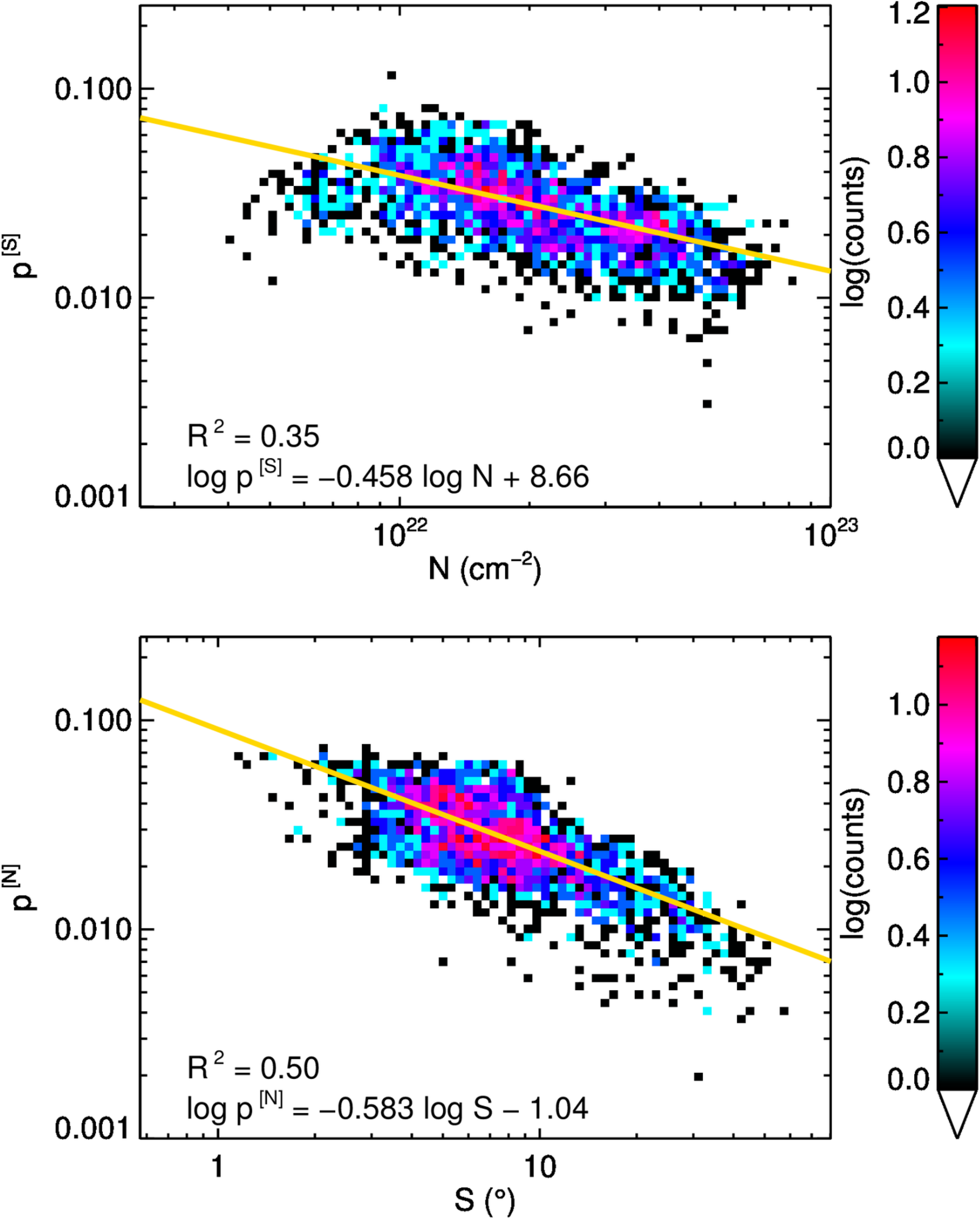}
\caption{Two-dimensional histograms showing decorrelated $p$~
as a function of $N$~and $S$~for ISRF-heated sightlines: $p^{[S]}$, the polarization fraction ($p$) with the dependence on $S$~removed vs.\ $N$~(top panel); and $p^{[N]}$, $p$~with the dependence on $N$~removed vs.\ $S$~(bottom panel).
The color is proportional to the logarithm of the number of data points within the bin.  The solid lines show the linear fits to the data (Section \ref{sect:p_N_dpsi_1corr}).  
}\label{fig_p_n_dpsi_decorr}
\end{figure}

\capstartfalse
\begin{deluxetable}{crrr}
\tabletypesize{\footnotesize}
\tablecaption{Fit parameters of $p(N,S)$~from Equation (\ref{eqn:p_dpsi_N_plane})
\label{tab:ref_slopes_plane}}
\tablewidth{0pt}
\tablehead{
\colhead{Diffuse Emission} & \colhead{$\alpha_{N}$} & \colhead{$\alpha_{S}$} & \colhead{${\lmfup K}$} \\
\colhead{Subtraction Method} & & &
}
\startdata
Intermediate & \alpbNerr & \alpbSerr & 8.42\,$\pm$\,0.3 \\
Conservative    & $-$0.41\,$\pm$\,0.01 & $-$0.59\,$\pm$\,0.01 & 6.92\,$\pm$\,0.3  \\
Aggressive     & $-$0.58\,$\pm$\,0.01 & $-$0.60\,$\pm$\,0.01 & 10.98\,$\pm$\,0.3 \\
\enddata
\tablecomments{The power-law exponents {\lmfup($\alpha_{N}$~and $\alpha_{S}$) and fitted constant (K)} listed in this table are calculated from a two-variable power-law fit to $N$~and $S$~as described in Section \ref{sect:p_N_dpsi_2corr}.}
\end{deluxetable}
\capstarttrue

We showed in Section \ref{sect:p_N_T} that the dependence of $p$~on $N$~can be removed~using Equation (\ref{p_decorr_N}) to create \pdecN.
Similarly we can normalize out the dependence of $p$~on $S$~by calculating
\begin{equation} \label{p_decorr_dpsi}
p_i^{[S]}\,=\,p_i\,\left(\frac{\bar{S}}{S_i}\right)^{\alpha_{S}}.
\end{equation}
 
The top panel of Figure \ref{fig_p_n_dpsi_decorr} shows that by 
removing the dependence of \dpsi~from $p$~the 
degree of correlation of $p^{[S]}$~with $N$~increases ($R^2\,=\,$\RsqpNdS~compared to \RsqpN).  Similarly, the {\lmfup bottom} panel shows that the 
correlation 
of $p^{[N]}$~with \dpsi~is better than the correlation of $p$~with $S$ ($R^2\,=\,$\RsqpSdN~compared to \RsqpS).  This indicates that both $N$~and $S$~contribute independently to the structure seen in $p$.
The fitted power-law exponents tend to be systematically shallower for the 
decorrelated \pdecdpsi\ and \pdecN~than for trends with $p$ (see the first row of Table \ref{tab:ref_slopes}), which might imply a weak underlying correlation between $N$~and $S$~(see Figure \ref{fig_p_n_dpsi}, bottom panel).

\subsection{Power-Law Fit $p(N,S)$} \label{sect:p_N_dpsi_2corr}

As noted in Section \ref{sect:p_N_dpsi_1corr}, Figure \ref{fig:p_n_dpsi} shows a color map of the median $p$~binned two-dimensionally in $S$~and $N$~for ISRF-heated sightlines.  The clear decrease of $p$~with both increasing $N$~and $S$~is suggestive of a joint power-law relationship.  Here we derive a function $p(N,S)$~that accounts for most of the structure seen in the $p$~map. Specifically, we adopt the joint power-law form
\begin{equation} \label{eqn:p_dpsi_N_plane}
\log p\left(N,S\right)\,=\,{\lmfup K}\,+\,\alpha_{N}\log N\,+\,\alpha_{S}\log S,
\end{equation}
where ${\lmfup K}$, $\alpha_{N}$~and $\alpha_{S}$ are the free parameters.

The exponents derived via a fit to Equation (\ref{eqn:p_dpsi_N_plane}) are $\alpha_{N}\,=\,$\alpbNerr~and~$\alpha_{S}\,=\,$\alpbSerr.  Just as in Sections \ref{sect:p_N_T} and \ref{sect:p_N_dpsi_1corr}, errors in fit parameters are derived via bootstrapping (Table \ref{tab:ref_slopes_plane}). We note that, as expected, $\alpha_{N}$~and $\alpha_{S}$~derived from the two-variable power-law fit to $N$~and \dpsi~are identical within the error bars to 
$\alpha_{p^{[S]}N}$ (the power-law fit to $p^{[S]}$~as a function of $N$)
and $\alpha_{p^{[N]}S}$ (the power-law fit to $p^{[N]}$~as a function of \dpsi), which were derived in Section \ref{sect:p_N_dpsi_1corr} (also see Table \ref{tab:ref_slopes}).  

We can remove the dependence of $p$~on $N$~and \dpsi~via
\begin{equation} \label{eqn:pdecorr_dpsi_N_plane}
p_{i}^{[N,S]}\,=\,\frac{p_i\,\bar{p}}{p\left(N_i,S_i\right)},
\end{equation}
where $p_{i}^{[N,S]}$~is the decorrelated $p$~for the \ith~data point, 
$p_i$~is the measured polarization fraction for the \ith~data point, 
  $p\left(N_i,S_i\right)$
is the value of the two-variable power-law fit for the \ith~data point,
and $\bar{p}\,=\,0.029$~is the median value of $p$.  A
{\lmfup comparison of the spatial distribution} of {\lmfup $p$~(middle panel) with} $p^{[N,S]}$~{\lmfup(bottom panel)} is shown in Figure \ref{fig_p_xy}.  We discuss potential causes of residual structure in the $p^{[N,S]}$~map in Section \ref{sect:pdecorr_scatter}.

\begin{figure*}
\epsscale{1.0}
\plotone{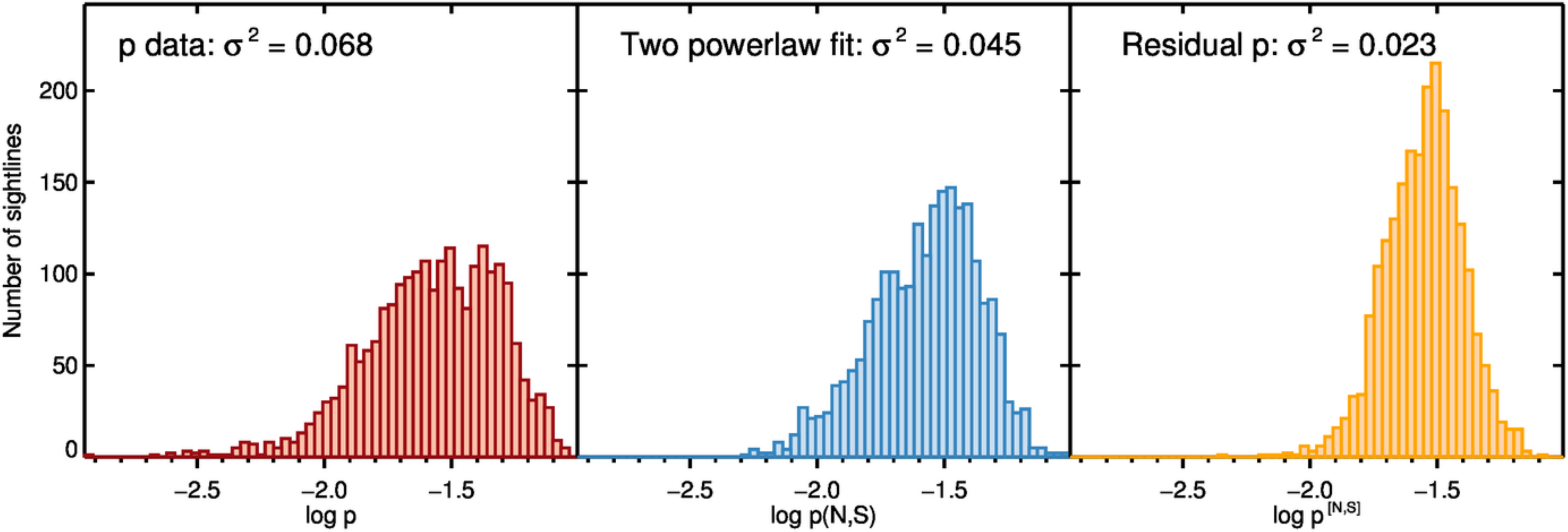}
\caption{Histograms of the logarithms of $p$~before decorrelation (left panel), $p\left(N,S\right)$~evaluated for $N$~and $S$~data using the model described in Equation (\ref{eqn:p_dpsi_N_plane}) (center panel), and  $p^{[N\,S]}$, the residual structure in $p$ after normalizing out $p\left(N,S\right)$~(right panel).  The variance ($\sigma^2$) of the distribution is given  at the top of each panel.  The quantity $p^{[N\,S]}$~
was normalized so that the median $p$~remained at 0.029 (see Equation (\ref{eqn:pdecorr_dpsi_N_plane})). 
 }\label{fig:p_histos}
\end{figure*}

Finally, we quantify the degree to which our two-variable power-law fit $p\left(N,S\right)$~can reproduce the 
observed dispersion in $p$. 
Figure \ref{fig:p_histos} shows histograms of: our calculated $\log p$ {\lmfup (left panel)}; $\log p\left(N,S\right)$,~the two-variable power-law fit values calculated for our $N$~and $S$~data points {\lmfup (center panel)}; and $\log p^{[N,S]}$, $p$~with the derived dependence on $N$~and $S$~removed using Equation (\ref{eqn:pdecorr_dpsi_N_plane}) {\lmfup (right panel)}.
For each of the three cases the median $p$~is 0.029.
Histograms of $\log p$~rather than $p$~are shown, because the fits are made in log space. The variances of $\log p$, $\log p\left(N,S\right)$, and $\log p^{[N,S]}$~are 0.068, {\lmfup 0.045}, and 0.023, respectively.  Our model $\log p\left(N,S\right)$~reproduces {\lmfup 66}$\%$~of the variance in the $\log p$~map, which shows that our two-variable power-law fit model captures most of the physical effects that determine variations in fractional polarization in Vela\,C.

\subsection{Uncertainties in the Power-Law Fit Exponents}\label{sect:exp_uncert}
The uncertainties of the exponents for the two-variable power-law fits $\alpha_N$~and $\alpha_S$~were 
estimated using the bootstrapping methods described in Section \ref{sect:p_N_T}.  
However, as discussed in Section
 \ref{sect:ref_reg}, the limiting uncertainty is our inability to precisely separate the contribution of background/foreground dust from the polarized emission of Vela\,C.  
  We repeated our analysis for maps of $p$~and \dpsi~made 
with the ``conservative'' and ``aggressive'' diffuse emission subtraction
methods, to gauge the systematic uncertainty of our derived power-law exponents.
Table \ref{tab:ref_slopes}
 gives power-law exponents derived from the individual correlations 
(Section \ref{sect:p_N_dpsi_1corr})
for the three different  
diffuse dust emission subtraction methods.  Table \ref{tab:ref_slopes_plane} lists 
the exponents $\alpha_{N}$~and $\alpha_{S}$~of the two-variable power-law fits 
again for all three diffuse emission subtraction methods.  Systematic uncertainties relating to the choice of subtraction method are seen to be $\sim$0.1 for $\alpha_{N}$~and $\sim$0.01 for $\alpha_{S}$.

\section{Discussion}\label{sect:discussion}

\subsection{Implications of the Dependence of $p$~on $N$~and $T$} \label{sect:disc_p_N_T}
In Section \ref{sect:p_N_T} we examined the dependence of the polarization fraction $p$~on
column density $N$~and dust temperature $T$ in Vela\,C.  We divided our Vela\,C sightlines
into two groups: those that show evidence of heating
from the compact \ion{H}{2} region \rcw, and those sightlines where the temperature decreases 
as $e^{-\betT N}$.  For the latter sightlines we suggested that the dust 
temperature is primarily set by exposure to the interstellar radiation field (ISRF),
with high $N$~sightlines having on average more shielding and 
therefore receiving less heating per unit mass from the ISRF.

For the ISRF-heated sightlines, we find that $p$~decreases with increasing $N$~and also that $p$~increases as $T$~increases.  
Depolarization for higher column density sightlines has been seen in many 
studies (see Section \ref{sect:p_N_dpsi_1corr}).
\cite{vail_2012}~used the ratio of $F$(850\um)/$F$(350\um) as a proxy for dust 
temperature in two massive star forming clouds.  
They found that the polarization tended to decrease with increasing
$F$(850\um)/$F$(350\um), implying that warmer dust grain populations tend to have a 
higher $p$.  This agrees with our result, but \cite{vail_2012} caution that 
variations in $F$(850\um)/$F$(350\um) could be due to changes in dust spectral index, rather than just dust temperature. 
Our study is the first to 
fit $p$~measurements within a molecular cloud as a function of both $N$~and $T$.  For the ISRF-heated sightlines, which are the majority of the sightlines, we find that the dependence of $p$~on $N$~is not separate from the dependence of $p$~on $T$, since $N$~and $T$~are highly correlated.

{\lmfup There are two general classes of explanations for our observations of $p$~vs.~$N$~and $T$ for the ISRF-headed sightlines.  We may have greater magnetic field disorder along high $N$~(and therefore lower $T$) dust columns, or we may have a decrease in the intrinsic polarization efficiency (see Section \ref{sect:intro}) for such sightlines.  In the first explanation the increased field disorder could arise because of a higher field disorder at high particle densities $n$, or because high $N$~sightlines pass through more cloud material and therefore may sample different field directions at different locations along the line of sight \citep{jones_1989}.  Regarding the second possibile explanation for our observed $p$~vs.~$N$~and $T$~trends, note that in the}
radiative torques (RATs) model of grain alignment, ``alignment torques" from an anisotropic radiation field are responsible for aligning the dust grain spin-axes with the local magnetic field \citep{lazarian_hoang_2007}.  Grains near the surfaces of molecular clouds (low $N$, high $T$) thus might be expected to show a higher average polarization fraction \citep{cho_lazarian_2005}. Alternatively, dust grain properties could change at high densities, e.g., grains could become rounder due to accretion of icy mantles \citep{whittet_2008}.

For our \rcw~heated sightlines there is 
a significant anticorrelation between $p$~and $N$ ($R^2$\,=\,{\lmfup 0.45}). However, for these heated sightlines there is no correlation between $N$~and $T$~and no strong correlation between $p$~and $T$ ($R^2\,=\,{\lmfup 0.03}$).  
This could indicate that the primary dependence of $p$~is on $N$, 
rather than $T$~{\lmfup and that the correlation of $p$~with $T$~only appears when there is a 
strong correlation of $T$~with $N$.}  
However, we caution that we have relatively few sightlines near \rcw~(\nVrcw~Nyquist-sampled sightlines compared to \nVisrfNT~ISRF-heated sightlines) so the lack of 
correlation between $p$~and $T$~could be caused by the angle of the magnetic field changing with respect to the line of sight, which would cause more spread in $p$.  Indeed Figure \ref{fig:maps_drapery} shows that near \rcw~there are significant changes in the inferred magnetic field orientation projected onto the plane of the sky. 

\subsection{Implications of the Two-variable Model $p(N,S)$} \label{sect:exponents}

In Section \ref{sect:p_N_dpsi_2corr} we fit a model that describes $p$ as a function with a 
power-law dependence on two variables, hydrogen column density $N$, and {\lmfup\dpsi, the dispersion in the polarization-angle on scales corresponding to our beam FWHM (2\farcm5, or 0.5\,pc)}. 
The derived power-law exponents are $\alpha_{N}$\,=\,\alpbbNerr~for the dependence on $N$, 
and $\alpha_{S}$\,=\,\alpbbSerr\ for the 
dependence on \dpsi~(see Section \ref{sect:exp_uncert}).  
Our $p(N,S)$~fit is able to reproduce most of the {\lmfup structure} seen in our 
$\log p$~maps.

The decrease in $p$ with increasing $S$~can be attributed to changes in the magnetic field 
direction within the volume probed by the \blastpol~beam. 
The mean magnetic field orientation, $\Phi$, is an average over {\lmfup both} the 
beam area (0.5\,pc) and along the length of the cloud in the line of sight direction, and is weighted by the density and intrinsic polarization efficiency (Section \ref{sect:intro}).
Large values of $S$~indicate a substantial change in $\Phi$ on the scale of a beam, which implies a significant change in the orientation of polarization within the beam.  This could be due to a sharp change in the magnetic field direction at some location within the cloud. Alternatively, it could indicate the overlap of two clouds,
well separated along the line of sight, each with a 
different $\Phi$.  In either case we should see an overall decrease in the measured polarization fraction, since some of the 
polarization components cancel.  \cite{planck2014-XX} note a decrease of $p$~with increasing $S$, both in their data and in corresponding MHD simulations (see Figure 19 of \citealt{planck2014-XX}).
However the \planck~study sampled $5\,\times\,10^{20}\,$cm$^{-2}\,<\,N\,<\,10^{22}\,$cm$^{-2}$ while our Vela\,C observations predominantly sample $10^{22}\,$cm$^{-2}\,<\,N\,<\,10^{23}\,$cm$^{-2}$. 
Also direct comparison with their derived power-law exponents is difficult, since they fit $S$~vs.\ $p$, thus minimizing the scatter in $S$, while we fit $p$~vs.\ $S$,~which minimizes scatter in $p$. Nevertheless they do find a significant anti-correlation of $p$~and $S$~in their data that is reproduced in their MHD simulations.  In these simulations there is by contrast only a weak correlation of $S$~with $N$\ (F. Levrier, private communication), just as we found in our data (Figure \ref{fig_p_n_dpsi}, lower panel). 

{\lmfup In Section \ref{sect:disc_p_N_T}, we discussed two classes of explanations for the observed $p$~vs.~$N$ trend.  The first class involves magnetic field disorder.  An example is the work of \cite{fg_2008}.  These authors were able to reproduce a decrease in $p$~with increasing $N$ via}
 synthetic polarization maps made from 
 supersonic, sub-Alfv\'{e}nic MHD 
 molecular cloud simulations, assuming uniform intrinsic polarization 
 efficiency.  Their power-law exponents $\alpha_N$~ranged from 0 to $-$0.5, 
 with models where the mean magnetic field was in the plane of the sky ($\gamma\,=\,$0\deg) having the steepest slope 
 and models where the mean field was parallel to the line of sight ($\gamma\,=\,$90\deg) having no dependence
 of $p$~on $N$.
 In this theoretical study, the decrease in polarization for 
 higher column density regions is due to an increase in the dispersion of the magnetic field direction for high density regions. {\lmfup An analytic model by \cite{jones_1989}, is similarly able to reproduce a falling $p$~vs.~$N$ for a medium having uniform intrinsic polarization efficiency.}

Our analysis shows only a weak correlation (or perhaps no correlation) between \dpsi~and $N$ (see Section \ref{sect:p_N_dpsi}).  
Thus on 0.5\,pc scales, we find no significant increase in the dispersion of $\Phi$~for sightlines of increasing column density.  
{\lmfup Such an increase might be expected if 
disorder in the magnetic field direction increased in high density regions (for example due to accretion-driven turbulence as in \citealt{hennebelle_2013}), or if the magnetic field were affected by large-scale gas motions near self gravitating filaments.  
In} the above-mentioned theoretical study by \cite{fg_2008}, the authors showed that rarefied cloud regions show little
variation in polarization direction whereas significant fluctuations in direction do occur within
dense condensations. 
In this case, one might expect a positive correlation between $S$~and $N$, which we do not see in our observations.
  
{\lmfup The second class of explanations for the observed decrease in $p$~with increasing $N$~involves changes in intrinsic polarization efficiency.  This idea derives support from the observations of \cite{whittet_2008}.  These authors}
measured the near-IR polarization 
of background stars in four nearby molecular clouds. 
{\lmfup For studies of polarization of background starlight the quantity that is analogous to fractional polarization of dust emission is referred to as the ``polarization efficiency'', defined as the fractional polarization of the starlight divided by the extinction optical depth at the same wavelength
$p{_\lambda}/\tau_{\lambda}$. \cite{whittet_2008} found that the polarization efficiency in their clouds} was consistent with a power-law dependence, 
$p{_\lambda}/\tau_{\lambda}\,\propto\,A_V^{-0.52}$. 
Because the inferred magnetic field direction is mostly uniform across the
region studied, they attributed all of the decrease in polarization efficiency with increasing $N$\ to changes in the 
intrinsic polarization efficiency.  It is interesting to 
note that our power-law exponent ($\alpha_{N}$\,={\lmfup \,\alpbbNerr}) is
similar to that found by \cite{whittet_2008}.
Other starlight polarization studies have found
power-law exponent values ranging from $-$0.34 to $-$1.0 \citep{goodman_1995,gerakines_1995,arce_1998,chapman_2011,alves_2014,cashman_2014,jones_2015}. 
Ground-based studies of polarized thermal dust emission yield similar results.  For example \cite{matthews_2002} examined $p$~vs.\ $I_{850}$~for three clouds in Orion B South and found $p\,\propto\,\left(I_{850\mu m}\right)^{\alpha}$, with $\alpha$~ranging from $-$0.58 to $-$0.95.

{\lmfup Which of the two general classes of explanations for the observed $p$~vs.~$N$~trend best explains our observations of Vela\,C?
Naively, the absence of a correlation between $S$~and $N$~would suggest that magnetic field disorder does not increase towards high $N$~sightlines, which would imply that variation in intrinsic polarization efficiency is the more likely explanation.  However, if the increased disorder in the field occurs on much smaller scales than 0.5\,pc, the scale probed by $S$, then $S$~is not sensitive to the random component of the field and so we would not expect a correlation between $N$~and $S$. We} emphasize that detailed statistical comparisons with simulations of magnetized clouds that include variations in intrinsic polarization efficiency are needed to fully understand the origin of the $p$~vs.\ $N$~anticorrelation (e.g., \citealt{soler_2013}).  Such comparisons are beyond the scope of the present paper.

\subsection{Analytic Models of $p$~vs.\ $N$} \label{sect:poleff_model}

\begin{figure}
\epsscale{1.2}
\plotone{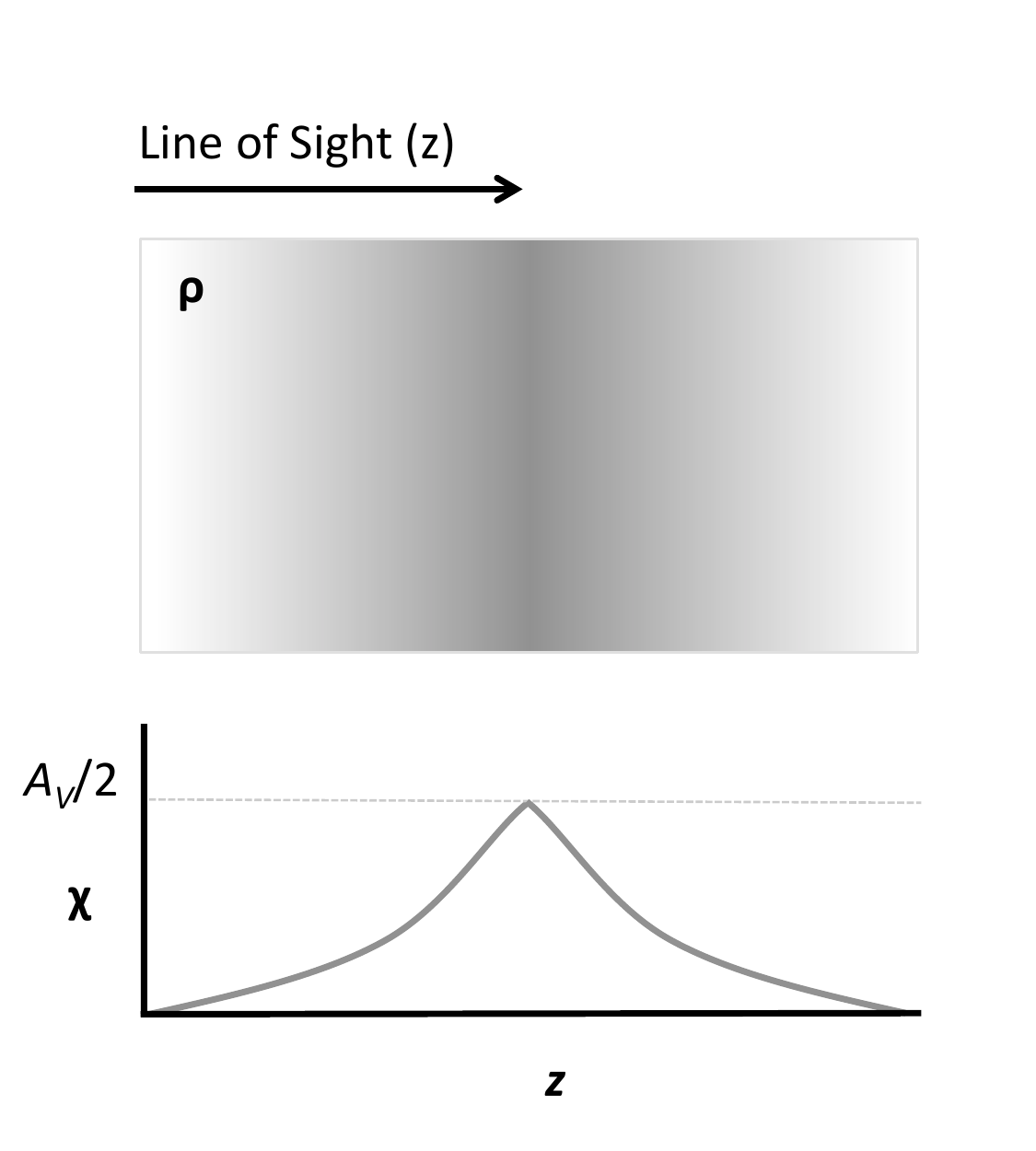}
\caption{Cartoon showing the parameterized depth into the cloud $\chi$~for a slab model of a molecular cloud.  {\lmfup The slab lies parallel to the plane of the sky}. For a given position in the cloud along the line of sight ($z$) $\chi$~is equal to the integrated visual dust extinction to the nearest cloud surface.  The maximum value of $\chi$~for a sightline of total visual extinction $A_\mathrm{V}$~is $A_\mathrm{V}/2$. \label{fig:chi_cartoon}}
\end{figure}

\begin{figure}
\epsscale{1.2}
\plotone{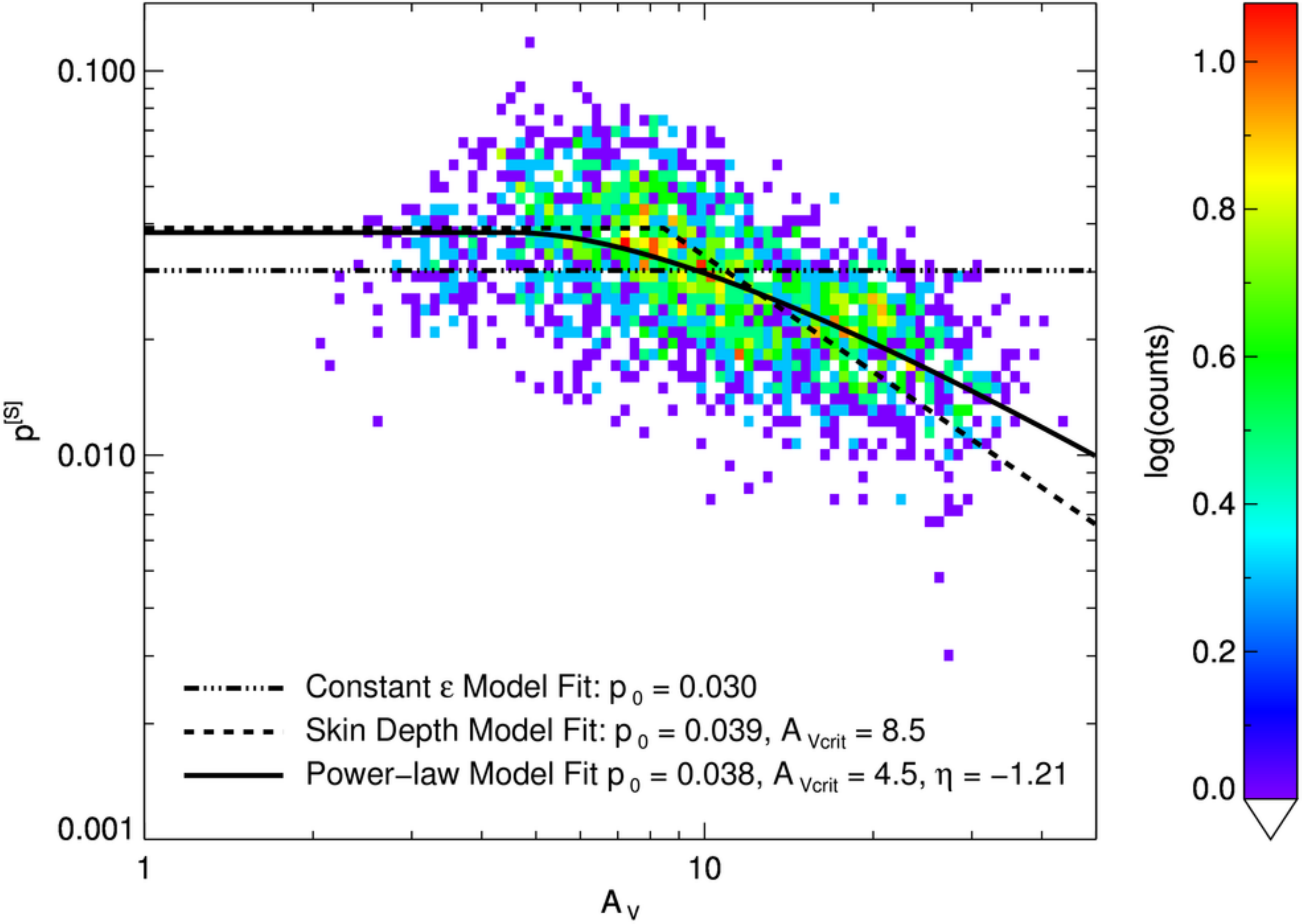}
\caption{BLASTPol measurements of the polarization fraction with the dependence of $S$~normalized out ($p^{[S]}$)~vs.\ $A_\mathrm{V}$.
The solid line shows the results of a least-squares fit to the power-law decay intrinsic polarization
efficiency ($\epsilon$) model {\lmfup with derived parameters $p_0\,=\,0.038$, $A_{\mathrm{V\,crit}}\,=\,4.5$\,mag~and
$\eta\,=\,-1.21$}. The dashed-dotted line shows a fit to the constant $\epsilon$~model 
with {\lmfup best-fit parameter {\lmfup$p_0$\,=\,$0.030$}}. The dashed line shows a fit to the ``skin depth" $\epsilon$~model, {\lmfup where the best-fit parameters are $p_0$\,
{\lmfup =\,$0.039$
$A_{\mathrm{V\,crit}}\,=\,8.5$\,mag.}}\label{fig:model_pN}}
\end{figure}

In Section \ref{sect:exponents} we advanced various explanations for the anticorrelation between $p$~and $N$~in Vela\,C.  Here we consider an extreme case where all of the dependence of $p$~on $N$~is due to reduced intrinsic polarization efficiency in shielded regions.  
{\lmfup Our goal is to quantify the ability of our measurements to trace magnetic fields deep inside the Vela\,C cloud under this pessimistic assumption.}
If most of the polarized emission comes from the outer diffuse layers of the
cloud then our derived magnetic field orientations will not be sensitive
to changes in the magnetic field direction within dense structures embedded deep in the cloud.  

{\lmfup We model the efficiency of the dust along a given cloud sightline in emitting polarized radiation with $\epsilon$, where epsilon is normalized such that $\epsilon\,=\,\xi\,\cos^2\left(\gamma \right)\,I/A_{\mathrm{V}}$, where $\xi$~is the intrinsic polarization intensity as defined in Section \ref{sect:intro}, $A_{\mathrm{V}}$~is the total dust extinction in the $V$~band for that sightline, and $\gamma$~is the angle of the magnetic field with respect to the plane of the sky.}
For these models
 $\epsilon\,=\,\epsilon\left(\chi\right)$, where $\chi$~is the parameterized depth into the cloud, which is equal to the integrated visual extinction to the nearest cloud surface as indicated in Figure \ref{fig:chi_cartoon}. 
Note that these models make a number of assumptions: (a) that the cloud is isothermal; (b) that the dust emissivity does not change within the cloud; (c) that the magnetic field direction is uniform; and (d) that the geometry of the cloud is slab-like.  

{\lmfup As the $p$~vs.~$N$~and $p$~vs.~$S$~trends appear to be independent (see Section \ref{sect:p_N_dpsi}) we compare predictions from our models with $p^{[S]}$, the polarization fraction decorrelated from $S$~(Figure \ref{fig_p_n_dpsi_decorr}, upper panel).}
Figure \ref{fig:model_pN} {\lmfup shows the predicted $p$~from three models of $\epsilon\left(\chi\right)$} compared to our measurements of $p^{[S]}$~vs.	{\lmfup$A_{V}$}, where here $N$~has been converted to \av~assuming $N(H)/$\av$\,=\,2\,\times\,10^{21}$\,cm$^{-2}$~(\citealt{bohlin_1978} with $R_{\mathrm{V}}\,=$\,3.1).  
Because of the normalization of $p^{[S]}$~the overall level of polarization is somewhat arbitrary, but of the right order.  

Here we describe {\lmfup the} three plausible models of $\epsilon\left(\chi\right)$~{\lmfup shown in Figure \ref{fig:model_pN}}:

\noindent{\em Constant $\epsilon$~Model:} If the intrinsic polarization 
efficiency were constant throughout the cloud we would expect no dependence of $p$~on $N$. {\lmfup The best fit to this model is shown as a} dashed-dotted line in Figure \ref{fig:model_pN}.   

\noindent{\em Skin Depth Model:} Alternatively, we can consider a model where the intrinsic polarization efficiency 
is constant up to an extinction depth $\chi_{crit}$~and zero thereafter;
i.e., a diffuse layer near the cloud surface is responsible for all of the polarized emission and the dust at cloud depths above $\chi_{\mathrm{crit}}$~does not contribute to the polarized emission.
For a sightline of total extinction $A_\mathrm{V}$~the 
maximum value of $\chi$~is $A_\mathrm{V}/2$.
We express $\epsilon$~as
\begin{eqnarray}\label{eqn:eps_skin}
\epsilon\left(\chi\right) = \left\{ \begin{array}{rl}
 \epsilon_0\, , &\mbox{ for $\chi\leq\chi_{\mathrm{crit}}$;} \\
  0\, , &\mbox{ for $\chi>\chi_{\mathrm{crit}}$,}
       \end{array} \right.
\end{eqnarray}
{\lmfup where $\epsilon_0$~is a constant,} and from this we can calculate the total polarized intensity:
\begin{eqnarray}\label{eqn:bigP_skin}
P\,&=&\,2\int_{0}^{A_{\mathrm{V}}/2}\epsilon\left(\chi\right)\,d\chi\,\nonumber\\ 
   &=&\,\left\{ \begin{array}{ll}
 A_{\mathrm{V}}\epsilon_0,&\mbox{ for $A_{\mathrm{V}}\,\leq\,A_{\mathrm{V\,crit}}$;} \\
 A_{\mathrm{V crit}}\epsilon_0,&\mbox{ for $A_{\mathrm{V}}\,>\,A_{\mathrm{V\,crit}}$,}
       \end{array} \right.
\end{eqnarray}
where we define $A_{\mathrm{V crit}}\,=\,2\chi_{\mathrm{crit}}$. The percentage polarization for a given sightline is then
\begin{eqnarray}\label{eqn:p_model_skin}
p\,=\,\frac{P}{I}\,= \left\{ \begin{array}{ll}
 p_0\, , &\mbox{ for $A_{\mathrm{V}}\,\leq\,A_{\mathrm{V\,crit}}$;} \\
 p_0\,\frac{A_{\mathrm{V\,crit}}}{A_{\mathrm{V}}}\, , &\mbox{ for $A_{\mathrm{V}}\,>\,A_{\mathrm{V\,crit}}$.}
       \end{array} \right.
\end{eqnarray}
In the ``skin depth'' model, $p$~is constant for sightlines with $A_\mathrm{V}\,\leq\,A_{\mathrm{V crit}}$ and decreases with a power-law slope of $-$1.0 for sightlines with $A_\mathrm{V}\,>\,A_{\mathrm{V crit}}$.  The dashed line in Figure \ref{fig:model_pN} shows a fit to the skin depth model.

\noindent {\em Power-law Model:} Finally we consider a model where the polarization efficiency is constant up to 
$\chi_{\mathrm{crit}}$ and thereafter decreases as a power-law with coefficient $\eta$:
\begin{eqnarray}\label{eqn:pl_eps}
\epsilon\left(\chi\right) = \left\{ \begin{array}{ll}
 \epsilon_0, &\mbox{ for $\chi \leq \chi_{\mathrm{crit}}$;} \\
  \epsilon_0\,\left(\frac{\chi}{\chi_{\mathrm{crit}}}\right)^{\eta}, &\mbox{ for $\chi>\chi_{\mathrm{crit}}$}.
       \end{array} \right.
\end{eqnarray} 
This model simulates a constant $\epsilon$~for the diffuse outer cloud layers and a decreasing $\epsilon$~at greater cloud depths.  
The polarized intensity for a given sightline described by the power-law model is:
\begin{eqnarray} \label{eqn:bigP_model_pl}
  P\, & = &\left\{ \begin{array}{ll}
    \epsilon_0\,A_{\mathrm{V crit}}\,a, &\mbox{ for $a\,\leq\,1$;} \\
    \epsilon_0\,A_{\mathrm{V crit}}\left(1\,+\,\frac{1}{\zeta}
    \left[a^{\zeta}-1\right]\right),
    & \mbox{ for $a\,>\,1$,} \end{array}\right.
\end{eqnarray}
where $\zeta\,=\,\eta\,+\,1$ and $a\,\equiv\,A_\mathrm{V}/A_{\mathrm{V crit}}$. The corresponding fractional polarization is then
\begin{eqnarray} \label{eqn:peff_pl}
p\,& = &\left\{ \begin{array}{ll}
 p_0, &\mbox{ for $a\,\leq\,1$;} \\
 p_0\,a^{-1}\left(1\,+\,\frac{1}{\zeta}
    \left[a^{\zeta}-1\right]\right), &\mbox{ for $a\,>\,1$.}
       \end{array} \right.
\end{eqnarray}
The power-law $\epsilon$~model best fit parameters are
$p_0\,=\,0.038$, $A_{\mathrm{V crit}}$\,{\lmfup =\,4.5\,mag~and
$\eta\,=\,{\lmfup-1.21}$}~(Figure \ref{fig:model_pN} solid line). 
Our power-law model fit would imply that at cloud depths of about {\lmfup two} magnitudes or greater of visual extinction the intrinsic polarization efficiency decreases with depth into the cloud as ${\lmfup \sim\chi^{-1.21}}$.  

{\lmfup It can be seen that both the skin-depth and power-law model capture the negative slope of the $p^{[S]}$~vs.~$N$~curve for high $N$.  For the purposes of quantifying our ability to trace magnetic fields deep within the cloud, we will use the power-law model as it seems to more closely follow the data points in Figure \ref{fig:model_pN}.}
We also caution that these are all simple models, so the fits should be taken merely as indicative of the trends of $\epsilon$~with $\chi$.

\begin{figure}
\epsscale{1.15}
\plotone{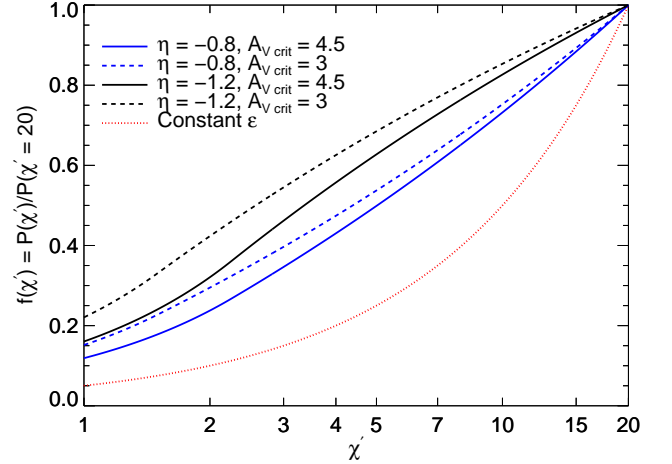}
\caption{{\lmfup Models for the fraction of the total polarized intensity for a sightline of $A_V\,=\,40\,$mag~($\chi_{\mathrm{max}}\,=\,20\,$mag), contributed by all dust at cloud depths $<\,\chi^{\prime}$~(see Equation (\ref{eqn:fP_chi})).}
The line color represents the power-law slope $\eta$~assumed: {\lmfup$-$0.8} (blue); and {\lmfup$-$1.2} (black).
Linestyle represents the $A_{\mathrm{V crit}}$~assumed: 3.0 {\lmfup mag} (dashed); and  
{\lmfup 4.5 mag} (solid).  The red dotted line shows the expected $f(\chi^{\prime})$~for dust of constant intrinsic polarization efficiency $\epsilon$.
\label{fig:model_rp_vs_chi}
}
\end{figure}

\begin{figure}
\epsscale{1.15}
\plotone{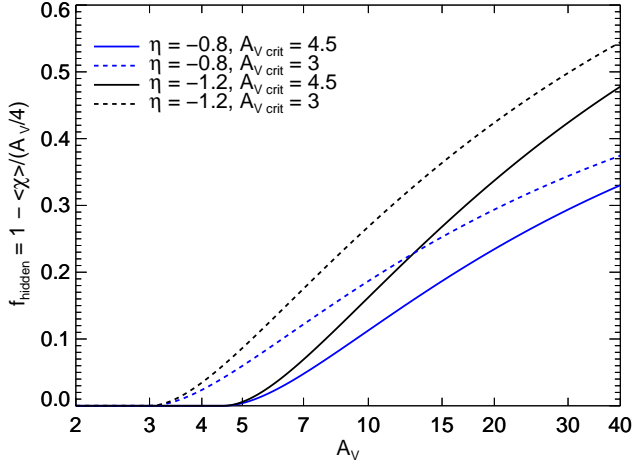}
\caption{
Fraction of the dust column that is hidden, i.e. not
traced by polarized emission as a function of $A_\mathrm{V}$ {\lmfup as described by Equation (\ref{eqn:fhidden})}.
The line color indicates $\eta$~and the linestyle indicates $A_{\mathrm{V crit}}$.\label{fig:chibar}
}
\end{figure}

Using Equation (\ref{eqn:bigP_model_pl}) for a sightline of total 
dust extinction $A_\mathrm{V}$~we can calculate  the fractional contribution to the polarized intensity from cloud material at  
depths of $<\,\chi^{\prime}$ 
\begin{equation}\label{eqn:fP_chi}
f(\chi^{\prime})\,=\,\frac{P\left(\chi^{\prime}\right)}{P(\chi_{\mathrm{max}})}
\end{equation}
where by definition $\chi_{\mathrm{max}}\,=\,A_\mathrm{V}/2$.  
Figure  \ref{fig:model_rp_vs_chi}~shows $f\left(\chi^{\prime}\right)$ for a sightline of total dust extinction 
$A_\mathrm{V}\,=\,{\lmfup 40\,mag}$ (about the largest found in \velac) over the range of $\chi^{\prime}\,=\,1$~to $\chi^{\prime}\,=\,\chi_{\mathrm{max}}\,=\,{\lmfup 20\,mag}$.
{\lmfup Dashed and solid lines represent different assumptions for 
$A_{\mathrm{V crit}}$, and line colors represent different} power-law slopes $\eta$ in Equation (\ref{eqn:bigP_model_pl}).  The solid black line is derived using the best-fit parameters.  
For comparison we also show the expected behavior for the constant $\epsilon$~model (red dotted line).  
For our best fit parameters ${\lmfup 27}\%$~of the polarized emission comes from the outer 2.2 magnitudes of extinction, or the outer
$2.2/{\lmfup 20}\,=$\,{\lmfup 11}$\%$ of the cloud. A further ${\lmfup 47}\%$~of the total polarized emission comes from 
$2.2\,\mathrm{mag}\,<\,\chi\,<\,10\,\mathrm{mag}$, which accounts for {\lmfup 39}$\%$ of the dust column.  The most deeply embedded
regions of the cloud ($10\,\mathrm{mag}\,<\,\chi\,<\,{\lmfup 20\,\mathrm{mag}}$) contribute {\lmfup 50}$\%$ of 
the total dust column but only {\lmfup 27}$\%$ of the total polarized emission.  
Figure 
\ref{fig:model_rp_vs_chi} also shows that steeper power-law slopes and lower $A_{\mathrm{V crit}}$~values would imply that more of the total polarized intensity measured comes from the outer diffuse cloud layers.

To estimate the fraction of the cloud that, from the perspective of contributing to polarization, is ``hidden'' we first calculate the  $\epsilon$~weighted mean cloud depth 
$\left<\chi\right>$:
\begin{equation}
\left<\chi\right>\,=\,\frac{\int_{0}^{A_V/2}\,\chi\,\epsilon\,d \chi}{\int_{0}^{A_V/2}\,\epsilon\,d\chi}\,
=\,\frac{{\lmfup A_{\mathrm{V crit}}}}{4}\,\frac{h\left(\chi\right)}{g\left(\chi\right)},
\end{equation}
where $g\left(\chi\right)\,=\,P/(\epsilon_0 A_{\mathrm{V crit}})$ (see Equation (\ref{eqn:bigP_model_pl}))~and
$h\left(\chi\right)$
is given by
\begin{eqnarray}\label{eqn:h_chi}
h\left(\chi\right)\,
   = \left\{ \begin{array}{ll}
    a, &\mbox{for $a\,\leq\,1$;} \\
    1\,+\,\frac{2}{\zeta\,+\,1}
    \left(a^{\zeta\,+\,1}\,-\,1\right),
    & \mbox{for $a\,>\,1$}. \end{array}\right.
\end{eqnarray}

If $\epsilon$~were constant throughout the sightline then 
$\left<\chi\right>$~would equal half of the maximum value of $\chi$ giving 
$\left<\chi\right>\,=\,A_\mathrm{V}/4$.  The fraction of the cloud that is hidden can then be roughly estimated as 
\begin{equation}\label{eqn:fhidden}
f_{hidden}\,=\,1.0-\frac{\left<\chi\right>}{A_\mathrm{V}/4},
\end{equation}
which is shown in Figure \ref{fig:chibar}.  For $A_\mathrm{V}\,=\,10${\lmfup\,mag}~(assuming
our best fit parameters) only ${\lmfup 16}\%$~of the cloud is hidden.
For a sightline of $A_\mathrm{V}\,=\,{\lmfup 40}${\lmfup\,mag}~about 
{\lmfup 48}$\%$~of the cloud is hidden.  
 
In summary, for ``moderate'' dust column sightlines ($A_\mathrm{V}\,<$\,10\,{\lmfup\,mag}) our polarization measurements sample most of the cloud ($f_{\mathrm{hidden}}\,<\,{\lmfup 16}\%$).  So for sightlines with dust columns of $A_\mathrm{V}\,\sim\,10$\,{\lmfup\,mag}~or less, the \blastpol~500\,\um~measurement of the magnetic field orientation should be representative of the density-weighted average magnetic field orientation along the sightline.  For higher dust column sightlines, the fraction of the cloud that is not well sampled by our polarization measurements increases ($f_{\mathrm{hidden}}\,\sim\,{\lmfup 34}\%$~for $A_\mathrm{V}\,=\,20$\,{\lmfup\,mag})
{\lmfup and for} our highest column sightlines ($A_\mathrm{V}\,\sim\,{\lmfup 40}$\,{\lmfup\,mag}) about {\lmfup half} of the dust contributes little to the polarization measured by \blastpol. {\lmfup For these latter sightlines} \blastpol~would not be sensitive to changes in magnetic field direction in the most deeply embedded cloud material.  Recall that our model assumes that all of the decrease in $p$~with $N$~is due to lower intrinsic polarization efficiency of material deep within the cloud.  If some of the decrease in $p$~with $N$~is due to increased field {\lmfup disorder} along high $N$~sightlines then the $\epsilon$~drop-off with $\chi$~would be shallower, which would decrease the fraction of the cloud that is hidden.  

As noted earlier, our model has many implicit assumptions (slab-geometry, uniform dust temperature, power-law dependence of $\epsilon\left(\chi\right)$). 
In particular, the assumption of isothermality is clearly incorrect.  Figure \ref{fig_n_vs_t} shows that for the ISRF-heated sightlines included in this analysis the average 
temperature decreases with increasing column density (and thus increasing $A_\mathrm{V}$). For the temperature extremes of 11\,K and 15\,K of the ISRF sightlines in Figure \ref{fig_n_vs_t}, we calculate that for the 
colder highest 
column density sightlines the dust on average emits half as much radiation per unit 
mass {\lmfup at 500\um} compared with dust on the warmer more diffuse cloud sightlines.  
It is quite likely that the more deeply embedded dust grains in Vela\,C are colder, which implies they will contribute less than warmer grains near the cloud surfaces to both the total intensity and the measured polarized intensity. 
{\lmfup We therefore expect that the average magnetic field orientation inferred from polarization data will be weighted more towards the orientation in the warmer regions of the cloud.   This will increase $f_{\mathrm{ hidden}}$: even assuming uniform intrinsic polarization efficiency if the outer half of the dust grains had $T\,=$\,15\,K and the inner half of the dust grains had $T\,=$\,11\,K then we find that $f_{\mathrm{ hidden}}\,=\,20\,\%$.  

To some degree this problem can be reduced by measuring polarization at millimeter wavelengths where the intensity of thermal dust emission is less sensitive to temperature.  For detailed statistical comparisons of {\lmfup submillimeter} polarization data with synthetic observations of molecular clouds derived from numerical simulations it will be important to not only model the effects of grain alignment in simulation postprocessing but also include a realistic cloud temperature structure. Due to the aforementioned uncertainties that are related to the assumptions in our model, our values of $f_{\mathrm{hidden}}$~should be taken only as crude estimates.}  

{\lmfup Despite these uncertainties, we note that}, our {\lmfup results are} consistent with the findings of \cite{cho_lazarian_2005}, who showed that dust grains can be aligned efficiently by radiative 
torques at cloud depths $\chi$~of up to\,10~magnitudes in visual extinction.  
\cite{bethell_2007} found that the exact depth to which grains are aligned depends on grain size and on the degree of anisotropy of the local radiation field.  Our model is also consistent with recent observations by \cite{alves_2014} who argue that their 
observations of submillimeter polarization of a starless core suggest
loss of grain alignment at column densities higher than $A_\mathrm{V}\,=\,30$\,{\lmfup mag}.  
 If $\epsilon$~changes appreciably with $\chi$~in the cloud then this might be revealed in {\lmfup the} frequency dependence of $p$.  Thus studying $p$~at higher frequencies, as can be done {\lmfup using \blastpol~data}, might provide further constraints.

\subsection{Possible Causes of the Residual Structure in $p^{[N,S]}$}\label{sect:pdecorr_scatter}

In Section \ref{sect:p_N_dpsi_2corr} we showed that we can account for most of the variations in $p$~that we observe in Vela\,C with a simple two-variable power-law fit $p(N,S)$.  Here, 
we consider a number of factors besides $N$~and $S$~which could contribute to the variance in $p$.  
The dispersion in the logarithm of the decorrelated fractional polarization $\log\left(p^{[N,S]}\right)$~is 
{\lmfup 0.15}, which corresponds to 
a variance in $p^{[N,S]}$~of {\lmfup 1.0\,$\times$\,10$^{-4}$}.  

If the variance in $p$ were entirely due to the measurement uncertainty, then we would 
expect the variance in $p$~to be described by:
\begin{equation} \label{eqn:stddev_p_stat}
\sigma_{\mathrm{p\,stat}}^{2}\,=\,\sqrt{\frac{\sum_i^n{\sigma_{p_i}^2}}{n}}\,=\,
{\lmfup 1.7}\,\times\,10^{-6},
\end{equation}
where $\sigma_{p_i}^2$~is the variance for each individual $p_i$~and 
$n$~is the total number of data points.  This value for $\sigma^2_{p\,stat}$\, is much smaller than the 
measured variance in $p^{[N,S]}$.  
Measurement uncertainties thus play a minor part in the observed variance of 
$p^{[N,S]}$. 

A more likely possibility is that the variance in $p^{[N,S]}$~seen in Figure \ref{fig:p_histos} and the bottom panel of Figure \ref{fig_p_xy} is the result of changes 
in the 
direction of the magnetic field with respect to the line of sight.  The observed polarization of a population of dust grains aligned with respect to a uniform magnetic field depends on $\gamma$,~the angle between the magnetic field direction and the plane of the sky:
\begin{equation}\label{eqn:p_gamma}
p\,=\,p_{\mathrm{max}} cos^2\gamma,
\end{equation}
where $p_{max}$~is the polarization that would be observed if the magnetic field were parallel to the plane of the sky ($\gamma\,=\,0$\deg).
Our inferred magnetic field maps (Figure \ref{fig:maps_drapery}) clearly show several large scale changes in 
magnetic field direction $\Phi$.  Corresponding large scale changes in $\gamma$~would add width to the $\log p$~distribution.  
In theory a detailed statistical comparison of $S$~and $p^{[N,S]}$~on different angular scales could be used to gain insight into the three-dimensional structure of the magnetic field. However, such a treatment is beyond the scope of the present paper.

\section{Summary}\label{sect:conclusions}
In this work we present 500\,\um\ maps of the Vela\,C giant molecular cloud from the 2012 flight of \blastpol.  Our polarization maps were calculated from Stokes $I$, $Q$ and $U$~maps with background/foreground diffuse polarized emission subtracted as described in Section \ref{sect:ref_reg}. These maps were used to calculate the inferred magnetic field orientation $\Phi$\ projected onto the plane of the sky.  Overall we see a change in the magnetic field orientation across the cloud, from perpendicular to the main cloud elongation direction in the south, to nearly parallel to the cloud elongation in the north.  We also see regions 
of sharp changes in the magnetic field direction, as traced by $S$, the average angular
dispersion on scales corresponding to our beam (2\farcm5 or 0.5\,pc scales).  

As a first step in our analysis of the Vela\,C data we examine the dependence of polarization fraction $p$~as a function of column density $N$, dust temperature $T$, and 
local angular dispersion $S$ for sightlines in four of the five cloud regions defined in \cite{hill_2011}.  The goal of this work is to look for empirical trends that
can be compared to numerical simulations of molecular clouds.  These trends can be used to learn about the magnetic field properties and intrinsic polarization efficiency within the cloud.  As part of our analysis we separate our sightlines into those that appear to be primarily heated by the interstellar radiation field (ISRF) and the minority of sightlines that show evidence of heating from the compact \ion{H}{2} region \rcw.

Our main findings are as follows:
\begin{enumerate}
\item For the ISRF-heated sightlines we find that $p$~is anticorrelated with $N$~and correlated with $T$, i.e., the polarization fraction decreases with increasing column density, and increases with increasing dust temperature.  However, $N$~and $T$~are also highly anticorrelated with one another and normalizing out the power-law dependence of $p$~with $N$~removes the correlation with $T$.  {\lmfup In the absence of bright internal sources it is expected that the density structure of the cloud largely determines the observed $T$; therefore we choose $N$~as our independent variable in the subsequent analysis}. For the \rcw~heated sightlines where there is no correlation 
between $N$~and $T$, we see no correlation between $p$~and $T$ but there is still a strong anticorrelation between $p$~and $N$.
{\lmfup This suggests that 
for the \rcw -heated sightlines the important variable controlling $p$~is $N$.}  
\item We derive a two-variable power-law empirical model $p\,\propto\,N^{\alpha_N} S^{\alpha_S}$\ for the ISRF-heated sightlines, where $\alpha_N$\,=\,\alpbbNerr~and $\alpha_{S}$\,=\,\alpbbSerr.  This model can reproduce $\sim${\lmfup 66}$\%$~of the variance in $\log p$.  The decrease in $p$~with increasing $S$~is probably the result of changes in the magnetic field direction within the volume of the cloud sampled by the beam.  The decrease in $p$~with $N$~could be caused by increased {\lmfup disorder in} the magnetic field for high column density sightlines or changes in the intrinsic polarization efficiency (e.g., the fraction of aligned grains, or grain axis ratio) for deeply embedded cloud material. 
\item We do not find a strong correlation between $N$~and~$S$. {\lmfup This suggests that the disorder in the magnetic field does not increase significantly with density, which would in turn imply that the explanation for the decrease of $p$~with increasing $N$~is reduced intrinsic polarization efficiency for high $N$~sightlines.  However, this might not be the case.  It might be that there is more disorder in the magnetic field towards higher column density sightlines, but this disorder occurs on much smaller scales than 0.5\,pc, the scale probed by $S$, such that $S$~is not sensitive to the disordered magnetic field component.} 
\item As a limiting case we consider the implications if the decrease in $p$~with increasing $N$~is due solely to reduced intrinsic polarization efficiency along high column density sightlines.  In this case our \blastpol~measurements of the magnetic field orientation $\Phi$~would preferentially sample the material closer to the surface of the cloud and be less sensitive to changes in the field direction in the highly extinguished regions deep within the cloud.
{\lmfup We introduce a crude model in which the intrinsic polarization efficiency is uniform in the outer cloud layers and then drops with a power-law dependence on the parameterized cloud depth $\chi$.  From a fit of our observational data to this crude model we conclude that for sightlines having  $A_\mathrm{V}$\,$<$\,10\,mag, $\Phi$~is a reasonable measure of the average magnetic field direction along the line of sight, but for sightlines of $A_V$\,=\,40\,mag, much of the cloud (roughly the inner half) is not well traced by $\Phi$.}
This model might be a ``worst-case'' scenario because some of the decrease of $p$~with $A_\mathrm{V}$~could arise from effects of magnetic field geometry not included in the model (e.g., more structure in the magnetic field along high column density sightlines).  \item The remaining scatter in $p^{[N,S]}$, the polarization fraction with our derived power-law dependence on $S$~and $N$~normalized out, is too large to be explained by our measurement uncertainties in $p$, but could
be explained by variations in the angle of the magnetic field with respect to the plane of the sky.

\end{enumerate}
{\lmfup In this paper we have examined polarization trends for only one cloud.  Other clouds with different properties, in particular different average angle $\gamma$~of the magnetic field with respect
to the plane of the sky, might show different trends. 
To better constrain numerical simulations of star
formation, our two-variable power-law fit should be repeated for a wide variety of clouds, which will 
presumably encompass a range of $\gamma$~values.
Our study provides constraints for 
numerical simulations of molecular clouds; for at least one assumed value of $\gamma$~synthetic polarization observations of the simulations should be able
to reproduce (a) our two-variable power-law fit exponents and (b) the lack of correlation between $N$~and $S$~(on 0.5\,pc scales).   }

\acknowledgments
{\lmfup The authors like to thank the referee for a detailed and thoughtful review that has helped to strengthen the paper.} The \blastpol~collaboration acknowledges support from NASA (through grant numbers NAG5-12785, NAG5-13301, NNGO-6GI11G,  NNX0-9AB98G, and the Illinois Space Grant Consortium), the Canadian Space Agency (CSA), the Leverhulme Trust through the Research Project Grant F/00 407/BN, the Natural Sciences and Engineering Research Council (NSERC) of Canada, the Canada Foundation for Innovation, the Ontario Innovation Trust, the Dunlap Institute for Astronomy \& Astrophysics, and the US National Science Foundation Office of Polar Programs.  L.M.F. was supported in part by an NSERC Postdoctoral Fellowship.  B. D. is supported through a NASA Earth and Space Science Fellowship.  C.B.N. also acknowledges support from the Canadian Institute for Advanced Research. F.P.S. is supported by the CAPES grant 2397/13-7.  Z.-Y.L. is supported in part by NSF AST1313083 and NASA NNX14AB38G. F.P. thanks the Spanish Ministry of Economy and Competitiveness (MINECO) under the Consolider-Ingenio project CSD2010-00064. The authors would also like to thank Diego Falceta-Gon{\c c}alves for making available line integral convolution code, which was used in making the drapery image shown in Figure \ref{fig:maps_drapery}. Finally, we thank the Columbia Scientific Balloon Facility (CSBF) staff for their outstanding work.

\appendix

\section{Detailed Description of Null Tests}\label{sect:null_append}
We restrict consideration to the 250\,\um\ maps for the purpose of the null tests because the larger number of detectors in this band allows coverage of the map area to remain complete even when splitting the data set in two. Furthermore, if the asymmetric beam shape is a significant source of systematics, these will be manifested more strongly at 250\,\um, as the beam in this band is the least symmetric of the three. We would expect any regions that pass the null tests at 250\,\um\ will also pass in the longer-wavelength bands. 

TODs were split into single raster scans, representing one complete raster of \blastpol~over the target map area at one half-wave plate position.
Once the total data set was segmented using one of the four criteria described in Section \ref{sect:null_tests}, separate maps of Stokes $I$, $Q$, and $U$~were made with {\tt TOAST}~{\lmfup(Section \ref{sect:toast})} for each of the two categories. The diffuse emission removal described in Section \ref{sect:ref_reg} was then performed for each map. For each null test criterion, we examined three metrics for evaluating systematic disagreement between the data segments:
\begin{enumerate}
\item	Polarization fraction ($p$): Independent maps of p were produced using the $I$, $Q$, and $U$~maps from each half of the data ($p_A$~and $p_B$, where “A” and “B” generically represent the left and right sets, the early and late sets, etc.) A $p$~residual map, $\Delta p$~was calculated where $\Delta p = (p_A - p_B)/2$, the absolute value of which is absolute separation of each of $p_A$~and $p_B$~from the mean $p$, $(p_A + p_B)/2$.
{\lmfup The quantity} $\Delta p$ was taken to represent the uncertainty in $p$ due to systematic sources of error, and we looked for regions in the full-data map where the calculated p is greater than $3\Delta p$~for each of the 4 null tests described in Section \ref{sect:null_tests}.
\item	Polarization angle ($\psi$): Analogously, two independent maps of $\psi$~were calculated for each of the null tests (again, $\psi_A$ and $\psi_B$, generically). Because a polarization measurement with 3$\sigma$~confidence in $Q$~and $U$~has an uncertainty in $\psi$~of about $10^{\circ}$, we looked for regions in the full-data map where the absolute difference between $\psi_A$~and $\psi_B$ was less than $20^{\circ}$. This standard is equivalent to requiring that the polarization-angle from each of the two data halves be consistent with the mean of $\psi_A$ and $\psi_B$.
\item	Polarized intensity $(P = \sqrt{Q^2 + U^2})$: To examine systematic errors in $P$, we reproduced the procedure described in Section 3.4 of \cite{matthews_2014}. Briefly,  residual maps of $Q$~and $U$~were calculated as the difference between that parameter and its average value in the null test data halves. The $Q$~and $U$~residuals were then used to form a $P_{\mathrm{res}} = \sqrt{Q_{\mathrm{res}}^2 + U_{\mathrm{res}}^2}$. As in the $p$~metric described above, $P_{\mathrm{res}}$~was taken as the systematic uncertainty in $P$, and we required the full-data measurement of $P$~to be greater than $3 P_{\mathrm{res}}$.
\end{enumerate}

\section{Polarization Conventions}\label{sect:pol_conventions}

In this paper we discuss the polarized component of the 
dust emission ($P$) and the fractional polarization of dust emission ($p$), both of which can
be derived from the linear polarization Stokes parameters:
\begin{equation}
 P\,=\,\sqrt{Q^2\,+U^2},
\end{equation}
and 
\begin{equation}
 p\,=\,\frac{P}{I}.
\end{equation}

The associated angle of the polarization $\psi$~is
\begin{equation}
 \psi\,=\,\frac{1}{2} \arctan\left(U,Q\right),
\end{equation}
where the two argument arctan function computes $\arctan(Q/U)$~while avoiding the ambiguity when $Q$\,=\,0 MJy\,sr$^{-1}$.
{\lmfup The polarization angle} $\psi$~is defined from -90\deg~to 90\deg.  Our conventions for $Q$\ and 
$U$~are such that 0\deg~corresponds to North in Galactic coordinates and $\psi$~
increases East of
{\lmfup Galactic} North (counterclockwise).  This follows the IAU conventions 
\citep{hamaker_1996}, but differs from the {\tt HEALPix}\footnote{\url{http://healpix.jpl.nasa.gov}; \cite{gorski_2005}.} convention adopted for Planck data, where {\lmfup $\psi$~increases West of Galactic North}~
(clockwise).  The apparent angle of 
the magnetic field projected on to the plane of the sky $\Phi$ is
\begin{equation}
\Phi\,=\,\psi\,+\frac{\pi}{2}.
\end{equation}
It is important to note that $\Phi$ is {\lmfup a} tracer of the cloud magnetic field direction
that is weighted by the efficiency of polarized dust emission averaged over the \blastpol\ 
beam and along the line of sight. 

The variances of $P$, $p$~and $\psi$ are defined in \cite{planck2014-XIX} (Equations (B2)- (B4)).
\bibliography{ms}

\newcommand{\noop}[1]{}
\begin{thebibliography}{}
\expandafter\ifx\csname natexlab\endcsname\relax\def\natexlab#1{#1}\fi

\bibitem[{{Alves} {et~al.}(2014){Alves}, {Frau}, {Girart}, {Franco}, {Santos},
  \& {Wiesemeyer}}]{alves_2014}
{Alves}, F.~O., {Frau}, P., {Girart}, J.~M., {et~al.} 2014, \aap, 569, L1

\bibitem[{{Andersson} {et~al.}(2015){Andersson}, {Lazarian}, \&
  {Vaillancourt}}]{andersson_2015}
{Andersson}, B.-G., {Lazarian}, A., \& {Vaillancourt}, J.~E. 2015, \araa, 53,
  501

\bibitem[{{Arce} {et~al.}(1998){Arce}, {Goodman}, {Bastien}, {Manset}, \&
  {Sumner}}]{arce_1998}
{Arce}, H.~G., {Goodman}, A.~A., {Bastien}, P., {Manset}, N., \& {Sumner}, M.
  1998, \apjl, 499, L93

\bibitem[{{Bethell} {et~al.}(2007){Bethell}, {Chepurnov}, {Lazarian}, \&
  {Kim}}]{bethell_2007}
{Bethell}, T.~J., {Chepurnov}, A., {Lazarian}, A., \& {Kim}, J. 2007, \apj,
  663, 1055

\bibitem[{{Bohlin} {et~al.}(1978){Bohlin}, {Savage}, \& {Drake}}]{bohlin_1978}
{Bohlin}, R.~C., {Savage}, B.~D., \& {Drake}, J.~F. 1978, \apj, 224, 132

\bibitem[{Cabral \& Leedom(1993)}]{cabral_1993}
Cabral, B., \& Leedom, L.~C. 1993, in Proceedings of the 20th Annual Conference
  on Computer Graphics and Interactive Techniques, SIGGRAPH '93 (New York, NY,
  USA: ACM), 263--270

\bibitem[{{Cantalupo} {et~al.}(2010){Cantalupo}, {Borrill}, {Jaffe}, {Kisner},
  \& {Stompor}}]{cantalupo2010}
{Cantalupo}, C.~M., {Borrill}, J.~D., {Jaffe}, A.~H., {Kisner}, T.~S., \&
  {Stompor}, R. 2010, \apjs, 187, 212

\bibitem[{{Cashman} \& {Clemens}(2014)}]{cashman_2014}
{Cashman}, L.~R., \& {Clemens}, D.~P. 2014, \apj, 793, 126

\bibitem[{{Chapman} {et~al.}(2011){Chapman}, {Goldsmith}, {Pineda}, {Clemens},
  {Li}, \& {Kr{\v c}o}}]{chapman_2011}
{Chapman}, N.~L., {Goldsmith}, P.~F., {Pineda}, J.~L., {et~al.} 2011, \apj,
  741, 21

\bibitem[{{Chauvenet}(1863)}]{chauvenet_1863}
{Chauvenet}, W. 1863, {A manual of spherical and practical astronomy}

\bibitem[{{Cho} \& {Lazarian}(2005)}]{cho_lazarian_2005}
{Cho}, J., \& {Lazarian}, A. 2005, \apj, 631, 361

\bibitem[{{Crutcher}(2012)}]{crutcher_2012}
{Crutcher}, R.~M. 2012, \araa, 50, 29

\bibitem[{{Dobashi} {et~al.}(2005){Dobashi}, {Uehara}, {Kandori}, {Sakurai},
  {Kaiden}, {Umemoto}, \& {Sato}}]{dobashi_2005}
{Dobashi}, K., {Uehara}, H., {Kandori}, R., {et~al.} 2005, \pasj, 57, 1

\bibitem[{{Dotson}(1996)}]{dotson_1996}
{Dotson}, J.~L. 1996, \apj, 470, 566

\bibitem[{{Falceta-Gon{\c c}alves} {et~al.}(2008){Falceta-Gon{\c c}alves},
  {Lazarian}, \& {Kowal}}]{fg_2008}
{Falceta-Gon{\c c}alves}, D., {Lazarian}, A., \& {Kowal}, G. 2008, \apj, 679,
  537

\bibitem[{{Galitzki} {et~al.}(2014){Galitzki}, {Ade}, {Angil{\`e}}, {Benton},
  {Devlin}, {Dober}, {Fissel}, {Fukui}, {Gandilo}, {Klein}, {Korotkov},
  {Matthews}, {Moncelsi}, {Netterfield}, {Novak}, {Nutter}, {Pascale},
  {Poidevin}, {Savini}, {Scott}, {Shariff}, {Soler}, {Tucker}, {Tucker}, \&
  {Ward-Thompson}}]{galitzki_2014}
{Galitzki}, N., {Ade}, P.~A.~R., {Angil{\`e}}, F.~E., {et~al.} 2014, in Society
  of Photo-Optical Instrumentation Engineers (SPIE) Conference Series, Vol.
  9145, Society of Photo-Optical Instrumentation Engineers (SPIE) Conference
  Series, 0

\bibitem[{{Gandilo} {et~al.}(2016){Gandilo}, {Ade}, {Angil{\`e}}, {Ashton},
  {Benton}, {Devlin}, {Dober}, {Fissel}, {Fukui}, {Galitzki}, {Klein},
  {Korotkov}, {Li}, {Martin}, {Matthews}, {Moncelsi}, {Nakamura},
  {Netterfield}, {Novak}, {Pascale}, {Poidevin}, {Santos}, {Savini}, {Scott},
  {Shariff}, {Diego Soler}, {Thomas}, {Tucker}, {Tucker}, \&
  {Ward-Thompson}}]{gandilo_2016}
{Gandilo}, N.~N., {Ade}, P.~A.~R., {Angil{\`e}}, F.~E., {et~al.} 2016, \apj, in
  press, arXiv:1512.06745

\bibitem[{{Gerakines} {et~al.}(1995){Gerakines}, {Whittet}, \&
  {Lazarian}}]{gerakines_1995}
{Gerakines}, P.~A., {Whittet}, D.~C.~B., \& {Lazarian}, A. 1995, \apjl, 455,
  L171

\bibitem[{{Goodman} {et~al.}(1995){Goodman}, {Jones}, {Lada}, \&
  {Myers}}]{goodman_1995}
{Goodman}, A.~A., {Jones}, T.~J., {Lada}, E.~A., \& {Myers}, P.~C. 1995, \apj,
  448, 748

\bibitem[{{G{\'o}rski} {et~al.}(2005){G{\'o}rski}, {Hivon}, {Banday},
  {Wandelt}, {Hansen}, {Reinecke}, \& {Bartelmann}}]{gorski_2005}
{G{\'o}rski}, K.~M., {Hivon}, E., {Banday}, A.~J., {et~al.} 2005, \apj, 622,
  759

\bibitem[{{Greenberg}(1968)}]{greenberg_1968}
{Greenberg}, J.~M. 1968, {Interstellar Grains}, ed. B.~M. {Middlehurst} \&
  L.~H. {Aller} (the University of Chicago Press), 221--30

\bibitem[{{Griffin} {et~al.}(2002){Griffin}, {Bock}, \& {Gear}}]{grif02}
{Griffin}, M.~J., {Bock}, J.~J., \& {Gear}, W.~K. 2002, \ao, 41, 6543

\bibitem[{{Griffin} {et~al.}(2003){Griffin}, {Swinyard}, \& {Vigroux}}]{grif03}
{Griffin}, M.~J., {Swinyard}, B.~M., \& {Vigroux}, L.~G. 2003, in IR Space
  Telescopes and Instruments. Edited by John C. Mather . Proceedings of the
  SPIE, Vol. 4850, 686--697

\bibitem[{{Hall}(1949)}]{hall_1949}
{Hall}, J.~S. 1949, Science, 109, 166

\bibitem[{{Hamaker} \& {Bregman}(1996)}]{hamaker_1996}
{Hamaker}, J.~P., \& {Bregman}, J.~D. 1996, \aaps, 117, 161

\bibitem[{{Heiles}(2000)}]{heiles_2000}
{Heiles}, C. 2000, \aj, 119, 923

\bibitem[{{Hennebelle} \& {Andr{\'e}}(2013)}]{hennebelle_2013}
{Hennebelle}, P., \& {Andr{\'e}}, P. 2013, \aap, 560, A68

\bibitem[{{Hildebrand}(1983)}]{hildebrand_1983}
{Hildebrand}, R.~H. 1983, \qjras, 24, 267

\bibitem[{{Hildebrand}(1988)}]{hildebrand_1988}
---. 1988, \qjras, 29, 327

\bibitem[{{Hill} {et~al.}(2010){Hill}, {Longmore}, {Pinte}, {Cunningham},
  {Burton}, \& {Minier}}]{hill_2010}
{Hill}, T., {Longmore}, S.~N., {Pinte}, C., {et~al.} 2010, \mnras, 402, 2682

\bibitem[{{Hill} {et~al.}(2009){Hill}, {Pinte}, {Minier}, {Burton}, \&
  {Cunningham}}]{hill_2009}
{Hill}, T., {Pinte}, C., {Minier}, V., {Burton}, M.~G., \& {Cunningham}, M.~R.
  2009, \mnras, 392, 768

\bibitem[{{Hill} {et~al.}(2011){Hill}, {Motte}, {Didelon}, {Bontemps},
  {Minier}, {Hennemann}, {Schneider}, {Andr{\'e}}, {Men'shchikov}, {Anderson},
  {Arzoumanian}, {Bernard}, {di Francesco}, {Elia}, {Giannini}, {Griffin},
  {K{\"o}nyves}, {Kirk}, {Marston}, {Martin}, {Molinari}, {Nguyen Luong},
  {Peretto}, {Pezzuto}, {Roussel}, {Sauvage}, {Sousbie}, {Testi},
  {Ward-Thompson}, {White}, {Wilson}, \& {Zavagno}}]{hill_2011}
{Hill}, T., {Motte}, F., {Didelon}, P., {et~al.} 2011, \aap, 533, A94

\bibitem[{{Hiltner}(1949)}]{hiltner_1949}
{Hiltner}, W.~A. 1949, Science, 109, 165

\bibitem[{{Jones}(1989)}]{jones_1989}
{Jones}, T.~J. 1989, \apj, 346, 728

\bibitem[{{Jones} {et~al.}(2015){Jones}, {Bagley}, {Krejny}, {Andersson}, \&
  {Bastien}}]{jones_2015}
{Jones}, T.~J., {Bagley}, M., {Krejny}, M., {Andersson}, B.-G., \& {Bastien},
  P. 2015, \aj, 149, 31

\bibitem[{{Kowal} {et~al.}(2007){Kowal}, {Lazarian}, \&
  {Beresnyak}}]{kowal_2007}
{Kowal}, G., {Lazarian}, A., \& {Beresnyak}, A. 2007, \apj, 658, 423

\bibitem[{{Lazarian}(2007)}]{lazarian_2007}
{Lazarian}, A. 2007, \jqsrt, 106, 225

\bibitem[{{Lazarian} \& {Hoang}(2007)}]{lazarian_hoang_2007}
{Lazarian}, A., \& {Hoang}, T. 2007, \mnras, 378, 910

\bibitem[{{Li} {et~al.}(2006){Li}, {Griffin}, {Krejny}, {Novak}, {Loewenstein},
  {Newcomb}, {Calisse}, \& {Chuss}}]{li_2006}
{Li}, H., {Griffin}, G.~S., {Krejny}, M., {et~al.} 2006, \apj, 648, 340

\bibitem[{{Liseau} {et~al.}(1992){Liseau}, {Lorenzetti}, {Nisini}, {Spinoglio},
  \& {Moneti}}]{liseau_1992}
{Liseau}, R., {Lorenzetti}, D., {Nisini}, B., {Spinoglio}, L., \& {Moneti}, A.
  1992, \aap, 265, 577

\bibitem[{{Matthews} {et~al.}(2002){Matthews}, {Fiege}, \&
  {Moriarty-Schieven}}]{matthews_2002}
{Matthews}, B.~C., {Fiege}, J.~D., \& {Moriarty-Schieven}, G. 2002, \apj, 569,
  304

\bibitem[{{Matthews} {et~al.}(2001){Matthews}, {Wilson}, \&
  {Fiege}}]{matthews_2001}
{Matthews}, B.~C., {Wilson}, C.~D., \& {Fiege}, J.~D. 2001, \apj, 562, 400

\bibitem[{{Matthews} {et~al.}(2014){Matthews}, {Ade}, {Angil{\`e}}, {Benton},
  {Chapin}, {Chapman}, {Devlin}, {Fissel}, {Fukui}, {Gandilo}, {Gundersen},
  {Hargrave}, {Klein}, {Korotkov}, {Moncelsi}, {Mroczkowski}, {Netterfield},
  {Novak}, {Nutter}, {Olmi}, {Pascale}, {Poidevin}, {Savini}, {Scott},
  {Shariff}, {Soler}, {Tachihara}, {Thomas}, {Truch}, {Tucker}, {Tucker}, \&
  {Ward-Thompson}}]{matthews_2014}
{Matthews}, T.~G., {Ade}, P.~A.~R., {Angil{\`e}}, F.~E., {et~al.} 2014, \apj,
  784, 116

\bibitem[{{McKee} \& {Ostriker}(2007)}]{mckee_2007}
{McKee}, C.~F., \& {Ostriker}, E.~C. 2007, \araa, 45, 565

\bibitem[{{Moncelsi} {et~al.}(2014){Moncelsi}, {Ade}, {Angil{\`e}}, {Benton},
  {Devlin}, {Fissel}, {Gandilo}, {Gundersen}, {Matthews}, {Netterfield},
  {Novak}, {Nutter}, {Pascale}, {Poidevin}, {Savini}, {Scott}, {Soler},
  {Spencer}, {Truch}, {Tucker}, \& {Zhang}}]{moncelsi_2014}
{Moncelsi}, L., {Ade}, P.~A.~R., {Angil{\`e}}, F.~E., {et~al.} 2014, \mnras,
  437, 2772

\bibitem[{{Montier} {et~al.}(2015){Montier}, {Plaszczynski}, {Levrier},
  {Tristram}, {Alina}, {Ristorcelli}, \& {Bernard}}]{montier_2015}
{Montier}, L., {Plaszczynski}, S., {Levrier}, F., {et~al.} 2015, \aap, 574,
  A135

\bibitem[{{Mouschovias} \& {Ciolek}(1999)}]{mous_1999}
{Mouschovias}, T.~C., \& {Ciolek}, G.~E. 1999, in NATO Advanced Science
  Institutes (ASI) Series C, Vol. 540, NATO Advanced Science Institutes (ASI)
  Series C, ed. C.~J. {Lada} \& N.~D. {Kylafis}, 305

\bibitem[{{Murphy} \& {May}(1991)}]{murphy_1991}
{Murphy}, D.~C., \& {May}, J. 1991, \aap, 247, 202

\bibitem[{{Myers} {et~al.}(2014){Myers}, {Klein}, {Krumholz}, \&
  {McKee}}]{myers_2014}
{Myers}, A.~T., {Klein}, R.~I., {Krumholz}, M.~R., \& {McKee}, C.~F. 2014,
  \mnras, 439, 3420

\bibitem[{{Netterfield} {et~al.}(2009){Netterfield}, {Ade}, {Bock}, {Chapin},
  {Devlin}, {Griffin}, {Gundersen}, {Halpern}, {Hargrave}, {Hughes}, {Klein},
  {Marsden}, {Martin}, {Mauskopf}, {Olmi}, {Pascale}, {Patanchon}, {Rex},
  {Roy}, {Scott}, {Semisch}, {Thomas}, {Truch}, {Tucker}, {Tucker}, {Viero}, \&
  {Wiebe}}]{netterfield_2009}
{Netterfield}, C.~B., {Ade}, P.~A.~R., {Bock}, J.~J., {et~al.} 2009, \apj, 707,
  1824

\bibitem[{{Novak} {et~al.}(2009){Novak}, {Dotson}, \& {Li}}]{novak_2009}
{Novak}, G., {Dotson}, J.~L., \& {Li}, H. 2009, \apj, 695, 1362

\bibitem[{{Palmeirim} {et~al.}(2013){Palmeirim}, {Andr{\'e}}, {Kirk},
  {Ward-Thompson}, {Arzoumanian}, {K{\"o}nyves}, {Didelon}, {Schneider},
  {Benedettini}, {Bontemps}, {Di Francesco}, {Elia}, {Griffin}, {Hennemann},
  {Hill}, {Martin}, {Men'shchikov}, {Molinari}, {Motte}, {Nguyen Luong},
  {Nutter}, {Peretto}, {Pezzuto}, {Roy}, {Rygl}, {Spinoglio}, \&
  {White}}]{palm_2013}
{Palmeirim}, P., {Andr{\'e}}, P., {Kirk}, J., {et~al.} 2013, \aap, 550, A38

\bibitem[{{Pascale} {et~al.}(2008){Pascale}, {Ade}, {Bock}, {Chapin}, {Chung},
  {Devlin}, {Dicker}, {Griffin}, {Gundersen}, {Halpern}, {Hargrave}, {Hughes},
  {Klein}, {MacTavish}, {Marsden}, {Martin}, {Martin}, {Mauskopf},
  {Netterfield}, {Olmi}, {Patanchon}, {Rex}, {Scott}, {Semisch}, {Thomas},
  {Truch}, {Tucker}, {Tucker}, {Viero}, \& {Wiebe}}]{pascale_2008}
{Pascale}, E., {Ade}, P.~A.~R., {Bock}, J.~J., {et~al.} 2008, \apj, 681, 400

\bibitem[{{\sorthelp{Planck Collaboration IntS}}{Planck Collaboration Int.
  XIX}(2015)}]{planck2014-XIX}
{\sorthelp{Planck Collaboration IntS}}{Planck Collaboration Int. XIX}. 2015,
  \aap, 576, A104

\bibitem[{{\sorthelp{Planck Collaboration IntT}}{Planck Collaboration Int.
  XX}(2015)}]{planck2014-XX}
{\sorthelp{Planck Collaboration IntT}}{Planck Collaboration Int. XX}. 2015,
  \aap, 576, A105

\bibitem[{{\sorthelp{Planck Collaboration IntZH}}{Planck Collaboration Int.
  XXXIII}(2016)}]{planck2014-XXXIII}
{\sorthelp{Planck Collaboration IntZH}}{Planck Collaboration Int. XXXIII}.
  2016, \aap, in press, arXiv:1411.2271

\bibitem[{{\sorthelp{Planck Collaboration IntZJ}}{Planck Collaboration Int.
  XXXV}(2016)}]{planck2015-XXXV}
{\sorthelp{Planck Collaboration IntZJ}}{Planck Collaboration Int. XXXV}. 2016,
  \aap, in press, arXiv:1502.04123

\bibitem[{{Press} {et~al.}(1992){Press}, {Teukolsky}, {Vetterling}, \&
  {Flannery}}]{nr_boot}
{Press}, W.~H., {Teukolsky}, S.~A., {Vetterling}, W.~T., \& {Flannery}, B.~P.
  1992, in Numerical Recipes in C: The Art of Scientific Computing, Second
  Edition (Cambridge: Cambridge University Press)

\bibitem[{{Roussel}(2013)}]{roussel_2013}
{Roussel}, H. 2013, \pasp, 125, 1126

\bibitem[{{Roy} {et~al.}(2011){Roy}, {Ade}, {Bock}, {Brunt}, {Chapin},
  {Devlin}, {Dicker}, {France}, {Gibb}, {Griffin}, {Gundersen}, {Halpern},
  {Hargrave}, {Hughes}, {Klein}, {Marsden}, {Martin}, {Mauskopf},
  {Netterfield}, {Olmi}, {Patanchon}, {Rex}, {Scott}, {Semisch}, {Truch},
  {Tucker}, {Tucker}, {Viero}, \& {Wiebe}}]{roy_2011}
{Roy}, A., {Ade}, P.~A.~R., {Bock}, J.~J., {et~al.} 2011, \apj, 730, 142

\bibitem[{{Soler} {et~al.}(2013){Soler}, {Hennebelle}, {Martin},
  {Miville-Desch{\^e}nes}, {Netterfield}, \& {Fissel}}]{soler_2013}
{Soler}, J.~D., {Hennebelle}, P., {Martin}, P.~G., {et~al.} 2013, \apj, 774,
  128

\bibitem[{{Vaillancourt} \& {Matthews}(2012)}]{vail_2012}
{Vaillancourt}, J.~E., \& {Matthews}, B.~C. 2012, \apjs, 201, 13

\bibitem[{{Wardle} \& {Kronberg}(1974)}]{wardle_1974}
{Wardle}, J.~F.~C., \& {Kronberg}, P.~P. 1974, \apj, 194, 249

\bibitem[{{Whittet} {et~al.}(2008){Whittet}, {Hough}, {Lazarian}, \&
  {Hoang}}]{whittet_2008}
{Whittet}, D.~C.~B., {Hough}, J.~H., {Lazarian}, A., \& {Hoang}, T. 2008, \apj,
  674, 304

\bibitem[{{Yamaguchi} {et~al.}(1999){Yamaguchi}, {Mizuno}, {Saito},
  {Matsunaga}, {Mizuno}, {Ogawa}, \& {Fukui}}]{yamaguchi_1999}
{Yamaguchi}, N., {Mizuno}, N., {Saito}, H., {et~al.} 1999, \pasj, 51, 775

\end{thebibliography}

\end{document}